\documentclass[aps,rmp,preprintnumbers,reprint,amsmath,amssymb,graphicx,
longbibliography]{revtex4-1}

\usepackage{bm}
\usepackage{graphicx}
\usepackage{color}
\usepackage{cancel}
\usepackage{slashed}
\usepackage{enumitem}

\newcommand{\lsi}{\raise0.3ex\hbox{$<$\kern-0.75em\raise-1.1ex\hbox{$\sim$}}}
\newcommand{\gsi}{\raise0.3ex\hbox{$>$\kern-0.75em\raise-1.1ex\hbox{$\sim$}}}
\newcommand{\lsim}{\mathop{\lsi}}
\newcommand{\gsim}{\mathop{\gsi}}

\newcommand{\ve}{\varepsilon}

\newcommand{\f}{\phi}
\newcommand{\af}{\bar{\phi}}
\newcommand{\al}{\bar{l}}
\newcommand{\eq}{\text{eq}}

\newcommand{\mt}{\widetilde{m}_1}                  
\newcommand{\mb}{\overline{m}}
\newcommand{\mpl}{M _ { \rm Pl } } 
\newcommand{\BL}{$B\!-\!L$}

\newcommand{\M}{\mathcal{M}}

\renewcommand{\vec}[1]{{\bf #1}}
\newcommand{\vol}{V}
\def\gsim{\mbox{~{\raisebox{0.4ex}{$>$}}\hspace{-1.1em}
        {\raisebox{-0.6ex}{$\sim$}}~}}
\def\lsim{\mbox{~{\raisebox{0.4ex}{$<$}}\hspace{-1.1em}
        {\raisebox{-0.6ex}{$\sim$}}~}}
\newcommand{\omspec}{\omega  _ { \rm spec }}

\newcommand{\Gsph}{\Gamma _ { \rm sph } } 
\newcommand{\higgs}{\phi  }
\newcommand{\nfam}{n_f}
\newcommand{\NCS}{N _ { \rm CS } }

\newcommand{\gw}{\gamma  _ { \rm w} }
\newcommand{\lw}{\ell _ { \rm w} }
\newcommand{\vw}{v _ { \rm w} }
\newcommand{\ea}{{\it et al.}}

\begin{document}

\title{Baryogenesis from the weak scale to the grand unification scale}

\preprint{DESY 20-141}

\author{Dietrich B\"odeker}
\affiliation{Fakult\"at f\"ur Physik, Universit\"at Bielefeld,  33501 Bielefeld, 
Germany}
\author{Wilfried Buchm\"uller} 
\affiliation{Deutsches Elektronen-Synchrotron DESY, 22607 Hamburg, Germany}


\begin{abstract}
The current status of baryogenesis is reviewed, with an emphasis on
electroweak baryogenesis and leptogenesis. The first detailed studies
were carried out for SU(5) grand unified theory (GUT) models where $CP$-violating decays of
leptoquarks generate a baryon asymmetry. These GUT models were excluded
by the discovery of unsuppressed, $(B+L)$-violating sphaleron processes
at high temperatures. Yet a new possibility emerged: electroweak
baryogenesis. Here sphaleron processes generate a baryon asymmetry during
a strongly first-order phase transition. This mechanism has been studied in
detail in many extensions of the standard model. However, constraints
from the LHC and from low-energy precision experiments exclude most of the
known models, leaving composite Higgs models of electroweak symmetry
breaking as an interesting possibility. Sphaleron processes are also
the basis of leptogenesis, where $CP$-violating decays of heavy
right-handed neutrinos generate
a lepton asymmetry that is partially converted to a baryon asymmetry.
This mechanism is closely related to that of GUT baryogenesis, and simple
estimates based on GUT models can explain the order of magnitude of
the observed baryon-to-photon ratio. In the one-flavor approximation
an upper bound on the light-neutrino masses has been derived that is consistent with the cosmological
upper bound on the sum of neutrino masses. For quasidegenerate
right-handed neutrinos the leptogenesis temperature can be lowered
from the GUT scale down to the weak scale, and $CP$-violating oscillations
of GeV sterile neutinos can also lead to successful leptogenesis.
Significant progress has been made in
developing a full field-theoretical description of thermal
leptogenesis, which demonstrated that interactions with gauge bosons of the thermal
plasma play a crucial role. Finally, recent ideas on how the
seesaw mechanism and $B-L$ breaking at the GUT scale can be probed by
gravitational waves are discussed.
\end{abstract}


\maketitle

\tableofcontents{}

\section{Introduction}
\label{sec:intro}
The current theory of particle physics, the standard model (SM), is a
low-energy effective theory that is valid at the Fermi scale of weak
interactions $\Lambda_\text{EW} \sim
100~\text{GeV}$. Theoretical ideas beyond the SM extend up to the
scale of grand unified theories (GUTs) $(\Lambda_\text{GUT} \sim 10^{15}~\text{GeV})$,
possibly including new gauge interactions at intermediate scales and 
supersymmetry. Once quantum gravity effects are relevant, the
Planck scale and the string scale also enter. At the LHC, the SM
has been tested up to TeV~energies, with no hint of new particles
or interactions. Thus far the only evidence for physics beyond the SM
is nonzero neutrino masses that are deduced from neutrino
oscillations, and that can be explained by extensions of the SM
ranging from the weak scale to the GUT scale. Moreover, there is
evidence for dark matter and dark energy that, however, might have a purely
gravitational origin.

During the past 40 years impressive progress has been made in
early Universe cosmology, which is closely related to particle
physics. This has 
led to a standard model of cosmology with the key elements of inflation,
baryogenesis, dark matter, and dark energy. However, the associated
energy scales are uncertain. The energy density during the
inflationary phase can range from the scale of strong interactions
to the GUT scale,
dark matter particles are
considered with masses between $10^{-22}~\text{eV}$ and
$10^{\text{18}}~\text{GeV}$, dark energy may simply be a cosmological
constant constrained by anthropic considerations,
and the energy scale of baryogenesis can vary between the scale
of strong interactions
and the GUT scale.

This review is concerned with a single number, the ratio of the number
density of baryons to photons in the universe, which has been measured
most precisely in the cosmic microwave backgound (CMB) 
\citep{Planck:2018nkj}:
\begin{equation}
  \eta_B \equiv \frac{n_B}{n_\gamma} = (6.12 \pm 0.04)\times 10^{-10}\ ,
\end{equation}
which is consistent with the most recent analysis of primordial
nucleosynthesis (except for the ``lithium problem'')
\citep{Fields:2019pfx}.
Since the existence of antimatter in the Universe is excluded by the
diffuse $\gamma$-ray background \citep*{Cohen:1997ac}, the ratio $\eta_B$ is also
a measure of the matter-antimatter asymmetry:
\begin{equation}
\frac{n_B-n_{\bar{B}}}{n_\gamma} =\frac{n_B}{n_\gamma} =\eta_B \ .
\end{equation}
From the seminal work of \citet{Sakharov:1967dj} we know that
the baryon asymmetry can be
generated by physical processes and that it is related to the
violation of $CP$, the product of charge conjugation ($C$) and space
reflection ($P$), and to baryon-number violation in the fundamental theory. 

Our knowledge about the early Universe rests on only a few numbers: the
abundances of light elements (explained by nucleosynthesis), amplitude, and
slope of the scalar power spectrum of density fluctuations and
the tensor-to-scalar ratio (determined by the CMB), and the
contributions of dark energy,  matter, and baryonic matter to the
energy density of the Universe, which, normalized to the critical
energy $\rho_c = 3H_0^2/(8\pi G)$, read\footnote{The Hubble parameter is determined
  as $H_0 = 67.36 \pm 0.54 \text{km~s$^{-1}$~Mpc$^{-1}$} \equiv
    h \times 100~\text{km~s$^{-1}$~Mpc$^{-1}$}$ \cite{Planck:2018vyg};
    in a flat universe, as
      predicted by inflation, one has $\Omega_{\Lambda} + \Omega_m = 1$.} 
$\Omega_{\Lambda} = 0.6847
\pm 0.0073$, $\Omega_{m}h^2 = 0.1428 \pm 0.0011$, and
$\Omega_bh^2 = 0.02237 \pm 0.00015$ \citep{Planck:2018vyg},
with $\eta_B = 2.74 \times 10^{-8}~\Omega_bh^2$ \cite{Fields:2019pfx}.
One can always make a theory for a single number like
$\eta_B$. Hence, to
make progress it is important to develop a consistent
picture of the evolution of the Universe that correlates the few
available numbers in the framework of a theoretically consistent
extension of the standard model. In the review we emphasize this
point of view following the work of Sakharov.

This review focuses on electroweak baryogenesis (EWBG)
\citep*{Kuzmin:1985mm}, which is tied to
the Higgs sector of electroweak symmetry breaking, and on
leptogenesis \citep{Fukugita:1986hr}, which is closely related to neutrino physics. An
attractive feature of EWBG is that in principle all ingredients are
already contained in the SM. However, our knowledge of the electroweak
theory implies that a more complicated Higgs sector is needed for
EWBG, and the stringent constraints from the LHC and low-energy
precision experiments have led to extended models where scales of electroweak
symmetry breaking are considered well above a TeV. On the other hand,
leptogenesis originally started out at the GUT scale. But the desire to probe
the mechanism at current colliders led to the construction of models
where the energy scale of leptogenesis is lowered down to the weak scale.
A further interesting mechanism is Affleck-Dine barogenesis 
\citep{Affleck:1984fy}, which makes use of the coherent motion of scalar
fields in extensions of the SM with low-energy supersymmetry. In the
absence of any hints of supersymmetry at the LHC, we do not further
discuss the Affleck-Dine mechanism in this review.

In the following, we first recall the theoretical foundations of
baryogenesis in Sec. \ref{sec:foundations}: Sakharov's conditions, 
sphaleron processes, and some elements of thermodynamics in an
expanding Universe. We then move on to electroweak baryogenesis
in Sec. \ref{sec:electroweak}. We first review the electroweak phase
transition and the charge transport mechanism, and we illustrate the
current status of the field with a number of representative examples,
corresponding to weakly coupled as well as strongly coupled models
of electroweak symmetry breaking. Section~\ref{sec:leptogenesis} deals
with leptogenesis. After recalling the basics of lepton-number
violation and kinetic equations, we consider thermal leptogenesis
at different energy scales and also leptogenesis from sterile-neutrino
oscillations. We then describe interesting recent progress towards a
complete description of the nonequilibrium process of leptogenesis 
on the basis of thermal field theory. Finally, we discuss
an example in which by correlating inflation, leptogenesis, and dark matter
one arrives at a prediction for primordial gravitational waves emitted from a
cosmic-string network. After a discussion of other models of
baryogenesis in Sec. \ref{sec:others}, we present a summary and outlook
in Sec. \ref{sec:summary}. Different aspects of the theoretical
work on baryogenesis over 50~years have previously been described in
a number of comprehensive reviews; see \citet{Kolb:1990vq},
\citet*{Rubakov:1996vz}, \citet{Dine:2003ax},  
and \citet*{Buchmuller:2005eh}.

\section{Theoretical foundations}
\label{sec:foundations}

\subsection{Sakharov's conditions for baryogenesis}
\label{sec:Sakharov}
\citet{Sakharov:1967dj} wrote his famous paper on
baryogenesis two years after the discovery of $CP$ violation in
$K^0$ decays \citep{Christenson:1964fg}
and one year after the discovery of the cosmic microwave background
\citep*{Penzias:1965wn}, which had been predicted as a remnant of a hot
phase in the early Universe 20 years earlier \citep{Gamow:1946eb}.

Sakharov's paper contains three necessary conditions for the
generation of a matter-antimatter asymmetry from microscopic processes: 
\begin{enumerate}[label=(\arabic*)]
\item {\it Baryon-number violation.}---As we know today,
  after an inflationary phase one cannot have $B \neq 0$ as an initial
  condition of the hot early Universe, and if baryon number were
  conserved a state with $B=0$ could not evolve into a state with $B\neq 0$.
\item {\it $C$ and $CP$ violation.}---If the fundamental interactions were
  invariant under $C$ and the product of parity and
  charge conjugation $CP$, the reaction rate for the two processes, related
  by the exchange of particles and antiparticles, would be the
  same. Hence, no baryon asymmetry could be generated.
\item {\it Departure from thermal equilibrium.}---Sakharov considered an
  initial state of the Universe at high temperature. Thermal
  equilibrium would then mean that the system is stationary, so an
  initially vanishing baryon number would always be zero. A departure
  from thermal equilibrium defines an arrow of time. In a nonthermal
  system this can be provided by the time evolution of the scalar fields,
  as in Affleck-Dine baryogenesis.
\end{enumerate}

Sakharov considered a concrete model for baryogenesis. He proposed as the 
origin of the baryon asymmetry $CP$-violating decays of
``maximons,'' hypothetical neutral spin-zero particles with mass of the
order of the Planck mass $M_{\rm P} \sim 10^{19}~\text{GeV}$. Their
existence already leads to a departure from thermal equilibrium at an
initial temperature $T_i \sim M_{\rm P}$, where a small
matter-antimatter asymmetry is then generated. The $CP$ violation in
maximon decays is related to the $CP$ violation in $K^0$ decays, one of
the motivations for Sakharov's work, and an unavoidable consequence of
this model is that protons are unstable and decay. The proton lifetime
is predicted to be $\tau_p > 10^{50}\ {\rm yr}$, much
longer than in grand unified theories.

GUTs have played an important role in the
development of realistic models of baryogenesis
\cite{Dimopoulos:1978kv,Yoshimura:1978ex, Toussaint:1978br,Weinberg:1979bt}.
These theories naturally provide heavy particles, scalar and vector
leptoquarks, whose decays violate baryon and lepton number and can
therefore be the source of a baryon asymmetry. However, the simplest
GUT models based on SU(5) conserve $B-L$, the difference of baryon and
lepton numbers. Hence, leptoquark decays can create only a $B+L$ asymmetry,
with a vanishing asymmetry for $B-L$. As emphasized by
\citet*{Kuzmin:1985mm}, at temperatures above the
electroweak phase transition (B+L)-violating sphaleron processes are in
thermal equilibrium. Hence, any nonzero (B+L)-asymmetry is washed
out. The simplest GUT beyond SU(5) is based on SO(10), which
includes right-handed neutrinos and a $B-L$ gauge boson. With
$B-L$ broken at the GUT scale, right-handed neutrinos with masses
below the GUT scale are ideal agents for generating a
$B-L$ asymmetry, and therefore a baryon asymmetry, again because of the
sphaleron processes. This is
the leptogenesis mechanism proposed by \citet{Fukugita:1986hr}.

Electroweak baryogenesis is a process far from thermal equilibrium,
with a strongly first-order phase transition, nucleation and
propagation of bubbles, $CP$-violating interactions on the wall
separating the broken and unbroken phases, and a crucial change of the
sphaleron rate across the wall. On the contrary, leptogenesis is a
process close to thermal equilibrium, with the departure being a deviation 
of the density of the right-handed neutrinos from their equilibrium
distribution. Hence,
the time evolution of the nonequilibrium process is well under control and a full
quantum field-theoretical treatment is possible. Successful electroweak
baryogenesis imposes constraints on masses and couplings of Higgs
bosons, whereas successful leptogenesis is connected with properties
of the neutrinos.

\subsection{Sphaleron processes}
\label{sec:sphaleron}

In the standard model both baryon and lepton number are conserved
according to the classical equations of motion. 
However, quantum effects give rise to the chiral anomaly and  
violate baryon-number conservation~\cite{tHooft:1976rip}, 
\begin{align}
   \partial _ \mu  J _ B ^ \mu  
	=  \frac { n _ f } { 32 \pi ^ 2 }
    g ^ 2 F ^ a _ { \mu  \nu  } \widetilde F ^{ a \mu  \nu  }    
    ,
   \label{abj} 
\end{align}
where $ n _ f  = 3 $ is the number of families,
$ F ^  a _ { \mu  \nu  } $ is the weak SU(2) field strength, and  
$ \widetilde F ^ { a \mu  \nu  } \equiv ( 1/2 ) \varepsilon
^{ \mu  \nu  \rho  \sigma  } F ^ a _ { \rho  \sigma  } $. 
In Eq.~(\ref{abj}) we have neglected the U(1) hypercharge gauge field contribution
(as later discussed).
The same relation holds for the lepton-number current $ J _ L ^ \mu $,
so that $  B - L $ is conserved in the standard model.

The change of baryon number 
is thus linked  to the following dynamics of gauge fields:
\begin{align}
	B ( t ) - B ( 0 ) = \nfam Q ( t ) \ ,
\end{align}  
with
\begin{align}
	Q ( t ) \equiv
	\int _ 0 ^ t dt'\int d ^ 3 x 
	\frac { g ^ 2 } { 32 \pi  ^ 2 }
	F ^ a _ { \mu  \nu  } \widetilde F
		^{ a \mu  \nu  } \ .
	\label{Q} 
\end{align} 
Because of the coupling constants in Eq.~(\ref{Q}), a change of baryon
number of the order of unity must be accompanied by a large gauge field. 
In particular, such processes do not show up in a weak-coupling
expansion and are nonperturbative in nature. 

\begin{figure}[t]
\hspace{-4mm}
\includegraphics[width=0.45\textwidth]{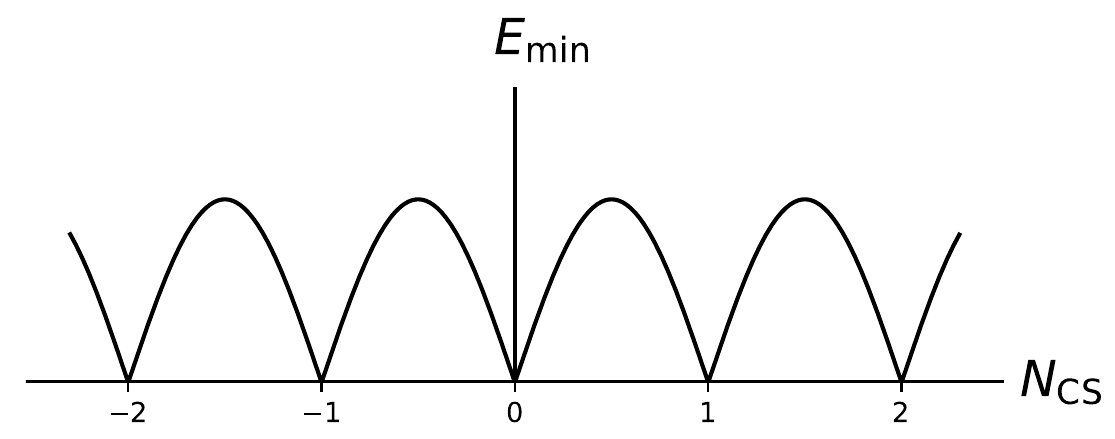}
\caption{Sketch of the minimal field energy for a given value of the
Chern-Simons number $ \NCS $.
The energy approaches the minima with nonzero slope \citep*{Akiba:1988ay}.}
\label{f:period}
%
\end{figure}

Baryon-number changing processes are closely connected to the topology
of the SU(2) gauge plus Higgs fields. 
To see this, note that 
the integrand of Eq.~(\ref{Q}) can be written as a total derivative 
$ \partial _ \mu  K ^ \mu  $,
with
\begin{align}
	K ^ \mu  
	= 
	\frac 1 { 32 \pi  ^ 2 }
	\varepsilon  ^{ \mu  \nu  \rho  \sigma  }
		g ^ 2 \left ( F ^ a _ { \nu  \rho  }	A ^ a _ \sigma  
		- \frac 13 g \varepsilon  ^{ abc }	
		A ^ a _ \nu  A ^ b _ \rho  A ^ c _ \sigma  
		\right )
	\label{K} 
	.
\end{align} 
An Abelian gauge field requires nonzero field strength 
to obtain nonvanishing $ K ^ \mu  $. This is not the
case for the non-Abelian field due to the second term in Eq.~(\ref{K}).
If one can neglect 
the integral $ \int d ^3 x \nabla \cdot \vec K $,
e.g., with periodic boundary conditions or if 
$ \vec K $ vanishes at spatial infinity, then
\begin{align} 
	\label{abjint} 
	Q ( t ) = \NCS  ( t ) - \NCS  ( 0 ) \ ,
\end{align}
with the Chern-Simons number
\begin{align} 
	\NCS  =
	\int 
	d ^ 3 x K ^ 0
	.
	\label{ncs} 
\end{align} 
In the vacuum, the Higgs field can be chosen to be
constant and at the minimum 
of its potential $ A  _ \mu  $ is a pure gauge.
In the $ A ^ a _ 0 = 0 $ gauge, $ \NCS  $ is the
gauge field winding number, which is an
integer.
It is invariant under ``small'' gauge transformations, i.e., 
gauge transformations continuously connected to the identity. 
To change $ \NCS  $ by $ \pm 1 $, one has to 
go over an energy barrier. Figure \ref{f:period}  shows the minimal 
static energy of the gauge-Higgs fields as a function 
of $ \NCS $.
The minima of the energy differ by large gauge transformations
and all describe the vacuum state. 
A vacuum-to-vacuum transition along this path 
would change  baryon and lepton number  by a multiple of $ n _ f $
\cite{tHooft:1976rip}.
The barrier is given by a static solution to the equations
of motion, the so-called sphaleron \cite{Klinkhamer:1984di},
which has half integer $ \NCS $ and an energy of the order of $ m _ W /\alpha _ W  $,
$ \alpha  _ W =g ^ 2 /4 \pi$. 
Thus, at low energies an $\NCS$-changing transition
can occur only via tunneling. The amplitude of such a process is 
proportional to $ \exp ( - 16  \pi ^ 2  /g ^ 2 ) $, which is small
and has no observable consequences.

However, at high temperatures there can be thermal fluctuations that
take the system over the sphaleron barrier. 
The baryon number is no longer conserved and the value of $ B $ 
will relax to its equilibrium
value $  B _ { \rm eq }  $.\footnote{When $B-L$ is nonzero,
  $ B _ { \rm eq } $  does not vanish.} 
For a sufficiently small deviation from equilibrium, this  
is determined by a linear equation (without Hubble expansion)
\begin{align}
   \frac d { dt }   B = - \gamma ( B  - B _ { \rm eq } ) 
	\label{diss} \ .
\end{align}
The dissipation rate $ \gamma  $ depends only on the temperature and 
the value of conserved charges like $B-L$.
Furthermore, at a first-order electroweak phase transition it depends
on whether one is in the symmetric or the broken phase. 
When the dissipation rate $ \gamma  $ is larger than the 
Hubble parameter, the baryon number is in equilibrium.

The dissipation rate
$ \gamma  $ can be related to the properties of thermal fluctuations
of baryon number around its equilibrium value $ B _ { \rm eq } $: 
If  $ B - B _ { \rm eq } $ has made a fluctuation to a nonzero value,
this will tend to zero  at the rate $ \gamma  $. 
Therefore, the 
time-dependent correlation function of the fluctuation
reads
\begin{align}
	\langle 
		[ B ( t ) - B _ { \rm eq } ] 
		[ B ( 0 ) - B _ { \rm eq } ] 
	\rangle 
			= \langle [ B  - B _ { \rm eq } ] ^ 2
	\rangle 
	 e ^{ - \gamma  | t | } \ ,
	\label{bcor} 
\end{align}
which implies
\begin{align}
   \langle [ B     ( t )
   -  B   ( 0 ) ] ^ 2 
   \rangle
   = 
	2   \langle [ B  - B _ { \rm eq } ]^ 2 \rangle 
   \left ( 1 - e ^{ - \gamma  | t | }  \right ) 
	\label{bfluc}
   . 
\end{align} 
For $ t \ll \gamma ^{ -1 }  $ this grows approximately linearly
with time:
\begin{align}
   \langle [ B ( t )   - B ( 0 )  ] ^ 2 
   \rangle
   \simeq 
	2
   \langle [ B  - B _ { \rm eq } ]     ^ 2 \rangle 
   \gamma  | t  | 
   \label{brownian} \  . 
\end{align} 
The mean square fluctuation on the right-hand side
is determined by equilibrium thermodynamics and is proportional to 
the volume of the system $ V $. 
It has to be evaluated at fixed $ B - L $.
The leading-order computation is straightforward but requires some 
care ~\cite{Khlebnikov:1996vj}.
In the temperature range between the electroweak transition and 
the equilibration temperature of the right-handed electron Yukawa interaction,
it takes the value 
\begin{align}
	\langle [ B  - B _ { \rm eq } ]     ^ 2 \rangle 
	=
	V T ^ 3 
		\frac{2 \nfam ( 5\nfam + 3 N _ s) }{3 ( 22\nfam +13N _
  s) } \ ,
	\label{B2} 
\end{align} 
where $ N _ s $ is the number of Higgs doublets.
At lower temperatures one has to take into account a nonzero Higgs expectation
value, and at higher temperatures there are additional conserved
charges,  which reduces the size of the fluctuation ~\cite{Rubakov:1996vz}.

Owing to Eq.~(\ref{abj}) the left-hand side of Eq.~(\ref{brownian}) is determined by the
dynamics of the gauge fields:
\begin{align}
   \langle [ B     ( t )
   -  B   ( 0 ) ] ^ 2 
   \rangle
   = 
	\nfam ^ 2	\left \langle Q ^ 2 ( t ) 
  \right \rangle \ .
	\label{dyn} 
\end{align} 
For Eqs.~(\ref{brownian}) and (\ref{dyn}) to be consistent, 
$ Q $ in Eq.~(\ref{abjint}) must satisfy 
\begin{align}  
	\left \langle Q ^ 2 ( t )
  \right \rangle 
	= 
  \Gsph  V | t |  
	\label{rand}
	.
\end{align}
This can be easily visualized 
with the help of  Fig.~\ref{f:period}. 
Most of the time the system sits near one of the minima, but every
once in a while there is a thermal fluctuation that
lets it hop to a neighboring one. 
This gives rise to a random walk 
leading to the behavior in Eq.~(\ref{rand}).
$ \Gsph $, the number of transitions per unit time and unit volume,
is known as the Chern-Simons diffusion rate or sphaleron rate.
It can be estimated as 
\begin{align}
	\Gsph \sim t _ { \rm tr } ^{ -1 } 
 \ell ^{ -3 } 
	\label{est} 
	,
\end{align}
where 
$ t _ { \rm tr } $  is the time 
of a single transition, and $ \ell $ is the spatial size of the 
corresponding  
field configuration.
The U(1) hypercharge gauge field does not contribute to the diffusive 
behavior in Eq.~(\ref{rand}), so we neglect it here.

The linear growth with time can only be valid 
on timescales that are large compared to $ t _ { \rm tr } $.
On the other hand, the linear growth can be valid only on timescales
that are small 
compared to $ \gamma  ^{ -1 }$. 
If there is a time window in which Eqs.~(\ref{brownian}) and (\ref{rand}) 
are both valid, which will be checked {\it a posteriori}, then one can 
match the two expressions to determine $ \gamma  $, which gives
\begin{align}
   \gamma  
	= \frac { \nfam ^ 2 \Gsph V  } 
	{ 2 \langle [ B - B _ { \rm eq }  ] ^ 2 \rangle } \ .
	\label{ga} 
\end{align} 
Using $ \langle [ B - B _ { \rm eq }  ] ^ 2 \rangle \sim V T ^ 3 $
one can therefore estimate $ \gamma  t _ { \rm tr } \sim ( \ell T ) ^{
-3 } $. Hence,
the window exists if the size of the relevant 
field configurations
is large compared to $ T ^{ -1 } $.

At  low temperatures the $ \NCS $-changing transitions
still proceed through tunneling.
The probability for thermal transitions over the 
sphaleron barrier is  
proportional to $ \exp ( - E _ { \rm sph }/T ) $
\citep*{Kuzmin:1985mm}.
They become dominant when $ E _ { \rm sph }/T \lsim 1/g ^ 2 $.  
The energy and the size of a sphaleron are of the order of 
$ E _ { \rm sph } \sim v/g $ and $ \ell _ { \rm sph } \sim 1/gv  $,
respectively. 
Therefore, the size of the sphaleron is larger than $ T ^{ -1 } $ 
when the thermal activation dominates, and the assumptions
leading to Eq.~(\ref{ga}) are valid in this case. 
For field configurations  with $ k \sim \ell ^ { -1 } \lsim  T $, 
the occupation
number, given by the Bose-Einstein distribution, is large 
[$ f _ { \rm B } ( k ) \simeq T/k \gsim  1 $], so such fields
can be treated classically. 

\begin{figure}[t]
\hspace{-9mm}
\includegraphics[width=0.4\textwidth]{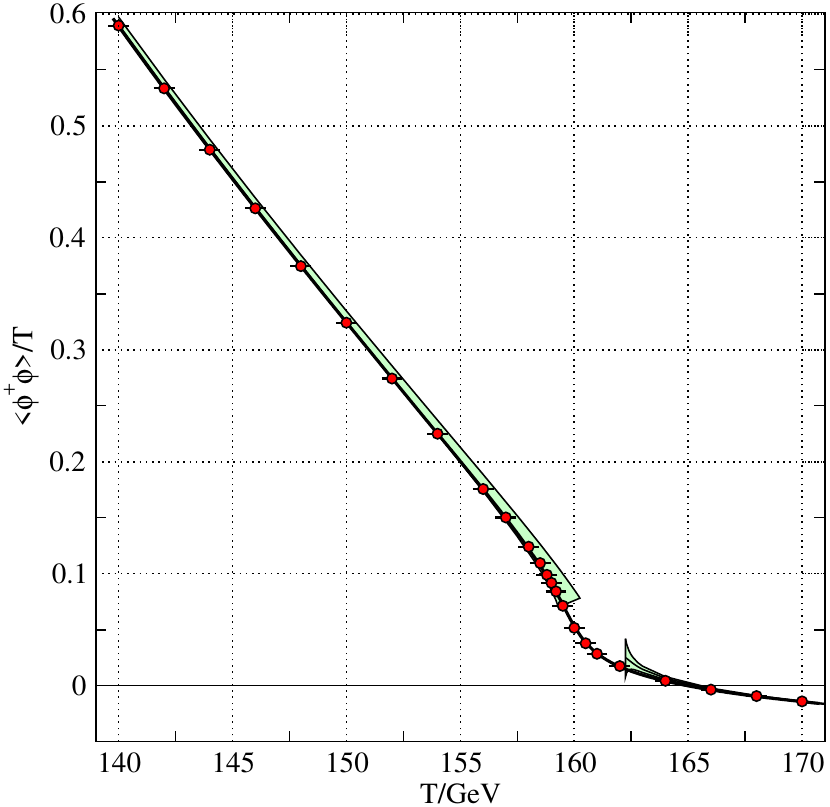}
\caption{Temperature dependence of the expectation value of 
$ \phi ^  \dagger \phi  $ in the standard
model. $ \phi  $ is the Higgs field.
The points are the results of lattice simulations, and the full line 
is an interpolation of the data.
The observable 
$  \langle \higgs ^  \dagger \higgs\rangle  $ 
can become negative, because it is ultraviolet divergent and  
additively renormalized. 
The shaded bands represent perturbative results for the 
broken and symmetric phases; their thickness estimates
unknown higher-order contributions.
From \citet{DOnofrio:2015gop}.
}
\label{f:higgs-ev}
\end{figure}

The Higgs expectation value decreases with increasing temperature:
see Fig.~\ref{f:higgs-ev} and Sec.~\ref{sec:phase}.
Therefore, the exponential suppression has already disappeared near 
the electroweak phase transition or crossover. 
The prefactor of the exponential corresponds to a one-loop
computation of the fluctuations around the sphaleron contribution.
The bosonic part was computed by
\citep*{Arnold:1996dy}, and 
the fermionic contributions were obtained by \cite{Moore:1995jv}.

In the symmetric phase the Higgs expectation value vanishes and 
there is no sphaleron solution.%
\footnote{Nevertheless, $ \Gsph $ in the symmetric phase is usually
referred to as the hot sphaleron rate.} 
The length scale for $\NCS$-changing field configurations can now
be easily determined. When the energy of a field configuration
is dominated by the electroweak 
magnetic field $ \vec B = \vec D \times \vec A$,
it can be estimated as
\begin{align}
	E \sim \ell ^ 3 \vec B ^ 2 \sim \ell \vec A ^ 2 \ .
\end{align} 
Using $ \vec E \sim \vec A/t _ { \rm tr } $,
the change of Chern-Simons number is then given by
\begin{align}
 Q \sim g ^ 2 t _ { \rm tr }  \ell ^ 3 \vec E \cdot \vec B \sim g ^ 2
  \ell ^ 2 \vec A ^ 2\ .
\end{align}   
If we require $ Q \sim 1 $ and 
$ E \lsim  T $ to avoid
Boltzmann suppression,
we obtain $ \ell \gsim ( g ^ 2 T ) ^{ -1 }  $. But $ ( g ^ 2 T ) ^{ -1 }
$ is the length scale beyond which static non-Abelian magnetic fields
are screened. 
Time-dependent fields can be screened on even shorter length scales. 
Therefore, the relevant length scale for $\NCS$-changing transitions
in the symmetric phase
is~\cite{Arnold:1987mh} 
\begin{align} 
	\ell \sim \frac 1 { g ^ 2 T } \ .
\end{align} 
The corresponding gauge field is of the order of $ \vec A \sim g T $. 
Therefore, both terms in the covariant derivative
$ \vec D - i g \vec A $ are of the same order of magnitude, and
the second term cannot be treated as a perturbation.
This leads to the breakdown of finite-temperature perturbation 
theory at this scale~\cite{Linde:1980ts}.  
Standard Euclidean (imaginary-time) lattice methods are not capable of
computing real-time dynamics. 
However, since the relevant fields have large occupation numbers they
can be approximated as classical fields, 
and $ \Gsph $ can be computed by solving classical field
equations of motion~\cite{Ambjorn:1990pu}, where some care
is needed to use the correct equations of motion. 

The time evolution of the fields responsible for  the sphaleron transitions
is influenced by plasma effects \citep*{Arnold:1987mh, Arnold:1996dy}.
The time-dependent gauge field has the nonvanishing electric field $
\vec E $, which induces a current because the plasma is a good conductor.
The relevant charges are the weak SU(2) gauge charges. 
The current is carried mostly by particles with hard momenta
of the order of $ T $ that are not described by classical fields.
Therefore, the classical field equations are not appropriate
for computing $ \Gamma  _ { \rm sph } $. 
However, one can use effective classical equations of motion
that should properly include the effect of the high-momentum particles.
The mean free path of the particles is smaller than
the length scale $ \ell $. Therefore, the current
can be written as $ \sigma  \vec E $ with the following
conductivity\footnote{In QCD the analogous quantity is called color conductivity.} of 
SU(2) charges $ \sigma $:
\begin{align}
	\sigma  = 
   \frac{4\pi}{3}\frac{m^ 2 _ { \rm D } }{Ng^2 T}\frac{1}{\log(1/g)}
	\sim \frac T { \log(1/g)} \ ,
	\label{sigma} 
\end{align} 
where $ N = 2 $ for SU(2)  and 
$m _ { \rm D }^ 2 =  (4N + 2 N_s + N_F)g^2 T^2/12$
is the Debye mass squared for  $N_F$ chiral fermions and
$N_s$ scalars in the fundamental representation.
In the $ A _ 0 = 0 $ gauge $ \vec E = - \dot { \vec A } $. 
Therefore, the current gives rise to a damping term in the equation of 
motion for $ \vec A $. 
Estimating  $ \vec D \times \vec B \sim \sigma  
\vec E $ gives 
\begin{align}
	 t _ { \rm tr } \sim \sigma  \ell ^ 2
	\sim [   g ^ 4 \log ( 1/g )T] ^{ -1 }   
	,
\end{align} 
which is much larger than $ \ell $. 
Thus, the gauge field is strongly damped and one can neglect 
$ \dot { \vec E } $ in the equation of motion, which 
becomes~\cite{Bodeker:1998hm}
\begin{align}
	\vec D \times \vec B = \sigma  \vec E + \boldsymbol { \zeta }
  \ .
	\label{langevin}
\end{align} 
$ \boldsymbol { \zeta } $ is also part of the current
of high-momentum particles. It is due to thermal fluctuations of 
all field modes with momenta larger than $ g ^ 2 T $, and
it is a Gaussian white noise that carries vector and group indices. 
It satisfies
\begin{align}
	\langle \zeta  ^ { ia}   ( x ) \zeta  ^ { j b }  ( x') \rangle
	=
	2 T \sigma  \delta  ^{ ij } \delta  ^{ ab } \delta  ( x-x') \ ,
	\label{noise} 
\end{align} 
so that Eq.~(\ref{langevin})  is a Langevin equation. 
The estimate (\ref{est}) then gives 
\begin{align}
	\Gsph \sim  
	  g ^{ 10 } \log ( 1/g ) T ^ 4 \ .
	\label{hot}  
\end{align} 
The numerical coefficient  can be computed by solving
Eq.~(\ref{langevin}) on a spatial lattice and determining $\Gsph$ from
Eq.~(\ref{rand}). 
The result can be written as 
\begin{align}
	\Gsph = \kappa  \frac { 2 \pi T } { 3 \sigma  } \alpha  ^ 5 T ^ 4 ,
	\label{kap}  
\end{align} 
with $ \kappa  = 10.8 \pm 0.7 $ \cite{Moore:1998zk} and $ \sigma  $
from Eq.~(\ref{sigma}).
The mean free path of hard particles is short compared to 
$ \ell $ by only a relative factor $ \log ( 1/g ) $.
Nevertheless, Eq.~(\ref{kap}) is still valid at
next-to-leading logarithmic order if 
$ \log ( 1/g ) $ in Eq.~(\ref{sigma}) is replaced by 
$  \log ( m _ { \rm D } /\gamma  ) + C $, where 
$\gamma  = ( N g ^ 2 T/4 \pi  ) \log(m _ { \rm D } /\gamma  ) $
and $ C \simeq  3.041 $  \cite{Arnold:1999uy,Moore:2000mx}. 

Close to the electroweak phase transition or crossover the 
thermal Higgs mass can become small, so that the Higgs field
can affect the dynamics at the scale $ g ^ 2 T $.
The effective theory described by Eq.~(\ref{langevin}) has been
extended to include the Higgs field ~\cite{Moore:2000mx}.
$ \kappa  $ also depends on the Higgs self-coupling and thus 
on the Higgs mass.
In the standard model, just above the crossover temperature 
one finds that \citep*{DOnofrio:2014rug}
$
  \Gamma_{\rm sph}/T^4 = (8.0 \pm 1.3) \times 10^{-7} \approx 
  (18 \pm 3) \alpha_W^5
$.
In the last form, factors of $\ln \alpha_W$ have been absorbed 
in the numerical constant. 
Without the Higgs field 
the rate is $\Gamma \approx  (25\,\pm\,2)\,\alpha_W^5 T^4$ 
\citep*{Moore:1997sn,Moore:2000ara}.

Beyond logarithmic accuracy, the current is not simply
a local conductivity times the electric field. 
To go beyond this approximation one has to solve the coupled
equations for the gauge fields and the high-momentum particles. 
Here fields with $ \vec k \sim g T $ are also important because
they mediate the scattering of the high-momentum particles, which
is small-angle scattering that changes the charge of the
particles. For these modes one cannot neglect the 
term $ \dot { \vec E } $, which leads to ultraviolet
divergences in the simulation prohibiting a continuum limit
\citep*{Bodeker:1995pp}.

\begin{figure}[t]
\hspace{-9mm}
\includegraphics[width=0.4\textwidth]{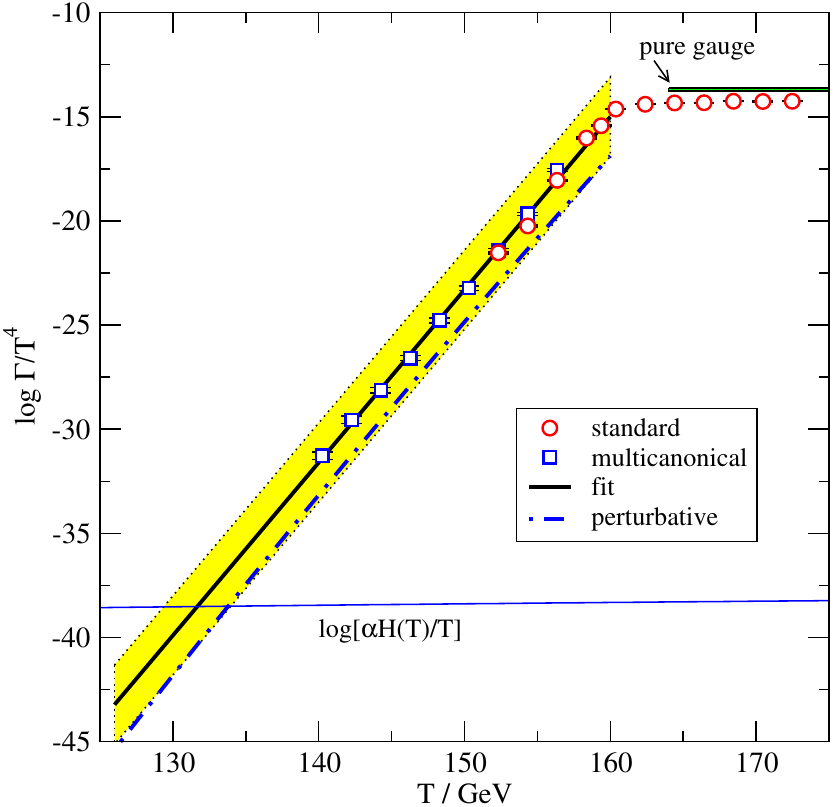}
\caption{
The Standard Model
sphaleron rate computed on the lattice  and the fit to the broken
phase rate [Eq.~(\ref{Gfit}], shown with a shaded error band.
At low temperatures the sphaleron rate is exponentially small, which
requires 
a special multicanonical  method for the simulation.
The perturbative result
(Burnier, Laine, and Shaposhnikov, 2006)
is the one-loop approximation to an expansion around the sphaleron solution.
Pure gauge refers to the rate in hot SU(2) gauge theory.  
The sphalerons  freeze out when $ \Gamma  $
crosses the appropriately  scaled Hubble parameter, which is shown as the
almost horizontal line.
From \citet*{DOnofrio:2014rug}.
}
\label{f:G}
%
\end{figure}

When the Higgs expectation value is sufficiently large, the 
sphaleron rate becomes exponentially suppressed and one can
perform a perturbative expansion around the sphaleron solution.
The signal in lattice simulations, on the other hand,
becomes small, which requires the use of
a special multicanonical method \cite{Moore:1998swa}. 
The current knowledge of the sphaleron rate in the standard 
model is summarized in Fig.~\ref{f:G}. 
In the temperature range $ 130 <T < 159~\text{GeV}$ it can be parametrized
as \citep*{DOnofrio:2014rug}
\begin{align} 
    	\log(\Gamma _ { \rm sph }/ T ^ 4)
	=(0.83\pm 0.01)\frac T{\rm GeV} - (147.7\pm1.9)  ,
	\label{Gfit} 
\end{align} 
which is the fit shown in Fig.~\ref{f:G}. 
The rate computed on the lattice is 
larger than the perturbative results \citep*{Burnier:2005hp} but consistent 
within errors. 
The corresponding values of the Higgs expectation value are depicted
in Fig.~\ref{f:higgs-ev}.

At  high temperatures, the  sphaleron rate again 
becomes smaller than the Hubble parameter, which happens at $ T \gsim  
 10 ^ { 12 } $ GeV ~\cite{Rubakov:1996vz}.

In theories with an extended Higgs sector, 
it is not obvious how to determine the freeze-out condition
from the SM results, because the sphaleron solution can be different. 
In the symmetric phase this is somewhat easier.
New particles interacting with the SU(2)  gauge fields 
would increase the Debye mass $ m _ { \rm D } $ appearing in 
Eq.~(\ref{sigma}), thus decreasing the hot sphaleron rate. On the
other hand, new particles would increase the Hubble parameter. 
Therefore the SM freeze-out temperature is an upper bound for the 
temperature below which $ \gamma  > H $. 

In QCD with vanishing quark masses, the axial quark number is 
classically conserved, but it is also violated by the
chiral anomaly. This process plays a role in both electroweak
baryogenesis and leptogenesis.
At finite temperature the Chern-Simons number of the gluon field 
can diffuse as in the electroweak theory in the symmetric phase, 
and the rate for anomalous axial quark number violation is 
again proportional to the Chern-Simons diffusion rate, which is 
then referred to as the strong-sphaleron rate. 
At high temperatures the quantum chromodynamics (QCD) coupling $ \alpha  _ { \rm s } $ 
is weak, and 
the dynamics of the gluon fields is described by 
Eqs.~(\ref{sigma})-(\ref{noise}) for SU(3) instead of SU(2).
At the electroweak scale $  \alpha  _ { \rm s } $  appears 
to be too large for the weak-coupling expansion to 
be valid. Using a different method, the strong-sphaleron rate at 
this scale was computed as  
 $ 
\Gamma_{\rm strong \: sph} \simeq 1.4 \times 10 ^{ -3 } T ^ 4 $
\cite{Moore:2010jd}.

Strong sphalerons are most likely the only sphalerons that can be 
created in experiments.
It has been argued that they could lead
to observable signals in relativistic heavy ion collisions through
the chiral magnetic effect
\cite{Kharzeev:2013ffa}.
In the simultaneous presence of a chiral imbalance and 
a magnetic field 
there is an electric vector current in the direction of the magnetic field.
Such a current separates electric charges 
and could lead to measurable charge asymmetries. 
The required imbalance of left- and right-handed (anti)quarks can  
be caused by random strong-sphaleron transitions. 
Furthermore, if the collision of the heavy ions is not head-on,
the remnants of the projectiles produce strong magnetic fields. 
There are ongoing experimental efforts to search for the chiral 
magnetic effect in heavy ion collisions.
A dedicated run has been performed at the Brookhaven Relativistic Heavy Ion Collider 
colliding Ru on Ru and Zr on Zr (results are expected in 2021). 
These two nuclei are isobars, i.e., they have the same number of nucleons 
but different numbers of protons ($Z=44$ for Ru and $Z =40$ for Zr).
Thus, the magnetic field is larger for Ru, so the 
charge asymmetries in Ru collisions should be 
larger than those in Zr
\citep{Wen:2018boz,Kharzeev:2020jxw}.

\subsection{Baryon and lepton asymmetries}
\label{sec:thermo}
Quarks, leptons and Higgs bosons interact via
Yukawa and gauge couplings and, in addition, via the nonperturbative
sphaleron processes.
In the temperature range
$100 < T < 10^{12}\ \text{GeV}$, 
which is of interest for baryogenesis, gauge interactions, including
the sphaleron interactions, are in equilibrium,
i.e., their rate is larger than the Hubble parameter.  
On the other hand, Yukawa interactions are in equilibrium only
in a more restricted temperature range that depends on the strength of the 
Yukawa couplings.
Thus, in different temperature ranges there are different
sets of charges that are conserved,
which leads to  the ``flavor effects'' that are
discussed in Sec.~\ref{sec:flavour}.
The corresponding partition function can be written as 
\begin{equation}
  Z(\mu,T,V) = \text{Tr} 
	\exp \Big \{\beta\Big (\sum_i\mu_iQ_i 
	 -H\Big)\Big \}\ ,
	\label{Z} 
\end{equation}  
where $\beta = 1/T$
and $H$ is the Hamiltonian. 
For each of the quark, lepton, and Higgs fields, 
there is an associated chemical potential $ \mu  _ i $;
the corresponding 
charge
operator is denoted by $Q_i$.
In the standard model, with one 
Higgs doublet $\phi$ and $n_f$ families one has $5n_f+1$ chemical 
potentials $ \mu  _ i $.%
\footnote{In addition to the Higgs doublet, the two left-handed
doublets $q_i$ and $\ell_i$ and the three right-handed singlets $u_i$, $d_i$,
and $e_i$ of each family each have an independent chemical potential.}

The processes that are in thermal equilibrium, the so-called spectator
processes, yield
constraints between the various chemical potentials \cite{Harvey:1990qw}. 
The $ N _ { \rm CS } $-changing transitions 
(see Sec.~\ref{sec:sphaleron}) change baryon and lepton
numbers in each family by the same amount. They affect only the
left-handed fermion fields, so that 
\begin{equation}\label{sphew}
\sum_i\left(3\mu_{qi} + \mu_{li}\right) = 0\;.
\end{equation}
One also has to take the SU(3) QCD
sphaleron processes into account 
\cite{Mohapatra:1991bz}. 
They change the chiral quark number (the number of right-handed
minus number of left-handed quarks) for each quark flavor by
the same amount, so that  
\begin{equation}\label{sphqcd}
\sum_i\left(2\mu_{qi} - \mu_{ui} - \mu_{di}\right) = 0\;.
\end{equation}
The Yukawa interactions that are in equilibrium yield relations
between the chemical potentials of the left-handed and right-handed fermions and
the Higgs:
\begin{align} 
\label{myuk}
-\mu_{qi}+\mu_{dj}  = \mu_{qi} -\mu_{uj} 
   = 
-\mu_{li} +\mu_{ei}  = \mu_\phi\ .
\end{align} 
The remaining independent chemical potentials are subject to another
condition, valid at all temperatures, that arises from the
requirement that the total hypercharge of the plasma vanish. 

In a weakly coupled plasma, the asymmetry between particle
and antiparticle number densities is given by 
\begin{equation}
  n_i-n_{ \bar i }
  = -\frac{\partial}{\partial\mu_i}\frac{T}{V}\ln{Z(\mu,T,V)}\ .
	\label{nnb}
\end{equation}
When computing the derivative in Eq.~(\ref{nnb}), all $ \mu  _ i $ have to
be treated as independent.
For massless particles one obtains 
\begin{equation}\label{number}
n_i-n_{ \bar i }  ={g_i T^3\over 6}
\left\{\begin{array}{rl}\beta\mu_i +{\cal
         O}\big(\left(\beta\mu_i\right)^3\big)\ 
&(\text{fermions})\ ,\\
2\beta\mu_i+{\cal O}\left(\big(\beta\mu_i\right)^3\big)\
         &(\text{bosons})\ ,
\end{array}\right.
\end{equation}
where $g_i$ denotes the number of internal degrees of freedom.
The following analysis is based on these relations for small chemical
potentials ($\beta \mu_i \ll 1$).

Using Eq.~(\ref{number}) and the known hypercharges one can write the
condition for hypercharge neutrality as
\begin{equation}\label{hypsm}
\sum_i\left(\mu_{qi} + 2 \mu_{ui} - \mu_{di} - \mu_{li} - \mu_{ei}
\right) =  2 \mu_{\phi}\ ,
\end{equation}
and the baryon-number 
and lepton-number densities  can be expressed in 
terms of the chemical potentials as follows:
\begin{align} 
  n _ B &= \frac { T ^ 2 } 6
	\sum_i \left(2\mu_{qi} + \mu_{ui} + \mu_{di}\right)
	\;, \\
  n _ { L_i } &= \frac { T ^ 2 } 6
   ( 	2\mu_{li} + \mu_{ei} ) \; . 
\end{align} 

Consider now the temperatures at which all Yukawa interactions are in equilibrium,
which is the case for $ T < 85 $ TeV \cite{Bodeker:2019ajh}, but still
above the electroweak transition.
The quark chemical potentials are  family independent, 
$ \mu  _ { qi } = \mu  _ q $, $ \mu  _ { u i } = \mu  _ u $, and
$ \mu  _ { di } = \mu  _ d $, and  
the asymmetries $L_i-B/n_f$ are conserved.
For simplicity, we assume that they are all equal, so that
the lepton chemical potentials are family-independent as well:
 $ \mu  _ { l i } = \mu  _ l $, $ \mu  _ { ei } = \mu  _ e $.
Using the sphaleron relation and the 
hypercharge constraint, one can express all chemical potentials, and 
therefore all asymmetries, in terms of a single chemical potential that 
may be chosen as $\mu_l$:
\begin{align}\label{exam1}
\frac{\mu_e}{\mu_l} &= {2n_f+3\over 6n_f+3}\ , \  
\frac{\mu_d}{\mu_l} = -{6n_f+1\over 6n_f+3}\ , \
\frac{\mu_u}{\mu_l} = {2n_f-1\over 6n_f+3}\ , \nonumber\\
\frac{\mu_q}{\mu_l} &= -{1\over 3}\  , \
\frac{\mu_{\phi}}{\mu_l} = 
	- {4n_f\over 6n_f+3} \ .
\end{align}
The corresponding baryon and lepton asymmetries are
\begin{equation}\label{nBnL}
  n_B = -{4n_f\over 3}\frac{T^2}{6}\mu_l\;, \quad
  n_L = {14n_f^2+9n_f\over 6n_f+3}\frac{T^2}{6}\mu_l\;.
\end{equation}
Equation \eqref{nBnL} yields the connection between the $B$, \mbox{$B-L$}, and $L$ 
asymmetries \cite{Khlebnikov:1988sr}
\begin{equation}\label{connection}
B = c_s (B-L)\ ,  \quad L= (c_s - 1)(B- L) \ ,
\end{equation}
where $c_s = (8n_f+4)/(22n_f+13)$.  
Near the electroweak transition the ratio $B/(B-L)$ is a function of
$ \langle \phi  \rangle /T $  
\cite{Laine:1999wv}.

The relations (\ref{connection}) between $B$, $B-L$, and $L$ numbers 
suggest that $B-L$ violation is needed\footnote{In the case of Dirac neutrinos,
which have extremely small Yukawa couplings, one can construct leptogenesis 
models where an asymmetry of lepton doublets is accompanied
by an asymmetry of right-handed neutrinos such that the total $L$ number
is conserved and the $B-L$ asymmetry vanishes \cite{Dick:1999je}.}
in order to generate a baryon asymmetry at high temperatures where sphaleron
processes are in thermal equilibrium.
Because the $B-L$ current has no anomaly, the value of $B-L$ at time $t_f$,
where the leptogenesis process is completed, determines the value of the
baryon asymmetry today:
\begin{equation}
B(t_0) = c_s (B-L)(t_f)\;.
\end{equation}
On the other hand, during the leptogenesis process the strength of ($B-L$)-violating,
and therefore $L$-violating interactions can only be weak. Otherwise, because 
of Eq. (\ref{connection}), they would wash out any baryon asymmetry. 
As we later see,
the interplay between these conflicting 
conditions leads to important constraints on the properties of the neutrinos.

The situation is different for electroweak baryogenesis. Here $B-L =
0$ and the change of the sphaleron rate across the bubble wall in a
first-order phase transition is essential for the generation of a
baryon asymmetry.

\section{Electroweak baryogenesis}
\label{sec:electroweak}
Electroweak baryogenesis is a sophisticated nonequilibrium process at
the electroweak phase transition \citep*{Cohen:1993nk}. 
We first describe the nature of the
phase transition and the basic idea of the charge transport mechanism.
We then illustrate the status of electroweak baryogenesis by some
representative examples, corresponding to a weakly as well as a
strongly interacting Higgs sector. Special emphasis is placed on
the implications of recent stringent upper bounds on the electron
electric dipole moment.

\subsection{Electroweak phase transition}
\label{sec:phase}
Electroweak baryogenesis requires a  first-order phase transition
to satisfy the Sakharov  condition of nonequilibrium. 
It has to be strongly first order, meaning that
in the low-temperature phase the sphaleron rate is sufficiently
suppressed and the just created asymmetry is not washed out;
see Sec.~\ref{sec:transport}.

At zero temperature the electroweak symmetry is broken by the
vacuum expectation value of the Higgs field $ \phi  $,
giving   mass to the electroweak gauge bosons and to fermions.
At high temperature the Higgs expectation value vanishes.
The symmetry that is broken by the expectation value
is a gauge symmetry, 
not a symmetry transforming physical states. 
Therefore, it is not guaranteed that there will be a phase transition associated
with the change of $ \langle \phi  \rangle $.
(Nevertheless, it is common nomenclature to speak about a  
symmetry-broken and a symmetric phase.)

The expectation value of $ \phi  $ is obtained by minimizing 
the effective potential
$ V _ { \rm eff } $, which can be defined as
   $ V _ { \rm eff } ( \phi  ) \equiv  -  P ( \phi  ) $, 
where $ P ( \phi  ) $ is the pressure in the presence of a constant
classical value $ \phi  $ of the Higgs field.
It includes the tree-level Higgs potential $ V_\text{tree} $.
To first approximation it  is 
given by the difference of $ V_\text{tree} $ and the pressure of an ideal gas
$ P _ { \rm ideal } $.
When the temperature is much larger than the particle mass $ M $,
the pressure of an ideal gas is, according to standard thermodynamics, 
\begin{align}
   P _ { \rm ideal }  
   = 
   T ^ 4 \left (  a  -b \frac { M ^ 2 } { T ^ 2 }  
   + c  \frac { M ^ 3 } { T ^ 3 } + O \left ( M ^ 4/T ^ 4 
  \right )
   \right )  \ ,
   \label{pressure}
\end{align} 
with positive constants $ a $ and $ b $. 
The $ O ( M ^ 2 /T ^ 2 ) $ contribution is negative because
a nonzero mass reduces the momentum of a particle with a 
given energy and thus the pressure.  
If the particle masses are proportional to the value of the
Higgs field, then a smaller $   \phi      $  leads to larger pressure. 
A phase with a smaller $ \phi    $ will push out one with a larger
value of the Higgs, so that the Higgs expectation value becomes zero.
Therefore,
at high temperature the electroweak symmetry is
unbroken.%
\footnote{There are models where some mass decreases when a scalar field
is increased. In this case it is possible that a symmetry gets broken
at high temperature~\cite{Weinberg:1974hy}.%
}
The region where this happens can be expected to 
be of the order of the weak gauge-boson mass.

Beyond the ideal gas approximation one can compute the 
effective potential as follows.  
One integrates out 
all field modes with  nonzero momentum in the imaginary-time  path integral: 
\begin{align}
	e ^{ - \beta  V  V _ { \rm eff }  ( \phi  )  } 
	=
	\int '{ \cal D }\Phi  \exp \left \{ -S _ E [ \Phi    ] \right
  \}  \ ,
	\label{pathintegral} 
\end{align} 
with the Euclidean, or imaginary-time, action ($t = -i\tau$)
\begin{align} 
	S _ E = - \int _ 0 ^ \beta  d \tau  \int d ^ 3 x { \cal L } \ .
\end{align} 
$ \Phi  $ denotes the set of all fields 
of our system, and the prime indicates that the integration over 
the zero-momentum modes $ \phi  $ is omitted.  
The partition function $ Z = \exp ( \beta  V  P )   $ is then
obtained by integrating Eq.~(\ref{pathintegral}) over $ \phi  $.  
This is done 
in the saddle-point approximation,
which gives the minimum condition for $ V _ { \rm eff } ( \phi  )$.
In the one-loop approximation Eq.~(\ref{pathintegral}) gives
$ -V _ { \rm eff }$ as the difference between $P _ { \rm ideal } $ and
the $ T =0 $ contribution to the effective potential
\cite{Coleman:1973jx}.%
\footnote{The effective potential defined by Eq.~(\ref{pathintegral}) 
is gauge fixing dependent. Physical quantities like the pressure,
and thus the value of $ V _ { \rm eff } $
at the minima,
are gauge fixing independent.}  

For illustration, first consider the case of a single real scalar 
field $\varphi $ with the Lagrangian 
\begin{align}
	{ \cal L } = \tfrac{1}{2} \partial _ \mu  \varphi  \partial ^ \mu  \varphi 
	- V_\text{tree} ( \varphi  ) 
\end{align}
and the potential
\begin{align}
	V_\text{tree} ( \varphi  ) = \frac { \mu   ^ 2 } 2 \varphi  ^ 2 + 
	\frac { \lambda } 4 \varphi  ^ 4\ ,
	\label{V} 
\end{align}   
with $ \mu  ^ 2 < 0 $, so that the minima  of the potential are at 
$ \varphi  = \pm v = \pm\sqrt{ - \mu  ^ 2 / \lambda  } $,
spontaneously breaking the symmetry $ \varphi  \to - \varphi  $.
Now the mass of a particle in the constant ``background'' field $\varphi$ is
$ M ^ 2 ( \varphi  ) = V''_ { \rm tree } 
( \varphi  ) = \mu  ^ 2 + 3 \lambda  \varphi  ^ 2$. 
Equation (\ref{pressure}) then gives
\begin{align}
	V _ { \rm eff } ( \varphi  ) 
	= 
     \frac 12 \left ( \mu  ^ 2 + \frac { \lambda  } 4 T ^ 2 \right ) 
	\varphi  ^ 2 + \frac \lambda  4 \varphi  ^ 4
	+ O ( M ^ 3/T ) \ ,
	\label{veffreal} 
\end{align} 	
where we have omitted the $ \varphi  $-independent terms.
At finite temperature there is a positive contribution $ \lambda  T ^ 2/4 $
to the coefficient of the quadratic term, the so-called 
thermal mass (squared). It drives the expectation value to smaller values.
When $ T  \gg 2 \sqrt{ - \mu  ^ 2 / \lambda  } $, the expectation 
value vanishes and the symmetry is restored.
\begin{figure}[t]
\hspace{-9mm}
\includegraphics[width=0.4\textwidth]{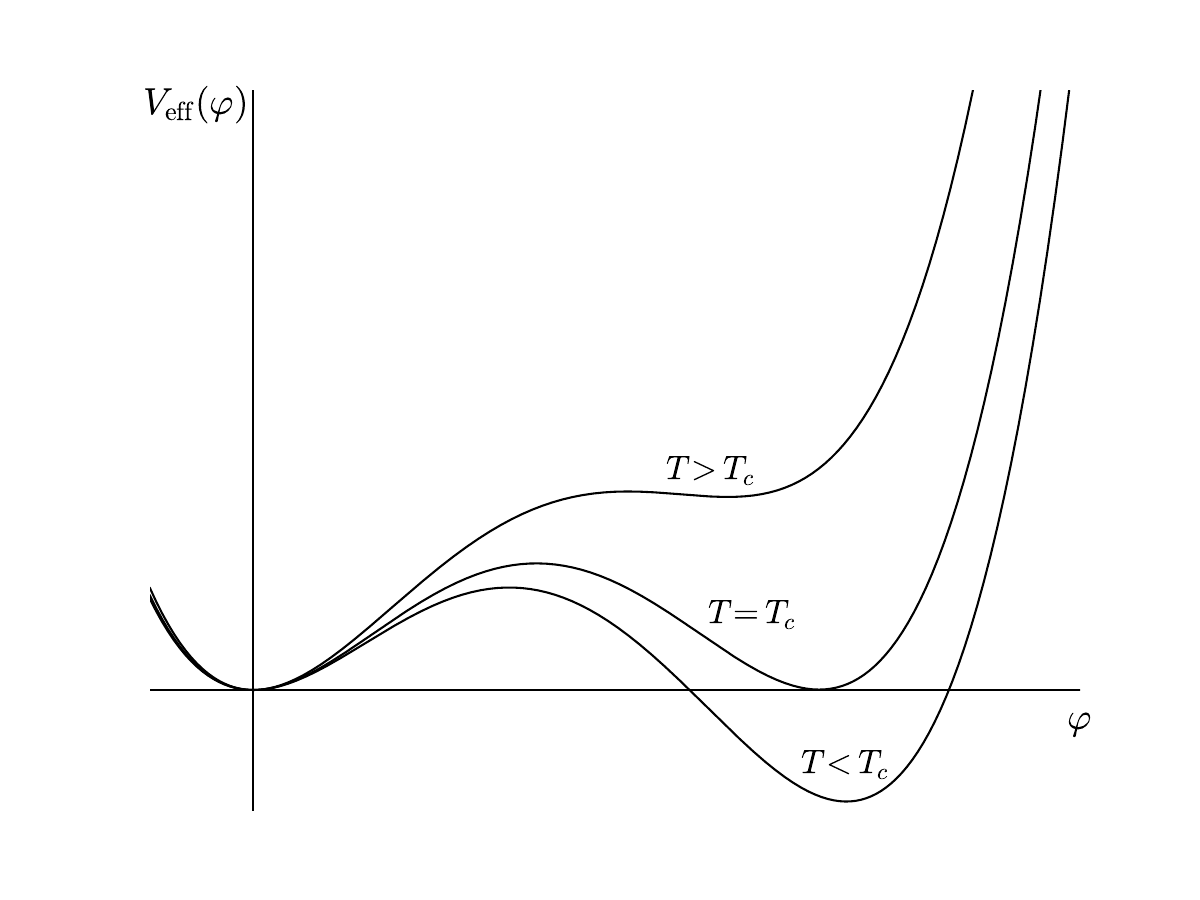}
\caption{Effective potential giving 
rise to a first-order  phase transition.
The value at $ \varphi  = 0 $ is subtracted.
}
\label{f:pt}
%
\end{figure}

One may worry that at small $ \varphi  $, with decreasing temperature
$ M ^ 2 ( \varphi  ) $
becomes zero and  
eventually negative, so that the $ O ( M ^ 3/ T ) $ 
term in  Eq.~(\ref{veffreal}) 
would give an imaginary part to the effective potential. 
However, it turns out that the loop-expansion parameter is $ \lambda  T/M $
\cite{Arnold:1992rz}.
Therefore, perturbation theory breaks down when $ M $ becomes too small
and is thus not reliable for determining the details 
of the phase transition. 
It is, in fact, second order, and the value of $ \varphi  $ changes 
continuously from zero above
 the critical temperature $ T _ c $ to a nonzero value below 
$ T _ c $. 

Next consider the SM with one Higgs doublet $ \phi  $. 
The tree-level potential is written as in Eq.~(\ref{V}) with $ \varphi  =
\sqrt{ 2 \phi  ^ \dagger \phi  } $.
Now all SM species contribute to the pressure and thus to
$ V _ { \rm eff } $. 
There is a qualitatively new effect compared to the previous example.
Since the electroweak gauge bosons obtain their mass from the Higgs
field and have no tree-level mass term, they contribute with 
$ M ^ 2 \sim g ^ 2 \phi  ^ \dagger \phi  $ in Eq.~(\ref{pressure}).
The $ M ^ 3 /T $ term in Eq.~(\ref{pressure})  then gives rise to 
a cubic term in the effective potential%
\footnote{The longitudinal gauge bosons receive a thermal mass, more precisely
the static screening mass, or Debye mass so that they do not contribute
to the cubic term. 
For simplicity, 
the resulting contribution is not shown in Eq.~(\ref{veffSM}). } 
\begin{align}
	V _ { \rm eff } ( \varphi  ) 
	= 
     \frac A2 \left ( T ^ 2 - T ^ 2 _ b \right ) 
	\varphi  ^ 2 
	- \frac B 3 \varphi  ^ 3 +  \frac \lambda  4 \varphi  ^ 4
	+ \cdots   .
	\label{veffSM} 
\end{align} 	
This potential would give a first-order phase transition, as illustrated
in Fig.~\ref{f:pt}.
At the critical temperature $ T _ c $ there are two degenerate minima.  
$ T _ b \equiv \sqrt{ -\mu  ^ 2 /A } $ is the temperature 
below which the potential barrier vanishes and 
the local minimum at $ \varphi  = 0 $ disappears. 

In the SM $ B $ is small. 
Therefore, the symmetry breaking minimum
$ \varphi  _ c $ is small, and so is the effective gauge-boson mass $ M $.
The loop-expansion parameter  $ g ^ 2 T /M $ is again large, so  
perturbation theory cannot be trusted. 
Using nonperturbative methods it was shown that for Higgs masses
larger than about 70-80 GeV, and thus in the SM,
there is no electroweak phase transition but a smooth crossover
\cite{Buchmuller:1994qy,Kajantie:1996mn,Csikor:1998eu}.
The Higgs expectation 
value changes continuously with temperature, as shown in
Fig.~\ref{f:higgs-ev}. Hence, during the transition the system stays
close to thermal equilibrium and Sakharov's third condition is
not satisfied.

A strongly first-order phase transition
is possible only in extensions of the SM. 
Since large field values imply large $ M ^ 2 ( \varphi  ) $,
the effective potential can be computed perturbatively. 
However, one may not be able to use the previously descrribed
high-temperature expansion,
in which case even the   one-loop effective
potential cannot be written in a simple analytic form.
A comprehensive discussion of the theoretical uncertainties was
recently given by \cite{Croon:2020cgk}.

Since the 
high-temperature phase is metastable as long as there is a potential
barrier separating the two minima, the Universe supercools to some
$ T < T _ c $; see \ Fig.~\ref{f:pt}. 
Bubbles of the symmetry-broken phase form through thermal fluctuations
with a probability that can be computed using 
a saddle-point approximation in statistical mechanics
\cite{Langer:1969bc}.
The probabilty of forming a bubble per time and volume is 
$ A \exp ( - \beta S _{\rm eff} [\phi_{\rm bubble}]) $, 
where the effective potential  $V _ {\rm eff} (\phi )
$ [see Eq.~(\ref{pathintegral})] has been replaced by the effective
action $S_{\rm eff} $ at the bubble configuration $ \phi_{\rm bubble}$ 
\cite{Linde:1980tt}.
It is the free energy of  a static configuration representing 
a barrier between the metastable state and a state with a bubble
of the low-temperature phase, similar to the sphaleron barrier;
cf.\ Sec.~\ref{sec:sphaleron}.
The temperature-dependent prefactor $ A $ is due to fluctuations
around the saddle point and can be computed perturbatively
\cite{Morrissey:2012db}.
Nonperturbative lattice computations of the nucleation rate
indicate that  perturbation theory 
slightly underestimates the strength of the phase transition
while overestimating the amount of supercooling 
\cite{Moore:2000jw}.

The bubbles nucleate roughly when the nucleation rate
equals $ H ^ 4 $, i.e., when one bubble nucleates per Hubble volume and 
time.%
\footnote{For a more precise criterion, see \citet{Bodeker:2009qy}.}
Since around the electroweak scale $ H \sim T _ c  ^ 2 /\mpl \sim 10^{ -17}T_c  $,
the rate is extremely small.
Once formed, the bubbles expand and begin to fill the entire Universe
with the low-temperature phase.
Important parameters of this process are the velocity $ \vw $ 
of the wall separating the two phases and their thickness $ \lw $
in the wall frame.
The bubble wall velocity is determined by the pressure difference 
between the two phases. 
The pressure consists of the vacuum contribution, i.e.~$-V_\text{eff}|_{T=0} $, and the pressure due to the plasma. 
When a particle mass depends on the value of the Higgs field 
it changes while the particle passes the wall. Therefore, 
there is a momentum transfer to the wall giving a contribution
to the pressure.
This includes a large contribution due to the magnetic-scale gauge fields
(see Sec.~\ref{sec:sphaleron}), which are suppressed in the
symmetry-broken phase and get pushed out by the wall 
\cite{Moore:2000wx}.
At the critical temperature the pressure difference between
the two phases vanishes. The system is static  and in thermal equilibrium.
Below $ T _ c $, the wall moves into the high-temperature phase,
the time dependence prevents the particle distribution
from equilibration, and one has to deal with a nonequilibrium problem.
One has to solve a set of Boltzmann equations which turns out
to be difficult. 
The wall velocity is quite model dependent: it can vary from 
$ \vw \ll 1 $ to $ \vw \sim 1 $ in the plasma rest frame.
For the SM it was found\footnote{ 
Assuming  a small Higgs mass $ m _ H < 90 $ GeV.%
}
that $ \vw \sim 0.4 $, and $ \lw T \sim 25 $ 
\cite{Moore:1995si}, while in the minimal supersymmetric standard model
(MSSM) $ \vw \lsim 0.1 $ 
\cite{John:2000zq}. 
Often the wall velocity is treated as a free parameter. 
A relatively simple case is ultrarelativistic bubbles with
$ \gw \equiv (1-\vw^2) ^{-1/2} \gg 1 $  \cite{Bodeker:2009qy}.
The reason is that the wall passes so fast that 
particles start scattering only when the wall has already passed.
There are models in which, based on this analysis, the bubble
wall can speed up indefinitely. However, additional radiation
off the particles passing the wall leads to a speed limit
\cite{Bodeker:2017cim,Hoche:2020ysm}.

\subsection{Charge transport mechanism}
\label{sec:transport}

When a phase-transition 
bubble wall sweeps through the plasma, it affects the motion of
the particles therein. 
The dominant effect is spin independent and contributes to
the pressure on the wall, as discussed in Sec.~\ref{sec:phase}.
Subleading but essential for baryogenesis is the $CP$-violating 
separation 
of particles with different spins.
On one side of the bubble wall there are more left-handed 
(negative helicity)
particles and their negative helicity antiparticles
than on the other side.
In the symmetric phase 
electroweak sphalerons are unsuppressed.
They act on left-handed particles and on right-handed antiparticles,
and thus wash out the baryon number $ B _ L $ carried by the left-chiral
fields describing left-handed particles and right-handed antiparticles.
If the weak-sphaleron rate is sufficiently suppressed on the other 
side of the wall, a net baryon number is generated;
for a comprehensive 
review of charge transport, see \citet{Konstandin:2013caa}.

One distinguishes the thin-wall limit $ \lw \sim T ^{ -1 } $
from the thick-wall limit $ \lw \gg T ^{ -1 }  $;
i.e., the de Broglie wavelength of a typical particle
$ T ^ { -1 } $ is small or of similar size relative to the 
wall thickness. 
In the former case, the particle-wall interaction is described
by quantum reflection and transmission
\citep*{Joyce:1994zn}.

In the thick-wall case 
the effect on the particles can be described as  a 
semiclassical force  \citep*{Joyce:1994zt}
that depends on their spin
\citep*{Cline:2000nw}.
Interactions  with the bubble wall 
give rise to space- and time-dependent mass terms,
which may contain a $CP$-violating phase.
For concreteness consider a single fermion field $ \psi $ with 
\begin{align}
	{ \cal L } _ { \rm mass } = 
	- \overline { \psi  _ L } m \psi  _ R
	- \overline { \psi  _ R } m ^ \ast  \psi  _ L\ 
	,
	\label{mass} 
\end{align} 
where $ m = | m | \exp ( i \theta  ) $.
Such a term can be due to interactions with varying
scalar fields like in Eq.~(\ref{lths})
in combination with the Yukawa interaction, or  
due to varying Yukawa couplings \citep*{Bruggisser:2017lhc}.
Bubble walls quickly grow to macroscopic sizes and thus can
be approximated as planar.
Let the wall move in the $ z $ direction.  
It is convenient to Lorentz boost
to the rest frame on the bubble so that  $ m $ depends only on $ z $.
One can expand in derivatives of $ m $, corresponding to an expansion
in $ ( \lw T ) ^{ -1 } $.
Keeping the first two terms, one obtains the semiclassical
force%
\footnote{%
The force was computed using the WKB approximation to the Dirac equation 
\citep{Kainulainen:2002th,Fromme:2006wx}
and from quantum field theory using Kadanoff-Baym equations 
\citep{Kainulainen:2002th,Prokopec:2003pj}.
}
\begin{align}
	\label{force} 
	F _ z =
	- \frac {\left ( |  m | ^ 2  \right )'  } { 2 E  } 
	+
	s \left [ 
				\frac { \left ( | m | ^ 2 \theta  ' \right ) '} 
				{ 2 E E _ z }  
				-
				\frac { | m | ^ 2 \left ( | m | ^ 2 \right ) ' \theta  '} 	
				{ 4 E  ^ 3 E _ z  } 
		\right ] \ ,
\end{align} 
with $ E = ( \vec p ^ 2 + | m | ^ 2 ) ^{ 1/2 }$, 
$ E _ z  = ( p _ z ^ 2 + | m | ^ 2 ) ^{ 1/2 }$,
and $ s = \pm 1 $ for spin
(as defined in the frame where the momentum transverse to the wall vanishes)
in the $ \pm z $ direction.
The prime denotes derivatives with respect to $z $.
The leading-order term is independent of spin. 
Because of the chiral nature of the mass term in Eq.~(\ref{mass}) 
there is a spin dependence, which first appears at second order in 
Eq.~(\ref{force}). 
Note that Eq.~(\ref{force}) holds for all four states of the fermion. 

\begin{figure}[t]
\begin{center}
\includegraphics[width=0.45\textwidth]{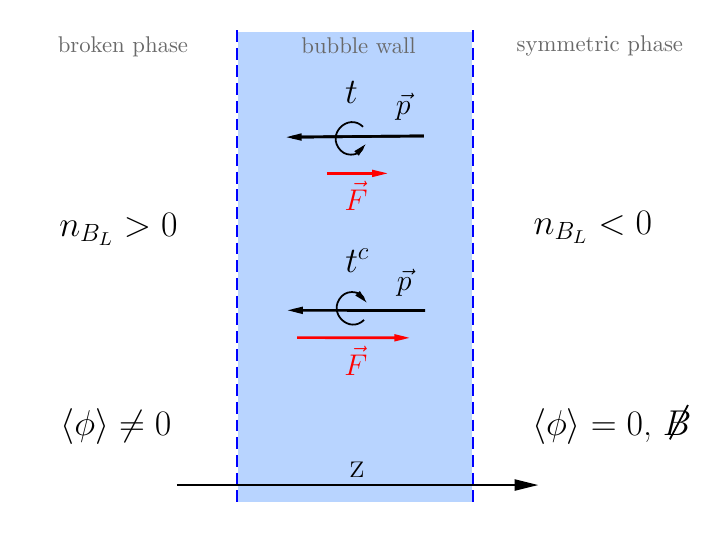}
\caption{
Sketch of the bubble wall (between the dashed lines)
in the rest frame of the wall.
More particles are crossing the wall from right to left. 
The force on tops with spin in the $ +z $ direction is smaller
than the force on antitops with spin in the $ -z $ direction. 
}
\label{f:forces}
\end{center}
\end{figure}

The forces on different (anti)particles are sketched in 
Fig.~\ref{f:forces} with top quarks as an example,
and with the square bracket in Eq.~(\ref{force}) assumed to be negative.
For all  (anti)tops the force is 
positive  and pushes them toward the symmetric phase. 
The spin-dependent term is negative for left-handed (anti)tops
and decreases the force acting on them, while it increases the 
force on right-handed (anti)tops. 
The left-handed tops carry positive $ B _ L $, while the right-handed 
antitops carry negative $ B _ L $. 
Therefore, the force changes the distribution of $ B _ L $ in space 
so that $ n _ { B _ L } $ becomes nonzero and $ z $ dependent.

The baryogenesis process is affected not only by the force
but also by scattering and the wall velocity.
In Fig.~\ref{f:forces} it is assumed that these effects
lead to $ n _ { B _ L } < 0 $ in the symmetric phase. 
Without electroweak sphalerons the 
total asymmetry vanishes: $n_B = n_{B_L} + n_{B_R} = 0$. 
In the symmetric
phase electroweak sphalerons are unsuppressed and diminish
$|n_{B_L}|$ leading to $n_B > 0$. 
Since electroweak sphalerons are not active in the broken
phase, this baryon asymmetry is frozen in after the phase transition
is completed.

For a quantitative description
the force is inserted into a Boltzmann equation, together with 
the collision terms describing particle scattering. 
For vanishing wall velocity the plasma is in local thermal
equilibrium. 
Thus, for small wall velocity one can make a fluid ansatz,
writing the phase-space densities as  
local equilibrium distributions with slowly varying 
chemical potentials, plus small perturbations $ \delta  f_ i $, 
representing deviations from kinetic equilibrium
\citep*{Joyce:1994zt}.
One then takes moments of the Boltzmann equations,
i.e., integrates over momentum with weights $ 1 $ and $ p _ z /E $.
The integrals of $ ( p _ z / E) \delta f _ i  $ represent corrections
$ \delta  v _ i $ to the local fluid velocity. 
One obtains a network of coupled differential equations for 
$ \delta  v _ i $ and $ \mu  _ i $. 
One must also include the effect of the weak and strong sphalerons.
The slowest interaction involves the weak-sphaleron transitions.
Therefore, the equations for the chemical potentials can
be computed by assuming baryon-number conservation, and 
finally the baryon asymmetry is computed from them.
The resulting asymmetry is directly proportional to the weak-sphaleron
rate in the symmetric phase.
While most works have assumed small wall velocity and expanded in
$ \vw $, baryogenesis with large $ \vw \sim 1 $ was recently studied
\cite{Cline:2020jre}. It was found, contrary to common lore, that 
baryogenesis with $\vw $ larger than the speed of sound is possible,
and that the generated asymmetry smoothly decreases with increasing 
$ \vw $.

During the entire process \BL~is unchanged because it is conserved
by the sphaleron processes. 
Therefore  the produced lepton asymmetry
is of the same order of magnitude as the baryon asymmetry. 
If a larger lepton asymmetry would be observed, this would rule out
electroweak baryogenesis as the sole origin of the baryon asymmetry
of the Universe. 

\subsection{Perturbative models}
\label{sec:models1}
In the SM the electroweak transition is only a smooth crossover but simple
extensions allow for a strongly first-order phase transition. The
first example to try is the two-Higgs-doublet model (2HDM), which has been
extensively studied in the literature: for a
review and references, see \citet{Branco:2011iw}.
\citet{Dorsch:2016nrg} thoroughly studied models of type II 
where leptons and down-type quarks couple to the Higgs doublet
$\Phi_1$ while up-type quarks couple to the second Higgs doublet
$\Phi_2$. The corresponding $\mathbb{Z}_2$ symmetry is softly broken
by a complex mass term $\mu^2$, and the scalar potential reads
\begin{equation}	
  \begin{split}
    V_{\rm tree}
    = 
-&\mu^2_1 \Phi_1^{\dagger}\Phi_1-
		\mu^2_2\Phi_2^{\dagger}\Phi_2-
		\frac{1}{2}\big(\mu^2\Phi_1^{\dagger}\Phi_2+{\rm H.c.}\big)\\
+&\frac{\lambda_1}{2}\big(\Phi_1^{\dagger}\Phi_1\big)^2+
		\frac{\lambda_2}{2}\big(\Phi_2^{\dagger}\Phi_2\big)^2\\
+&\lambda_3\big(\Phi_1^{\dagger}\Phi_1\big)\big(\Phi_2^{\dagger}\Phi_2\big)
+\lambda_4\big(\Phi_1^{\dagger}\Phi_2\big)\big(\Phi_2^{\dagger}\Phi_1\big)\\
+&\frac{1}{2}\big[\lambda_5\big(\Phi_1^{\dagger}\Phi_2\big)^2+{\rm h.c.}\big] .
   \end{split}
 \end{equation}
In addition to $\mu^2$ the quartic coupling $\lambda_5$ can be complex,
which leads to the complex vacuum expectation values
\begin{equation}
	\label{vevs}
	\langle\Phi_1\rangle = 
		\frac{1}{\sqrt{2}}
		\begin{pmatrix} 0 \\
				\,v\cos\beta\, 	
		\end{pmatrix}  ,\ 
	\langle\Phi_2\rangle = 
		\frac{1}{\sqrt{2}} 
		\begin{pmatrix} 0 \\
				\,v\sin\beta\,e^{i\theta}
		\end{pmatrix} .
              \end{equation}
In addition to the observed Higgs boson $h^0$, the model contains four
heavy Higgs bosons $H^0$, $A^0$ and $H^\pm$.
There are two field-redefinition-invariant phases that can be written as
\begin{equation}
	\delta_1 = {\rm arg}[(\mu^2)^2\lambda_5^*]\ ,\quad
	\delta_2 = {\rm arg}(v_1 v_2^*\, \mu^2 \lambda_5^*)\ . 
      \end{equation}
A benchmark scenario has been studied with $m_{H_0} =
200~\text{GeV}$ and $m_{A_0} = m_{H^\pm}$ at around $470~\text{GeV}$.
At the cost of some tuning the masses can be increased by about $100~\text{GeV}$.
The quartic couplings are large [$|\lambda_i| = \mathcal{O}(1)$] but satisfy
the perturbativity bound $|\lambda_i| \lesssim 2\pi$ and tree-level
unitarity, as well as constraints from flavor observables and the LHC.
For these parameters, the phase transition is strongly first order
($v_n/T_n \geq 1$), where $v_n$ is the jump of the Higgs expectation value at the
bubble nucleation temperature $T_n$. An interesting
aspect of the model is that, due to the large quartic scalar couplings, a
gravitational wave (GW) signal is predicted that would be observable at
LISA.\footnote{There is extensive literature on GWs from first-order
  phase transitions that lead to signals in the sensitivity range of
  LISA \cite{Caprini:2019egz}.}

\begin{figure}
\includegraphics[width=0.45\textwidth]{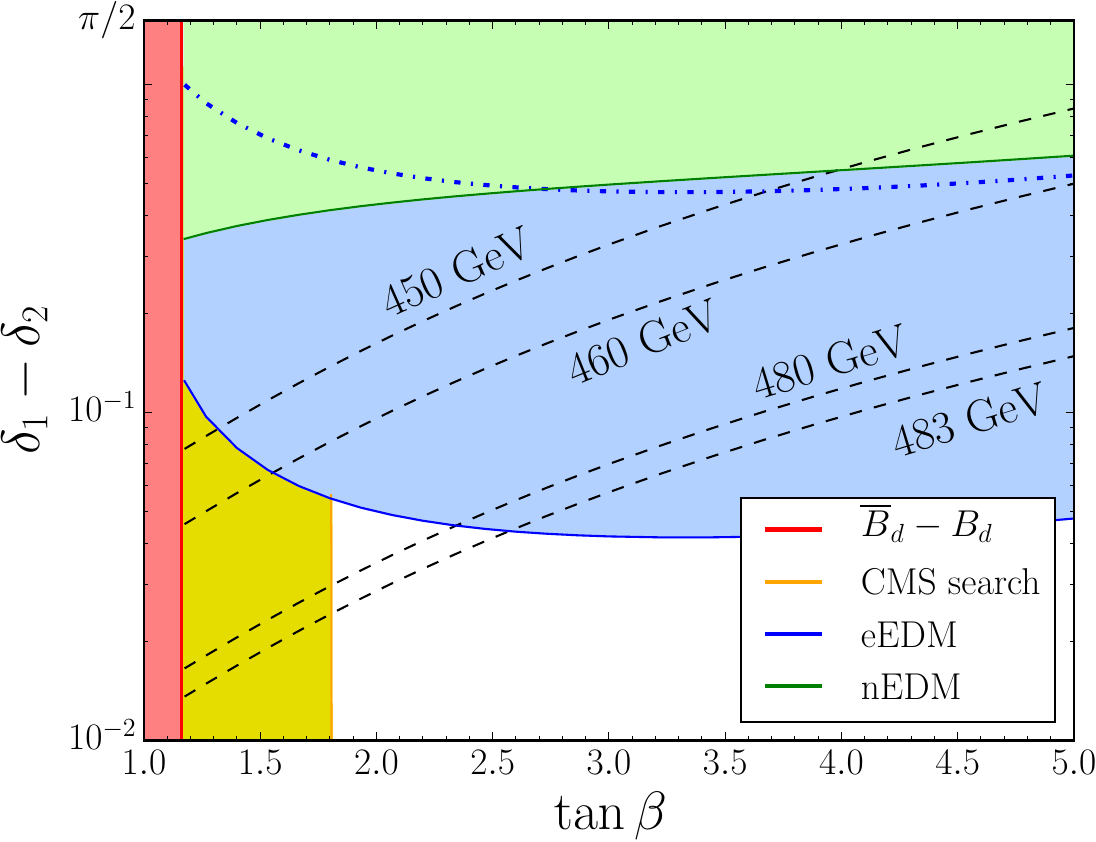}
\caption{EDM constraints for parameter benchmarks corresponding to
  different heavy-Higgs-boson masses in the \mbox{type-II} 2HDM.
  The dash-dotted line shows the electron EDM (eEDM) bound before
  the ACME experiment. The black dashed lines indicate
  the minimum $CP$-violating phase necessary for successful baryogenesis, with
  $m_{H^0} = 200~\text{GeV}$ and varying $m_{A^0} = m_{H^\pm}$.
  The green area is excluded by neutron EDM (nEDM) bounds, and the blue area is
  excluded by the electron EDM bound  from ACME (2014).
  From \citet{Dorsch:2016nrg}.}
\label{fig:EDM1}
\end{figure}
An attractive feature of electroweak baryogenesis models is also the
connection between the $CP$ violation needed for baryogenesis and low-energy
precision measurements. Particularly stringent are the following upper bounds on the
electron dipole moment (EDM) obtained by the ACME eperiment
\cite{Baron:2013eja,Andreev:2018ayy}:
\begin{equation}\label{ACME}
  \begin{split}
    \text{ACME~(2014)} :\quad |d_e|
    &< 8.7\times 10^{-29}\ e\ {\rm cm}\ ,\\
    \text{ACME~(2018)} :\quad |d_e|
    &< 1.1\times 10^{-29}\ e\ {\rm cm}\ .
\end{split}
\end{equation}
The ACME bound from 2014 is indicated in Fig.~\ref{fig:EDM1} by the
blue line, the lower boundary of the blue region. It is
consistent with all theoretical and phenomenological constraints on
the described model. The uncolored region represents the allowed
parameter region at that time. The ACME bound from 2018
improves this upper bound by a factor of $8.7$. This excludes
the parameter space of the model entirely.

For many years electroweak baryogenesis has also been studied in
supersymmetric 2HDM models (MSSM). In this case the quartic scalar couplings
are determined by gauge couplings. These models are now excluded due
to the lower bounds on superparticle masses obtained at the LHC. These
bounds and further theoretical constraints were described in detail by
\citet{Cline:2017jvp}, together with a discussion of some
nonsupersymmetric extensions of the standard model.

As an alternative to 2HDM models one can also consider a Higgs sector with
one SU(2)-doublet Higgs $\Phi$ and
an additional light SM singlet $s$, which
is partially motivated by composite Higgs models.
Electroweak
baryogenesis for such a setup was studied by \citet{Espinosa:2011eu}; 
see also \citet{Cline:2012hg}, \citet*{Bian:2019kmg}, and
\citet*{Carena:2019une}. 
The renormalizable part of the effective scalar potential reads
\begin{equation}
V_\text{tree} = V^{\textrm{even}}+V^{\textrm{odd}}\ ,
\end{equation}
with
\begin{equation}
\begin{split}  
V^{\text{even}}= &-\mu_h^2 |\Phi|^2+\lambda_h |\Phi|^4 \\ 
& - \frac{1}{2}\mu_s^2 s^2 +\frac{1}{4}\lambda_s s^4
+\frac{1}{2}\lambda_m s^2|\Phi|^2, \\ 
V^{\text{odd}}=&\frac{1}{2}\mu_m s |\Phi|^2 +\mu_1^3 s
+\frac{1}{3}\mu_3 s^3\ .
\end{split}
\end{equation}
The SU(2) doublet $\Phi$ contains the physical Higgs scalar
$h$. The potential $V^{\text{even}}$ is invariant with respect to the 
$\mathbb{Z}_2$ symmetry 
\begin{equation}
s\rightarrow -s\ ,
\end{equation}
which is softly broken by the potential $V^{\text{odd}}$. The vacuum
expectation value of $H$ implies mass mixing between $s$ and $h$.

\begin{figure}
\includegraphics[width=0.25\textwidth]{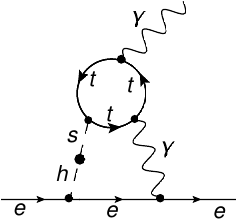}
\caption{Large contribution to the electron EDM from top loop and
  singlet-doublet mixing.
  From \citet{Espinosa:2011eu}.}
\label{fig:EDM2}
\end{figure}
An appropriate choice of quartic couplings and mass parameters
$\mu_i \sim 100~\text{GeV}$ lead
to a strongly first-order phase transition accompanied by
baryogenesis. The required CP violation is provided by a dimension-5
operator (see Fig.~\ref{fig:EDM2})
\begin{equation}
\mathcal{L}_{tHs} = \frac{s}{f} \Phi \bar{q}_{L3}(a+ib\gamma_5) t_R +
\text{h.c.}\ ,
	\label{lths} 
\end{equation}
which couples the top quark to the scalars $\Phi$ and $s$. During the
phase transition both scalars aquire an expectation value and
the profile of $s$ provides the $CP$-violating top-quark
scatterings. The compositeness scale has to be low ($f/b \sim
1~\text{TeV}$) so that strongly interacting resonances can be expected
in the LHC range. For the light singlet $s$, a mass is predited to be
comparable to the Higgs mass.

The mass mixing between $h$ and $s$ also generates an electron EDM;
see Fig.~\ref{fig:EDM2}. The analysis of \citet{Espinosa:2011eu}
was carried out while assuming the upper bound \cite{Hudson:2011zz}
\begin{equation}
d_e<1.05 \times 10^{-27}\ e\ \text{cm}\ .
\end{equation}
The improvement of this bound by 2 orders of magnitude by the
ACME experiment [Eq.~\eqref{ACME}] excludes the model in its original form.
A possible way out is to tune the parameters of the model such that
a two-step phase transition occurs, with $s\neq 0$ during baryogenesis and
$s = 0$ in the zero-temperature vacuum \cite{Kurup:2017dzf}. At zero temperature
the $\mathbb{Z}_2$ symmetry is then unbroken and the contribution to
the electron EDM vanishes. Choosing $m_s > m_h/2$, the Higgs-boson
decay width is unchanged and one obtains a ``nightmare scenario'' that
is difficult to test at the LHC \citep*{Curtin:2014jma}. For EWBG a
new source of $CP$ violation is needed, for instance, $CP$ violation in a
dark sector, which is transferred to the visible sector via a
new light vector boson \citep*{Carena:2018cjh}. However, such a
construction eliminates one of the main motivations for
electroweak baryogenesis: the connection between $CP$
violation measurable at low energies and the matter-antimatter asymmetry.

\begin{figure}
  \includegraphics[width=0.48\textwidth]{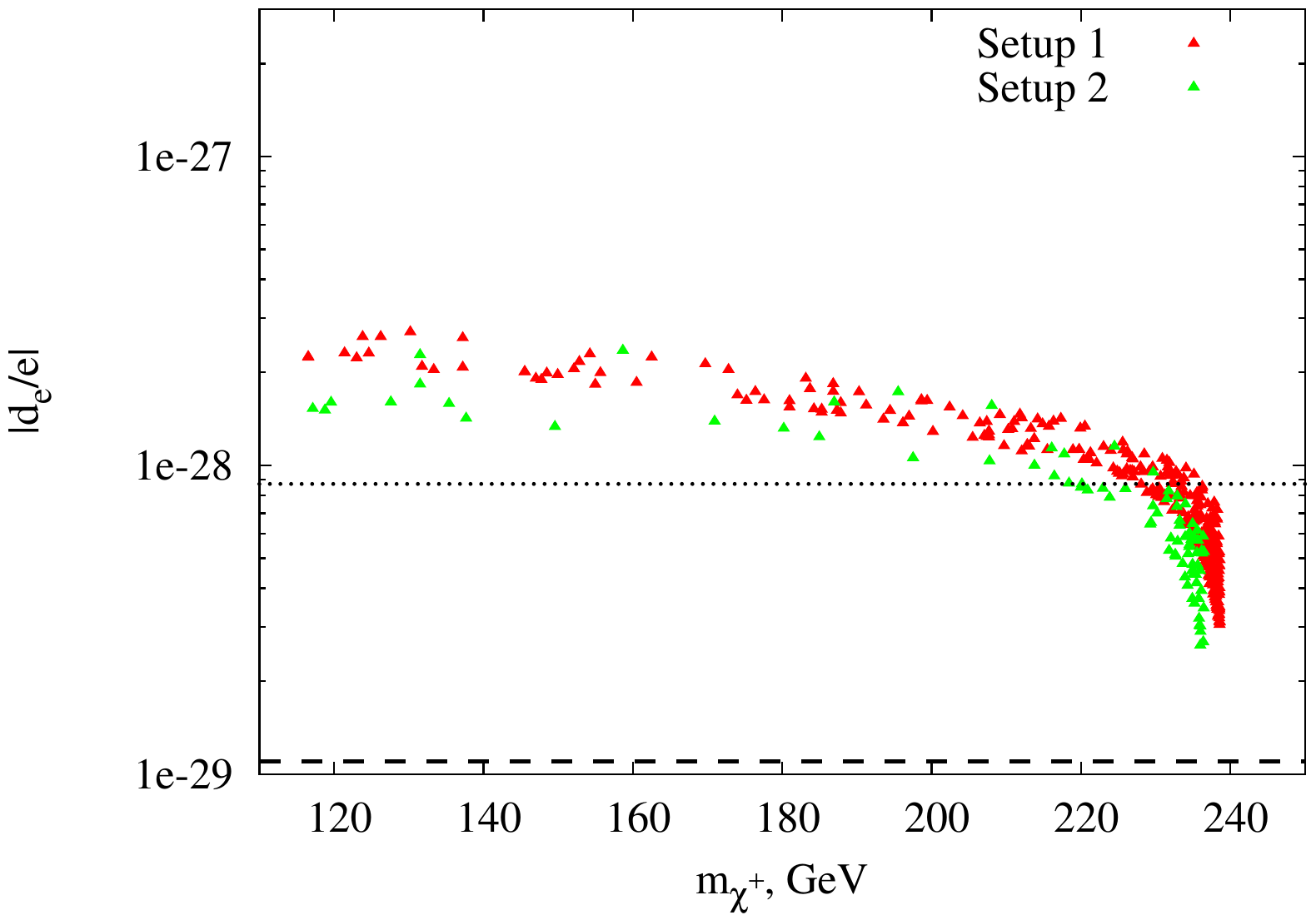}
  \caption{Electron EDM vs the lightest chargino mass in the split
    NMSSM for two parameter sets (setup 1, red dots; setup 2, green
    dots). The dotted line denotes the
    ACME (2014) upper bound $|d_e| < 8.7\times 10^{-29}\
    e\ \text{cm}$. We have added the dashed line that indicates the
    ACME (2018) upper bound $|d_e| < 1.1\times 10^{-29}\
    e\ \text{cm}$. Adapted from \citet*{Demidov:2016wcv}.}
\label{fig:sNMSSM}
\end{figure}
One may wonder whether EWBG can be more easily realized in models with
more scalar fields. An example is the split
next-to-minimal supersymmetric standard model (sNMSSM) \citep*{Demidov:2016wcv},
which contains two Higgs doublets $H_u$ and $H_d$ and an additional
singlet $N$. The corresponding superpotential reads
\begin{equation} 
 W=\lambda N H_u H_d+ \tfrac{1}{3} k N^3+\mu H_uH_d + r N \ . 
\end{equation}
Scalar quarks and leptons are removed from the low-energy spectrum but
gauginos have electroweak-scale masses. The dominant role in EWBG is
played by the scattering of charginos. At one-loop order they also
lead to an EDM for the electron. According to the
analysis of \citet*{Demidov:2016wcv}, a strongly first-order phase
transition and EWBG are compatible with the ACME (2014) bound on the
electron EDM. Figure~\ref{fig:sNMSSM} shows the electron EDM as a
function of the lightest chargino mass for two parameter benchmarks.
However, as the figure demonstrates, the stronger ACME (2018) bound
again excludes this model.

The connection between EWBG and electron EDM has also been analyzed in
a setup with the same particle content as the sNMSSM but without the
relations between the Yukawa couplings implied by supersymmetry  
\citep*{Fuyuto:2015ida}. As in the singlet-doublet model, a strongly first-order phase
transition is possible, and EWBG is driven by Higgsino and singlino
scatterings with masses $m_{\tilde{H}}$ and $m_{\tilde{S}}$,
respectively. At two-loop order an electron EDM is generated which
depends on the Higgsino masses. 
\begin{figure}
  \includegraphics[width=0.4\textwidth]{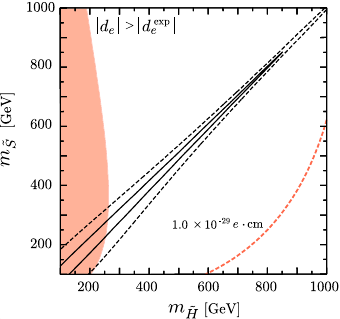}
 \caption{Contours of $\eta_B/\eta_B^\text{observed} =$~1~(solid
     lines) and 0.1~(dashed lines) along the
   $(m_{\tilde{H}},m_{\tilde{S}})$ plane. The orange area is
   excluded by the ACME (2014) bound. The orange dashed line
   corresponds to the anticipated sensitivity $|d_e| = 1.0\times
   10^{-29}\ e\ \text{cm}$, which essentially agrees with the ACME
   (2018) bound.
   From \citet*{Fuyuto:2015ida}.}
\label{fig:sNMSSM2}
%
\end{figure}
In Fig.~\ref{fig:sNMSSM2} a region of successful EWBG is shown along the 
$(m_{\tilde{H}},m_{\tilde{S}})$ plane for representative Higgsino
couplings. The orange area on the left is
excluded by the ACME (2014) bound, leaving a large range of viable
Higgsino and singlino masses. However, the ACME (2018) bound again
excludes this region. The electron EDM receives contributions from two
graphs that have charged and neutral gauge bosons in the loop,
respectively. Fine-tuning couplings, the contributions can cancel each
other out, which would eliminate the connection between
low-energy $CP$ violation and EWBG.

The upper bound on the electron EDM placed by the ACME experiment is
an impressive achievement. The experiment uses a heavy polar molecule,
thorium monoxide (ThO). In an external electric field it possesses
states whose energies are particularly sensitive to the electron
EDM. Moreover, the magnetic moment of these states is small, which
makes the experiment relatively impervious to stray magnetic fields.
A cryogenic beam source provides a high flux of ThO molecules.
In 2014 these techniques led to an upper bound on the electron EDM
more than 1 order of magnitude smaller than the best previous
measurements \cite{Baron:2013eja}. Four years later the upper limit
could be further improved by a factor of $8.7$ \cite{Andreev:2018ayy}.

Studies of electroweak baryogenesis with two Higgs doublets began in
the early 1990s \cite{Turok:1990zg,
McLerran:1990zh}, followed by other models with an extended Higgs
sector. Over the years the increasing lower
bound on the Higgs mass, and finally the discovery of a
$125~\text{GeV}$ Higgs boson, as well as bounds on the heavier
Higgs-boson
masses from flavor observables and the LHC strongly constrained
these models. Much progress was made in understanding the challenging
dynamics of electroweak baryogenesis, and the possible
connection to gravitational waves in the LISA frequency range was
explored. In a complementary way, upper bounds on dipole moments
played an increasingly important role since generic models of
electroweak baryogenesis connect low-energy $CP$ violation with the
baryon asymmetry of the Universe. As previously described, it appears that
finally these bounds have become so strong that they essentially
exclude all models of electroweak baryogenesis that can be treated
perturbatively. These developments over 30 years represent an example of
how the interplay of theory and experiment can guide us in our search
for physics beyond the standard model.

\subsection{Strongly interacting models}
\label{sec:models2}
\begin{figure}[b]
  \includegraphics[width=0.45\textwidth]{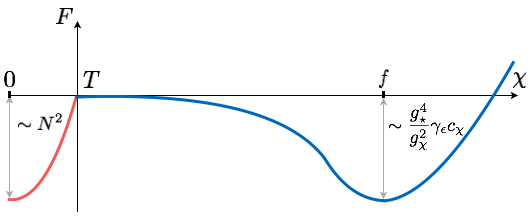}
  \caption{Schematic shape of the free energy as a function of the
    dilaton expectation value $\chi$. Red, hot region with $g_\chi
    \chi \lesssim T$; blue, cold region with $g_\chi
    \chi \gtrsim T$. From \citet{Bruggisser:2018mrt}.}
\label{fig:compHF}
\end{figure}

Thus far we have considered EWBG in perturbatively defined
renormalizable extensions of the SM. However, it is also possible that the
observed Higgs boson is a light state in a strongly interacting sector
of dynamical electroweak symmetry breaking.
This would qualitatively
change the electroweak phase transition as well as EWBG,
which can be treated by
means of an effective field theory \citep*{Grojean:2004xa}.
The light Higgs boson could emerge
from the spontaneous breaking of a global symmetry, such as SO(5) 
$ \rightarrow $ SO(4), 
together with a dilaton as pseudo-Nambu-Goldstone boson from
broken conformal symmetry in a strongly coupled hypercolor theory with partial
compositeness; for a review, see \citet{Panico:2015jxa}. In such a
framework EWBG was studied by
\cite{Bruggisser:2018mus,Bruggisser:2018mrt}
based on an effective Lagrangian with a minimal set of couplings and masses
\cite{Giudice:2007fh,Chala:2017sjk}. The analysis is based on the following
effective potential for the Higgs $h$ and the dilaton $\chi$:
\begin{equation}
  \begin{split}
    V_\text{eff} [h,\chi] &=  \left(\frac{g_\chi}{g_\star} \chi \right)^4
    \left[\alpha  \sin^2 \left(\frac{h}{f}\right)
    + \beta \sin^4 \left(\frac{h}{f}\right) \right] \\ 
&\quad + V_\chi(\chi) + \Delta V_T(h,\chi) \ ,
\end{split}
\end{equation}
where
\begin{equation}
\begin{split}  
\alpha[y]  &=  c_\alpha \sum_{i=1}^{N_f} \ g_\star^2 {N_c y_i^{2}[\chi]
  \over (4 \pi)^2} \ , \   
y_i[\chi]  \simeq  y_{0,i}  \left(\frac{\chi}{\chi_0}\right)^{\gamma_i}\ , \\
\beta[y] &= c_\beta \sum_{i=1}^{N_f} \ g_\star^2 {N_c  y_i^{2}[\chi] \over
  (4 \pi)^2}
\left({y\over g_\star}\right)^{p_\beta}\ . 
\end{split}
\end{equation}
The functions $y_i[\chi]$ connect left- and right-handed fermions,
$N_c=3$ is the number of QCD colors, $N_f$ is the number of quark
flavors, $\gamma_i$ are anomalous
dimensions, $f=0.8~\text{TeV}$ is the value of the condensate breaking SO(5), 
$g_*$ and $g_\chi$ are the couplings of heavy resonances and dilaton,
respectively, and $c_\alpha$ and $c_\beta$ are free parameters.
The effective potential has a discrete shift symmetry, $h \rightarrow
h + 2\pi f$, reflecting the Goldstone nature of the Higgs field, and it
is invariant with respect to scale transformations,
up to soft breaking terms contained in $V_\chi$, 
finite-temperature corrections in $\Delta V_T$, and the effect of nonzero anomalous
dimensions $\gamma_i$. The underlying strongly interacting theory has
$N$ hypercolors. The effective couplings of glueball-like and
mesonlike bound states are, respectively,
\begin{equation}
  g_\chi = \frac{4\pi}{N}\ \  (\text{glueball-like}) , \ 
    g_* = \frac{4\pi}{\sqrt{N}}\ \ (\text{mesonlike}) .
  \end{equation}
  Heavy resonances have masses $m_* = g_* f$. The dilaton can be
  glueball-like or mesonlike, 
  depending on the realization of conformal symmetry.
At high temperatures both Higgs and dilaton expectation values
vanish, and the free energy is determined by the number of hypercolors:
\begin{equation}
  F|_{\chi=0} \simeq  -\frac{\pi^2 N^2}{8} T^4 \ .
\end{equation}
Figure~\ref{fig:compHF} shows a sketch of the free energy together with
the zero-temperature dilaton potential.\footnote{Note that this figure
does not give a quantitative description of the two regions, in
particular, the phase transition that connects them.}
Around the critical temperature
\begin{equation}
  T_c \simeq \ 2 \left(\frac{g_\star^2}{4 \pi g_\chi N}\right)^{1/2}
  (2 \gamma_\epsilon c_\chi)^{1/4}f
\end{equation}
confinement and symmetry breaking phase transitions take place
that, due to the approximate conformal symmetry, can be strongly
first order. EWBG takes place by the scattering of quarks at the
bubble wall, where the $CP$ violation is enhanced by varying Yukawa
couplings \cite{Bruggisser:2018mus}. The model can account for the
observed baryon asymmetry and it predicts a GW signal that will be
probed by LISA \cite{Bruggisser:2018mrt}.

\begin{figure}
  \includegraphics[width=0.5\textwidth]{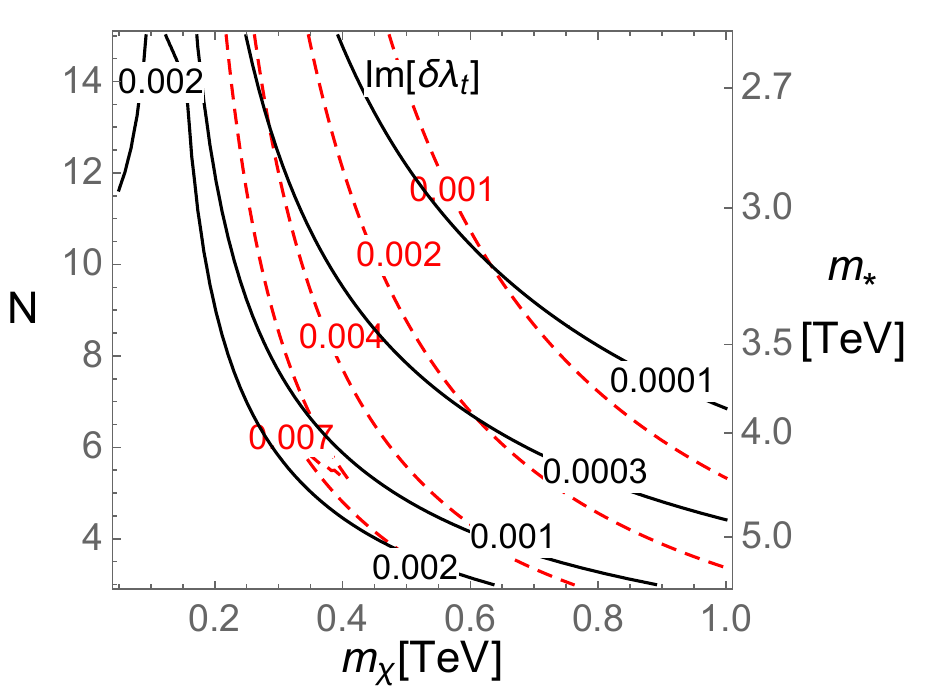}
  \caption{Contours of the $CP$-violating imaginary part of the top-Yukawa coupling
    in the $(m_\chi,m_\star)$ plane. The dashed lines correspond to a
    mesonlike dilaton, and the solid lines correspond to a glueball-like dilaton.
    From \citet{Bruggisser:2018mrt}.}
\label{fig:compHCPV}
\end{figure}

The $CP$-violating imaginary parts of quark-Yukawa couplings lead to an
electron EDM. Hence, the experimental EDM bounds constrain the viable
parameter space of the model. Figure \ref{fig:compHCPV} shows contours
of constant imaginary part for the top quark in the case of a
glueball-like dilaton as well as a mesonlike dilaton. The most stringent
bounds from the ACME experiment read
\begin{equation}
  \begin{split}
\text{ACME~(2014)}: \text{Im}[\delta\lambda_t] &\lesssim 2 \cdot
10^{-2}\ ,\\ 
\text{ACME~(2018)}: \text{Im}[\delta\lambda_t] &\lesssim 2 \cdot
10^{-3}\ .
\end{split}
\end{equation}
For a large number of hypercolors, $N=12$, corresponding to
resonance masses $m_* \gtrsim 3~\text{TeV}$, a mesonlike
(glueball-like) dilaton has to be heavier than $200~\text{GeV}$
($400~\text{GeV}$). Both scenarios can therefore be probed at the
LHC. Note that the effect of the ACME EDM bound on the Higgs sector
can be efficiently described by means of an effective field theory
\citep*{Panico:2018hal}.

The strong constraints from the LHC and the electron EDM give rise to
the question whether EWBG can be decoupled from low-energy physics.
Extending the scalar sector of the theory, it is indeed possible to
break the electroweak symmetry at a scale much higher than the Fermi
scale \citep*{Baldes:2018nel,Glioti:2018roy,Meade:2018saz}. In this way, $CP$ violation in EWBG is decoupled
from low-energy $CP$ violation. On the other hand, the need to connect
the high-scale vacuum expectation value to the Fermi scale requires
additional light scalars that are in reach of the LHC. Similarly, additional
light singlet fermions can lead to electroweak symmetry
nonrestoration at high temperatures. This can significantly relax the
upper bound from successful baryogenesis on a light dilaton in
composite Higgs models \cite{Matsedonskyi:2020mlz}.

\subsection{Summary: Electroweak baryogenesis}
\label{sec:summaryEW}
Electroweak baryogenesis is an appealing idea since it would allow to
connect the cosmological matter-antimatter asymmetry with physics at
the LHC and, moreover, with gravitational waves. The electroweak
phase-transition
and sphaleron processes are by now well understood. Since
in the standard model the phase transition is a smooth crossover,
extensions such as two-Higgs-doublet models or doublet-singlet models
are needed for electroweak baryogenesis. Results from the LHC strongly
constrain such models. Moreover, recent stringent upper bounds on the electron
electric dipole moment exclude most of the known models. This
led to the construction of models, where $CP$ violation in baryogenesis
and low-energy $CP$ violation are decoupled, and the electroweak phase
transition takes place at temperatures well above a TeV. On the other
hand, in strongly
coupled composite Higgs models electroweak baryogenesis is still
possible, which is compatible with all constraints from the LHC and low-energy
precision experiments. This underlines the importance of searching for
new heavy resonances and deviations from SM predictions for
Higgs couplings in the next run of the LHC.

\section{Leptogenesis}
\label{sec:leptogenesis}
In this section we first give an elementary introduction to the basics
of leptogenesis, namely, lepton-number violation and kinetic equations. We
then review thermal leptogenesis at the GUT scale as well as the weak
scale. Sterile-neutrino oscillations allow leptogenesis even at GeV
energies. Subsequently, we discuss recent progress toward a full
quantum field-theoretical description of leptogenesis. GUT-scale
leptogenesis is closely related to neutrino masses and mixings and,
on the cosmological side, it is connected with inflation and gravitational
waves.

\subsection{Lepton-number violation}
\label{sec:Lviolation}
The SM contains only left-handed neutrinos, and $B-L$ is
a conserved global symmetry. Hence, in the SM neutrinos are massless.
However, neutrino oscillations show evidence for nonzero
neutrino masses. These can be accounted for by introducing right-handed neutrinos
that can have Yukawa couplings with left-handed neutrinos. After
electroweak symmetry breaking these couplings lead to $B-L$ conserving
Dirac neutrino mass terms. As SM singlets, right-handed neutrinos can
have Majorana mass terms whose size is not constrained by the electroweak
scale. In the case of three right-handed neutrinos, the global $B-L$ symmetry can be  
gauged such that the Majorana masses result from the spontaneous breaking of
$B-L$. As in the SM, all masses are then generated by the spontaneous
breaking of local symmetries, which is the natural picture in theories
that unify strong and electroweak interactions. Since no $B-L$ gauge
boson has been observed thus far, the scale of $B-L$ breaking must be
significantly larger than the electroweak scale.
This leads to the seesaw mechanism
\citep*{Minkowski:1977sc,Yanagida:1979as,GellMann:1980vs,Ramond:1979py}
as a natural explanation of the smallness of the
observed neutrino mass scale, which is a key element of leptogenesis.

We now consider an extension of the standard model with three
right-handed neutrinos, whose masses and couplings are
described by the following Lagrangian (sum over $i,j$):
\begin{align}\label{SMnu}
  \mathcal{L} =&
  \overline{l_L}_i i\cancel D l_{L i} + \overline{e_R}_ii\cancel D e_{R i}
+ \overline{\nu_R}_i i\cancel\partial \nu_{R i} \\
              &-\big(h^e_{ij} \overline{e_R}_j  l_{L i} \tilde{\phi}
                + h^\nu_{ij}\overline{\nu_R}_jl_{L i} \phi
                + \frac{1}{2} M_{ij} \overline{\nu_R}_j \nu_{R i}^c
                + \text{h.c.}\big) ,
                \nonumber
\end{align}
where $\cancel D$ denotes SM covariant derivatives, 
$\nu_{R}^{c}=C\bar{\nu}_{R}^{T}$, $C$ is the charge conjugation matrix 
and $\widetilde{\phi} = i\sigma_2 \phi^*$. The vacuum expectation
value of the Higgs field ($\langle \phi \rangle = v_{\text EW}$)
generates Dirac mass terms  $m_e = h^e v_{\text EW}$ and
$m_D = h^\nu v_{\text  EW}$ for charged leptons and neutrinos, respectively. Integrating out the
heavy neutrinos $\nu_R$,  the light-neutrino Majorana mass matrix becomes
\begin{equation}\label{seesaw}
  m_\nu = -m_D\frac{1}{M}m^T_D \ .
  \end{equation}
The symmetric mass matrix is diagonalized by a unitary matrix $V$:
\begin{equation}
  V^T m_\nu V =
\begin{pmatrix} m_1 & 0 & 0 \\ 0 & m_2 & 0 \\ 
0 & 0 & m_3 \end{pmatrix} \ ,
\end{equation}
where $m_1$, $m_2$, and $m_3$ are the three mass eigenvalues. In the
following we mostly consider the case of normal ordering,
where $m_1 < m_2 < m_3$. A recent global analysis found for the
\begin{figure*}
\includegraphics[width=0.6\textwidth]{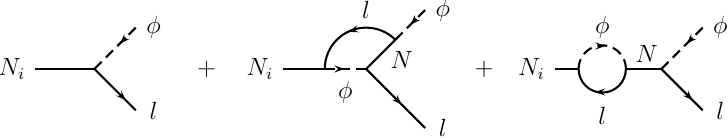}
\caption{Tree-level and one-loop diagrams contributing to heavy-neutrino decays}
\label{fig:oneloop}
\end{figure*}
largest and smallest splitting \cite{Esteban:2018azc}
\begin{equation}\label{atm}
\begin{split}  
m_{\text{atm}} &\equiv \sqrt{|m_3^2-m_1^2|} = 49.9\pm 0.3\ \text{meV}
\ , \\
m_{\text{sol}} &\equiv \sqrt{|m_2^2-m_1^2|} = 8.6\pm
0.1\ \text{meV} \ .
\end{split}
\end{equation}
The Majorana mass matrix $M$ can be chosen diagonal such that the light and heavy
Majorana neutrino mass eigenstates are
\begin{equation}\label{PMNS}
  \nu \simeq V^T\nu_L + \nu_L^c V^*\ , \quad N \simeq \nu_R +
  \nu_R^c\ .
\end{equation}
In a basis where the charged lepton matrix $m_e$ and the Majorana mass
matrix $M$ are diagonal, $V$ is the Pontecorvo-Maki-Nakagawa-Sakata
matrix in the leptonic charged current. $V$ can be written as
$V = V_\delta~\text{diag}(1,e^{i\alpha},e^{i\beta})$, where $V_\delta$
contains the Dirac $CP$-violating phase $\delta$ and
$\alpha$ and $\beta$ are Majorana phases.

Treating in the Lagrangian \eqref{SMnu} the Yukawa coupling $h^\nu$
and the Majorana masses $M$ as free parameters, nothing can be said
about the values of the light-neutrino masses. Hence, it is remarkable
that the correct order of magnitude is naturally obtained in GUT models.
The running of the SM gauge couplings points to a unification scale
$\Lambda_{\text{GUT}} \sim 10^{15}~\text{GeV}$. At this scale the GUT
group containing U(1)$_{B-L}$ is spontaneously broken and large Majorana
masses are generated ($M \propto v_{B-L} \sim 10^{15}~\text{GeV}$). As in
the SM, all masses are now caused by spontaneous symmetry breaking.
With Yukawa couplings in the neutrino sector having a similar pattern
as quarks and charged leptons, with the largest values being
$\mathcal{O}(1)$, one obtains for the largest light-neutrino mass
\begin{equation}\label{nuGUT}
  m_3 \sim \frac{v^2_{\text{EW}}}{v_{B-L}} \sim 0.01~\text{eV}\ ,
\end{equation}  
which is qualitatively consistent with the measured value $m_{\text{atm}}$.

The tree-level decay width  of the heavy Majorana neutrino $N_i$ reads
\begin{equation}
    \Gamma^0_{N_i} =\Gamma^0(N_i \rightarrow l\f) + \Gamma^0(N_i \rightarrow
    \al\af) = \frac{(h^{\nu\dagger}h^\nu)_{ii}}{8\pi} M_i ,
\end{equation}
and the $CP$ asymmetry in the decay is defined as
\begin{equation}\label{epsilon}    
    \ve_i = \frac{\Gamma(N_i \rightarrow l\f) - \Gamma(N_i \rightarrow \al\af)}
{\Gamma(N_i \rightarrow l\f) + \Gamma(N_i \rightarrow \al\af)}\ .
\end{equation}
We are often interested in the case of   hierarchical Majorana masses
$M_{2,3}\gg M_{1}\equiv M$.
One can then integrate out $N_2$ and $N_3$, which yields the following effective
Lagrangian for $N_1\equiv N$:
\begin{equation}\label{SMN}
\begin{split}
  \mathcal{L} =  & \frac{1}{2}\overline{N}i\cancel\partial N 
  -  h^\nu_{i1} N^TC l_{L i}\phi - \frac{1}{2}M N^T C N \\
      &  +\frac{1}{2}\eta_{ij} l_{L i}^T \phi Cl_{L j}\phi +
      \text{H.c.} \ ,
\end{split}
\end{equation}
where $\eta$ is the dimension-5 coupling
\begin{align}
\eta_{ij}=\sum_{k=2,3} h^{\nu}_{ik}\frac{1}{M_k} h_{kj}^{\nu T}\ .
\end{align}
Using this effective Lagrangian provides the advantage that vertex- and
self-energy contributions to the $CP$ asymmetry in the heavy-neutrino decay
are obtained from a single Feynman diagram;
see Sec.~\ref{sec:theoryLG}.

A nonvanishing $CP$ asymmetry in $N_i$ decays arises at one-loop
order. From Fig.~\ref{fig:oneloop} one obtains
\citep*{Covi:1996wh,Flanz:1994yx},
\begin{equation}\label{CPasymmetry}
  \ve_i = 
  -\frac{1}{8\pi}\sum_{i\neq k}\frac{\text{Im}\big(h^{\nu\dagger}h^\nu\big)^2_{ik}}
  {\big(h^{\nu\dagger}h^\nu\big)_{ii}}\ F\left(\frac{M_k^2}{M_i^2}\right) \ .
\end{equation}
In the case of hierarchical heavy neutrinos one obtains
\begin{equation}\label{hierarchical}
  F\left(\frac{M_k^2}{M_i^2}\right) \simeq -\frac{3}{2}\frac{M_i}{M_k}\ ,
\end{equation}
and the $CP$ asymmetry can be written as
\begin{equation}\label{CPasymmetry2}
\ve_i = 
-\frac{3}{16\pi}\frac{M_i}{v^2_{\text{EW}}\big(h^{\nu\dagger}h^{\nu}\big)_{ii}}
\text{Im}\big(h^{\nu\dagger}m_\nu h^{\nu *}\big)_{ii} \ .
\end{equation}
  For small mass differences, $|M_i-M_k| \ll M_i+M_k$, the $CP$
  asymmetry is dominated by the self-energy contribution\footnote{The
    self-energy part in Fig.~\ref{fig:oneloop} is part of the inverse
    heavy-neutrino propagator matrix.
    Unstable particles are defined as poles in $S$-matrix elements of
    stable particles whose residues yield their couplings. Such a
    procedure confirms the results of Eqs.~\eqref{hierarchical} and
    \eqref{degenerate} to leading order in the couplings \cite{Buchmuller:1997yu}.}  in
  Fig.~\ref{fig:oneloop} and enhanced \citep*{Covi:1996wh}:
  \begin{equation}\label{degenerate}
    F\left(\frac{M_k^2}{M_i^2}\right) \simeq -\frac{M_iM_k}{M^2_k -
      M_i^2}\ .
  \end{equation}
  Once mass differences become of the order of the decay widths, one
  reaches a resonance regime \cite{Covi:1996fm,Pilaftsis:1997jf} where resummations are necessary.

  Thus far we have considered the seesaw mechanism with right-handed
  neutrinos, often referred to as the type-I seesaw. Alternatively,
  light-neutrino masses can result from couplings to heavy SU(2) triplet
  fields \cite{Mohapatra:1979ia,Lazarides:1980nt,Mohapatra:1980yp,Wetterich:1981bx}, which is referred to as the
  type-II seesaw. In this case the complete light-neutrino mass matrix reads 
  \begin{equation}
  m_\nu = -m_D\frac{1}{M}m^T_D + m_\nu^\text{triplet}\ .
\end{equation}
Such matrices are obtained in left-right symmetric extensions of the
standard model; for a review, see \citet{Mohapatra:2006gs}.
Furthermore, one can consider the exchange
of heavy SU(2) triplet fermions, which is referred to as the type-III
seesaw \cite{Foot:1988aq}.

In addition to the Majorana mass matrix $M$ the charged lepton mass matrix 
$m_e = h^e v_{\text EW}$ can be chosen diagonal and real without loss
of generality. The Dirac neutrino mass matrix $m_D$ is then a general
complex matrix with nine complex parameters and therefore nine possible
$CP$-violating phases. Three of these phases can be absorbed into the
lepton doublets $l_L$, and hence six $CP$-violating phases remain physical.
These are known as high-energy phases, and the $CP$ asymmetries $\ve_i$ in $N_i$
decays depend on these phases. The light-neutrino mass matrix is
symmetric, with six complex parameters. As before, three of the phases can be
absorbed into the lepton doublets $l_L$, so three phases are
physical: the Dirac phase $\delta$ that is measured in neutrino
oscillations and two Majorana phases $\alpha_{1,2}$ that affect the
rate for neutrinoless double-$\beta$ decay
\citep*{Bilenky:1980cx,Schechter:1980gr}. There is no direct link
between the high-energy and low-energy $CP$-violating phases,
but interesting connections exist in particular models
\citep*{Branco:2011zb}.

 \subsection{Kinetic equations}
\label{sec:keqs}
Thermal leptogenesis is an intricate nonequilibrium process in the
hot plasma in  
the early Universe that involves decays, inverse decays, and scatterings of
heavy Majorana neutrinos $N$, left-handed leptons $l$ and $\al$, complex
Higgs scalars $\f$ and $\af$, gauge bosons, and quarks. 
A key role is played by weakly
coupled heavy Majorana neutrinos. In the expanding Universe they first
reach thermal equilibrium and then fall out of thermal
equilibrium, such that  $CP$- and lepton-number-violating processes 
lead to a lepton asymmetry and, via sphaleron processes, also a
baryon asymmetry. 

The main ingredients of the nonequilibrium process can be understood
by considering a simple set of Boltzmann equations, neglecting the
differences between Bose-Einstein and Fermi-Dirac distribution
functions, as in classical GUT baryogenesis
\cite{Kolb:1990vq,Kolb:1979qa,Harvey:1981yk}. 
Relativistic corrections and a full quantum field-theoretical
treatment are discussed in Sec. \ref{sec:theoryLG}.
For simplicity,
we restrict ourselves in the following to hierarchical heavy neutrinos where the
lightest one ($N$)  with mass $M$ dominates leptogenesis. We also sum
over lepton flavors in $N$ decays (one-flavor approximation).

We assume that at high temperatures $T \gg M$ the heavy
neutrinos are in thermal equilibrium, i.e.,
\begin{equation}
n_N = \tfrac{3}{4} n_\gamma \ ,
\end{equation}
where $n_\gamma$ is the photon number density, and the factor $3/4$
reflects the difference between the Bose and Fermi statistics. The heavy neutrinos
decay at a temperature $T_d$, which is determined by $\Gamma_N \sim
H(T_d)$, where $\Gamma_N$ and $H$ are  the decay width and
Hubble parameter, respectively. For leptogenesis one has
$T_d \lesssim M$ and the number density $n_N(T_d)$  slightly exceeds 
the equilibrium number density. This departure from thermal
equilibrium, together with the $CP$-violating partial decay widths
[see Eq.~\eqref{epsilon}], leads to the lepton asymmetry
\begin{equation}\label{maxasym}
\frac{n_l - n_{\al}}{n_\gamma} 
= \ve 
\frac{n_N}{n_\gamma} \sim \frac{3}{4} \ve \ .
\end{equation}

More realistically, one has to include inverse decays $l\f,\al\af
\rightarrow N$ in the calculation of the asymmetry. In general, the
time evolution of a system is governed by reaction densities. i.e., the
number of reactions $a + b + \ldots \rightarrow c + d + \ldots$ per time and volume:
\begin{align}\label{lowestorder}
  &\gamma(a + b + \ldots \rightarrow  c + d + \cdots) =\\
  &\int d\Phi f_a(p_a)  f_b(p_b)\ldots 
  |\M(a + b + \ldots\rightarrow c + d +\ldots)|^2 , \nonumber
\end{align}
where in first approximation $\M$ is a
zero-temperature \mbox{$S$-matrix} element and 
\begin{equation}
d\Phi = \frac{d^3p_a}{(2\pi)^3 2 E_a}
\ldots (2\pi)^4\delta^4(p_a+\cdots -p_c - \ldots) 
\end{equation}
is the phase-space volume element. Important thermal and quantum
  corrections to Eq.~\eqref{lowestorder} are discussed in Sec. \ref{sec:theoryLG}.

It turns out that in the considered scenario kinetic equilibrium is
a good approximation. In this case the distribution functions differ
from the corresponding equilibrium distribution functions only by the
normalization:
\begin{equation}
f_a(p) = \frac{n_a}{n^\eq_a}f_a^\eq(p) \ ,
\end{equation}
and reaction densities are proportional to equilibrium reaction
densities, 
\begin{equation}
\gamma(N\rightarrow l\f) = \frac{n_N}{n_N^\eq}\gamma^\eq(N\rightarrow
l\f)\ .
\end{equation}
Taking the expansion of the Universe into account, one then
obtains for the change of the heavy-neutrino number density with time
\begin{align}\label{Nchange1}
\dot{n}_N + 3Hn_N = &-\frac{n_N}{n_N^\eq}[\gamma^\eq(N\rightarrow l\f)
+ \gamma^\eq(N\rightarrow \al\af)] \nonumber\\
&+\gamma^\eq(l\f\rightarrow N)
+ \gamma^\eq(\al\af\rightarrow N) \ .
\end{align}
The reaction densities for neutrino decays into $CP$-conjugate final
states differ by the $CP$ asymmetry $\ve$:
\begin{equation}
  \begin{split}
 \gamma^\eq(N\rightarrow l\f) &= \frac{1+\ve}{2} \gamma_N\ , \\
\gamma^\eq(N\rightarrow \al\af) &= \frac{1-\ve}{2} \gamma_N\ , 
\end{split}
\end{equation}
and the reaction densities for decays and inverse decays are related by
$CPT$ invariance:
\begin{equation}\label{cpt}
  \begin{split}
\gamma^\eq(\al\af\rightarrow N) &= \gamma^\eq(N\rightarrow l\f)\ ,\\
\gamma^\eq(l\f\rightarrow N) &= \gamma^\eq(N\rightarrow \al\af) \ .
\end{split}
\end{equation}
Together with Eq.~\eqref{Nchange1}, Eq.~\eqref{cpt} yields the kinetic equation
for the heavy-neutrino number density:
\begin{align}\label{Nchange2}
\dot{n}_N + 3Hn_N = -\left(\frac{n_N}{n_N^\eq} -1\right)\gamma_N .
\end{align}
Integrating Eq.~\eqref{Nchange2} yields the time dependence of the
$N$-number density, which is determined by the expansion of the Universe and the
departure from thermal equilibrium.
\begin{figure}
\includegraphics[width=0.5\textwidth]{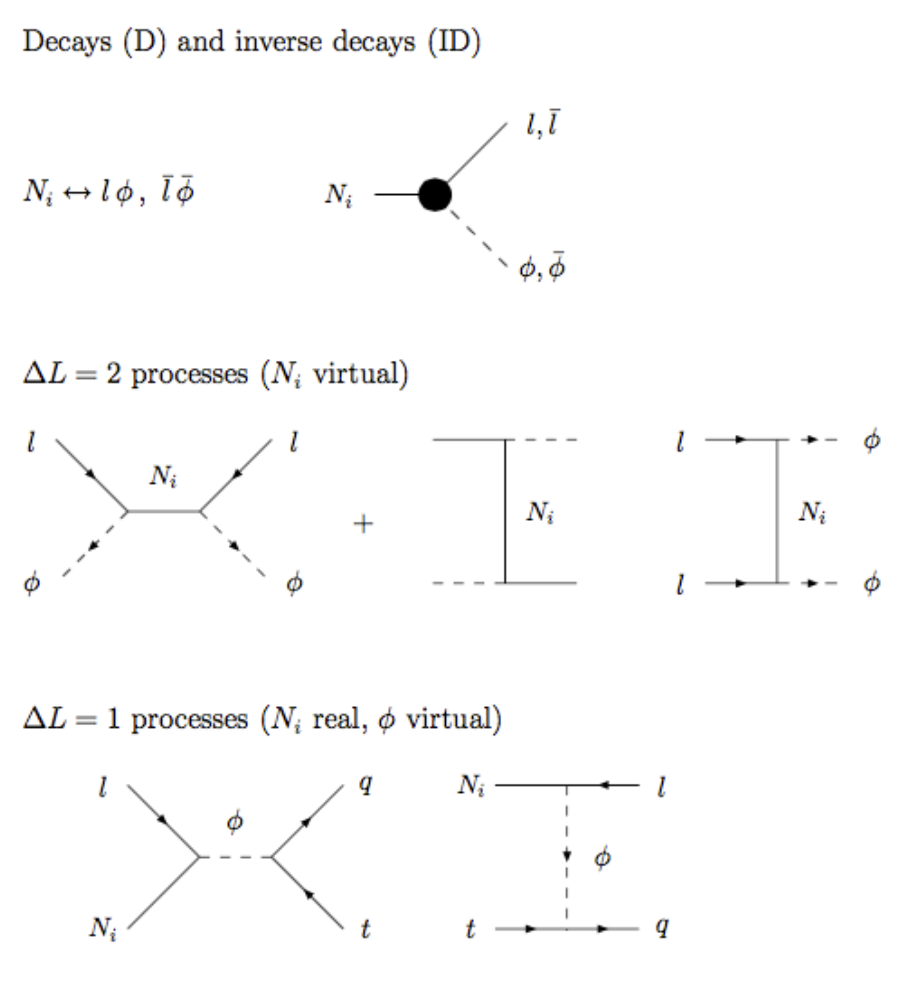}
\caption{Decays and inverse decays of heavy neutrinos, 
  \mbox{$\Delta L = 2$} processes with virtual intermediate heavy
  neutrinos, and $\Delta L = 1$ scattering processes.}
\label{fig:plasma_processes}
\end{figure}

The lepton asymmetry is generated by heavy-neutrino decays and inverse
decays as well as $2\rightarrow 2$ processes (see
Fig.~\ref{fig:plasma_processes}) with reaction densities such as
\begin{align}
\gamma(l\f\rightarrow \al\af) = \int d\Phi
f_l(p_1)f_\f(p_2)|\hat{\M}(l\f \rightarrow \al\af)|^2 \ .
\end{align}
Here $\hat{\M}$ is the matrix element for $l\f \rightarrow \al\af$, from
which the contribution of $N$ as the real intermediate state (RIS) has
been subtracted since this is already accounted for by decays and
inverse decays. Neglecting the effects of Fermi and Bose statistics
the distribution functions can be approximated as
\begin{align}
f_{l,\f}(p) = \frac{n_{l,\f}}{n^\eq_{l,\f}} f^\eq_{l,\f}(p) \simeq
  \frac{n_{l,\f}}{n^\eq_{l,\f}} e^{-\beta[E(p) - \mu_{l,\f}]} \ ,
\end{align}
where $\mu_l$ and $\mu_\f$ are the chemical potentials of the lepton and
the Higgs boson, respectively.
The change of the lepton-number density with time is given by
\begin{equation}\label{lchange}
 \begin{split}
\dot{n}_l + 3Hn_l &= \frac{n_N}{n_N^\eq}\gamma^\eq(N\rightarrow l\f)
-\frac{n_l}{n_l^\eq}\gamma^\eq(l\f \rightarrow N) \\
 &+\frac{n_{\al}}{n_{\al}^\eq}\gamma^\eq(\al\af\rightarrow l\f)
-\frac{n_l}{n_l^\eq}\gamma^\eq(l\f \rightarrow \al\af)  .
\end{split}
\end{equation}
The corresponding equation for $n_{\al}$ is obtained by interchanging
$l,\f$ with $\al,\af$. An important property of the decay and scattering
processes in the plasma is the unitarity of the zero-temperature $S$ matrix,
\begin{align}
\sum_i [|\M(l\f\rightarrow i)|^2 - |\M(i\rightarrow l\f)|^2] = 0 \ .
\end{align}
For $i = l'\f', \al\af$ with $E_l + E_\f = E_{l'} + E_{\f'} = E_{\al} +
E_{\af}$, this implies\footnote{This also holds for the RIS subtracted
  matrix elements.}
\begin{align}\label{unitarity}
\sum_{l\f,\al\af} [|\M(l\f\rightarrow \al\af)|^2 - |\M(\al\af\rightarrow
l\f)|^2] = 0 \ .
\end{align}
Expressing the lepton-number densites in terms of the $B-L$ number
density\footnote{Here we follow the usual treatment and ignore
sphaleron processes during the generation of the lepton asymmetry.
Sphaleron effects are then included by relating the final $L$ or
$B-L$ asymmetry to the baryon asymmetry using Eq.~\eqref{connection};
see Eq.~\eqref{etaB}. This amounts to neglecting ``spectator processes''
that can be taken into account in a more complete treatment
\cite{Buchmuller:2001sr,Nardi:2005hs,Garbrecht:2014kda}.}
\begin{equation}
n_l = n_l^\eq - \tfrac{1}{2}n_{B-L}\ , \quad   n_{\al} = n_l^\eq +
\tfrac{1}{2}n_{B-L}\ ,
\end{equation}
one obtains from Eqs.~\eqref{lchange} and \eqref{unitarity} the following
kinetic equation for the $B-L$ density:
\begin{equation}\label{BLchange1}
\dot{n}_{B-L} + 3Hn_{B-L} = -\ve\left(\frac{n_N}{n_N^\eq} \\
  -1\right)\gamma_N -\frac{1}{2}\frac{n_{B-L}}{n_l^\eq}\gamma_N\ .
\end{equation}
The generation of the $B-L$ asymmetry is driven by the departure of
the heavy neutrinos from equilibrium and the $CP$ asymmetry $\ve$, and
inverse decays also cause a washout of an existing $B-L$ asymmetry.
Note that only the reaction density for $N$ decays enters into Eq.~\eqref{BLchange1};
the reaction density for the two-to-two process in Eq.~\eqref{lchange}
drops out.

An important part of the $B-L$ washout is the $\Delta L =2$
processes $l l \rightarrow \f \f$ and $l \f \rightarrow \al \af$ with RIS
subtracted reaction densities\footnote{The RIS subtraction is a delicate issue. The
original, widely used prescription given by \citet{Kolb:1979qa}
and  \citet{Harvey:1981yk} turned out to be
incorrect, as observed by \citet{Giudice:2003jh}. A detailed discussion can be found
in Appendix~A of \citet*{Buchmuller:2004nz}.}
\begin{equation}\label{DL2}
\begin{split}
\gamma^\eq_\text{sub}(l\f \rightarrow \al\af) &= \gamma_{\Delta L=2,+} +
\tfrac{1}{2}\ve \gamma_N \ , \\
\gamma^\eq_\text{sub}(\al\af \rightarrow l\f) &= \gamma_{\Delta L=2,+} -
\tfrac{1}{2}\ve \gamma_N \ , \\
\gamma^\eq(l l \rightarrow \af\af) &= \gamma^\eq(\al \al \rightarrow \f\f)
= \gamma_{\Delta L=2,t}  \ .
\end{split}
\end{equation}
When one includes the $\Delta L=2$ washout processes, the kinetic equation for
the $B-L$ asymmetry becomes
\begin{equation}\label{BLchange2}
\begin{split}  
\dot{n}_{B-L} + 3Hn_{B-L} = &-\ve\left(\frac{n_N}{n_N^\eq}
  -1\right)\gamma_N \\
&-\frac{n_{B-L}}{n_l^\eq}\left(\frac{1}{2}\gamma_N
  +\gamma_{\Delta L =2}\right)\ ,
\end{split}
\end{equation}
where $\gamma_{\Delta L =2} = 2 \gamma_{\Delta L=2,+} +
2\gamma_{\Delta L=2,t}$.
Note that the full Boltzmann equation for the number density $n_{B-L}$
also depends on the number densities of charged leptons, quarks, and
Higgs boson, which satisfy their own Boltzmann equations. The corresponding
chemical potentials are all coupled by the sphaleron processes. 
A discussion of such ``spectator processes'' can be found in 
Sec.~\ref{sec:theoryLG} and in 
\citet{Buchmuller:2001sr}, \citet{Nardi:2005hs}, and \citet{Garbrecht:2014kda}. 
They  can
affect the final $B-L$ asymmetry by a factor of $\mathcal{O}(1)$.

Early studies of leptogenesis were partly motivated by trying to find
alternatives to electroweak baryogenesis, which did not seem to
produce a large enough asymmetry. Several extensions of the standard model
with hierarchical heavy-neutrino masses
were found that could explain the observed value of the baryon asymmetry
\citep*{Langacker:1986rj,Luty:1992un,Gherghetta:1993kn}
At that time models with keV-scale light neutrinos were still considered.
After washout processes were
correctly taken into account, it was realized that for hierarchical
mass matrices inspired by SO(10) GUTs neutrino masses below $1~\text{eV}$
were favored \cite{Buchmuller:1996pa}
Subsequently, atmospheric neutrino
oscillations were discovered, which led to a strongly rising interest
in leptogenesis and a large number of interesting models; for
reviews and references, see \citet{Mohapatra:2006gs} and \citet{Altarelli:2010gt}.
The minimal seesaw model for leptogenesis contains two right-handed neutrinos
\citep*{Frampton:2002qc}. This class of models was recently 
reviewed by \citet{Xing:2020ald}.

\subsection{Thermal leptogenesis}
\label{sec:thermalLG}
\subsubsection{One-flavor approximation}
\begin{figure}
  \includegraphics[width=0.41\textwidth]{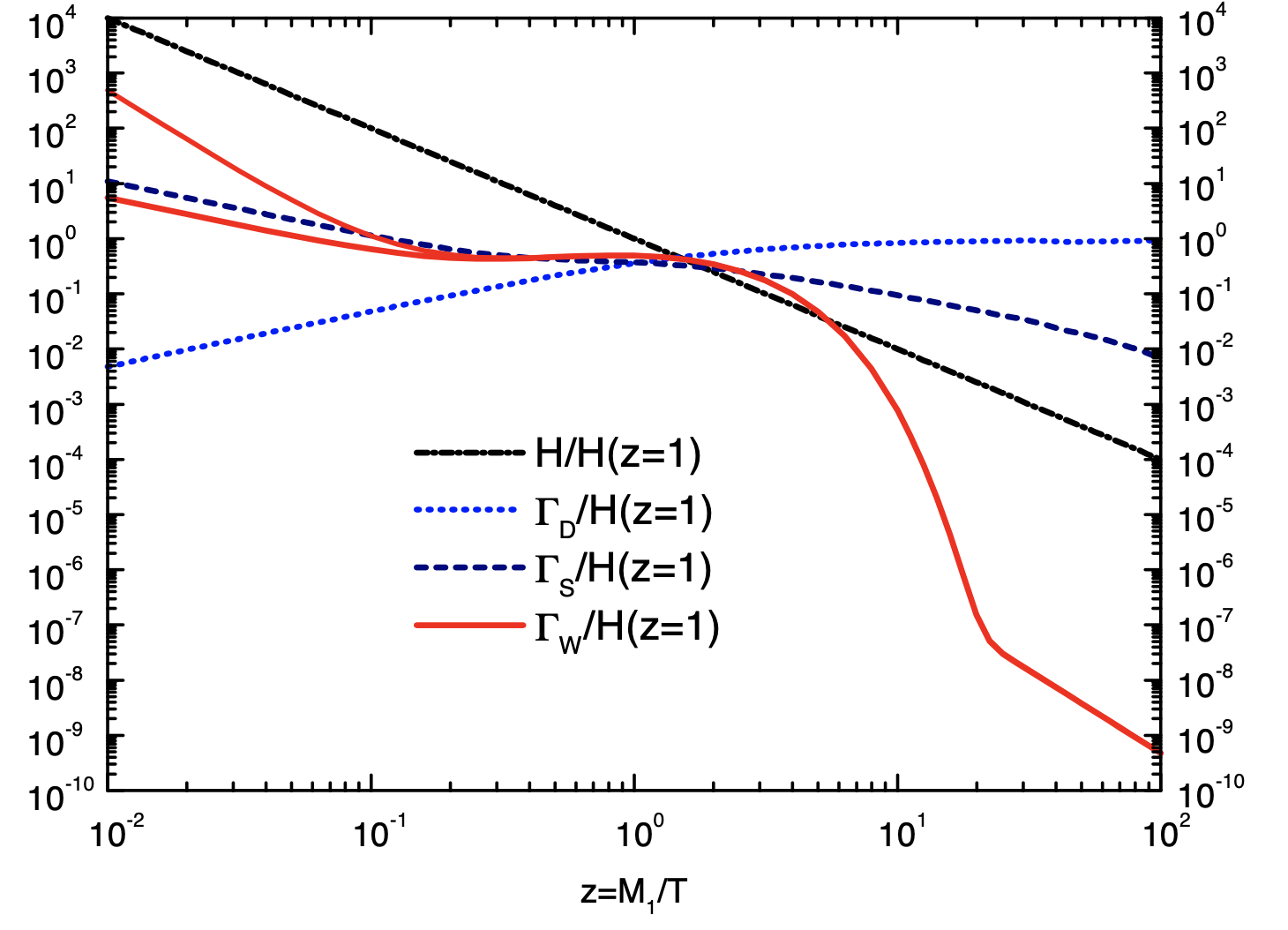}\hspace{1.0cm}
 \includegraphics[width=0.41\textwidth]{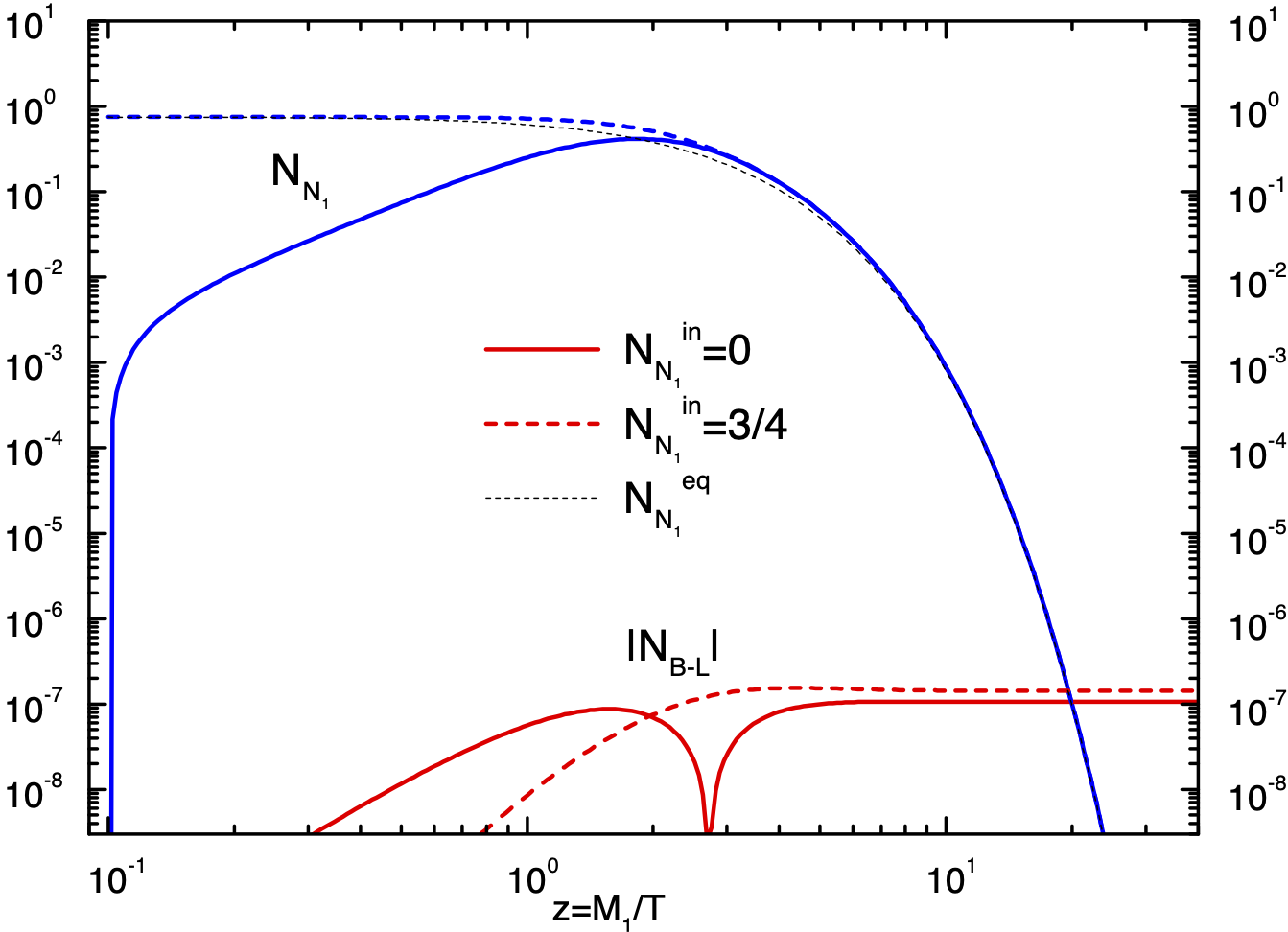}
  \caption{Top panel: decay, scattering and washout rates normalized to the
    Hubble parameter at $z=1$ compared to the Hubble parameter
    $H(z)$. The two branches of $\Gamma_W$ at $z \ll 1$ represent
    approximate upper and lower bounds. Bottom panel:
    evolution of $N_1$ abundance and $B-L$ asymmetry for both thermal
    and zero initial abundance. The neutrino parameters are $M_1 =
    10^{10}~\text{GeV}$, $\mt = 10^{-3}~\text{GeV}$, and $\bar{m} =
    0.05~\text{GeV}$. From \citet*{Buchmuller:2002rq}.}
\label{fig:leptogenesis}
\end{figure}

To understand the nonequilibrium process of thermal leptogenesis one
has to compare the reaction rates per particle with the Hubble
parameter as a function of temperature or, more conveniently,
$z=M/T$. The decay and washout rates are obtained by  dividing the
reaction densities by the relevant equilibrium number densities:
\begin{equation}
  \Gamma_N = \frac{1}{n_N^\eq} \gamma_N\ , \quad
  \Gamma_W = \frac{1}{n_l^\eq} \left(\frac{1}{2}\gamma_N +
    \gamma_{\Delta L=2}\right) \ .
	\label{G1f} 
\end{equation}
At low temperatures ($z >1$), decays and inverse decays dominate
  $N$ production and $B-L$ washout, whereas at high temperatures ($z <
  1$) $2\rightarrow 2$ scatterings with rate $\Gamma_S$ are equally
  important; see Fig.~\ref{fig:plasma_processes} and Sec.~\ref{sec:theoryLG}
for details.
  All rates have to be evaluated as functions of
$z$ by performing a thermal average over the corresponding matrix
elements \cite{Luty:1992un,Plumacher:1996kc,Biondini:2017rpb}. They are
compared to the Hubble parameter
in the upper panel of  Fig. \ref{fig:leptogenesis}. For $z < 1$, all
processes are out of thermal equilibrium ($\Gamma_{D,W,S} < H$).
Around $z \sim 1$, the various processes come into thermal equilibrium.
Heavy neutrinos now decay and, since their number density slightly
exceeds the equilibrium number density, a $B-L$ asymmetry is generated
in these decays. As long as washout processes are in equilibrium, the
asymmetry is partly washed out again. At $z > 1$, $N$ production is
kinematically suppressed, the washout processes eventually  get out of
equilibrium at some $z_L$, and the $B-L$ asymmetry is frozen in.

In the kinetic equations \eqref{Nchange2} and \eqref{BLchange2}
the Hubble parameter appears. It is convenient to separate the time
dependence of the leptogenesis process from the expansion of the
Universe. This can be achieved by considering the ratio of a number density
$n_X$ to the entropy density $Y_ X = n_X/s$, or the product of $n_X$ and
the comoving volume occupied by one particle, such as a photon, i.e.,
$N_X = 2n_X/n_\gamma$, at some time before the onset of leptogenesis.
For the standard model in the high-temperature phase, assuming
one relativistic heavy-neutrino species, one has $s = 217\pi^4/[90\zeta(3)]
n_{\gamma}$ \cite{Kolb:1990vq}, and therefore
\begin{equation}
  Y_X(z) = \frac{45\zeta(3)}{217\pi^4} N_X(z)\ , \quad z < 1\ .
\end{equation}
Changing variables and defining the rescaled reaction rates
$D=\Gamma_N/Hz$ and $W = \Gamma_W/Hz$, the kinetic equations
\eqref{Nchange2} and \eqref{BLchange2} take the simple form
\begin{equation}\label{kes}
  \begin{split}
    \frac{dN_N}{dz} &= - D(N_N - N_N^\eq)\ , \\
    \frac{dN_{B-L}}{dz} &= -\ve D(N_N - N^\eq) - W N_{B-L}\ .
  \end{split}
  \end{equation}
    
  The maximal $B-L$ asymmetry to which leptogenesis can lead is
    determined by the $CP$ asymmetry in $N$ decays out of equilibrium, as
    described by Eq.~\eqref{maxasym}. The coupling of the heavy
    neutrinos to the thermal bath implies a suppression of the final asymmetry
    \mbox{$N_{B-L}^\text{f} = N_{B-L}(z \gg 1)$}, which is expressed in terms of an efficiency
    factor $\kappa$ \cite{Barbieri:1999ma} as
   \begin{equation}
    N_{B-L}^\text{f} = -\tfrac{3}{4} \ve \kappa_\text{f}\ ,
  \end{equation}
  where the factor $3/4$ is taken from the Fermi statistics. During the
  evolution of the Universe the $B-L$ asymmetry in a comoving volume
  element remains constant, whereas the number of photons increases.
  The measured  baryon-to-photon ratio at recombination is then given by
  \begin{equation}\label{etaB}
    \eta_B = \frac{n_B}{n_\gamma} = \frac{3}{4}\frac{c_s}{f}\ve
    \kappa_\text{f}  \simeq \eta_B \simeq 0.96\times 10^{-2} \ve \kappa_\text{f}
    \ .
  \end{equation}
  Here $c_s$ is the fraction of $B-L$ asymmetry converted into a
  baryon asymmetry by sphaleron processes
 [see Eq.~\eqref{connection}], and the dilution factor $f$
  is the increase of the number of photons in a comoving volume
  element. In the standard model with one heavy neutrino one has
  $c_s = 28/79$ and $f = 2387/86$.
     
\begin{figure*}
%
  \includegraphics[width=0.35\textwidth,angle=90]{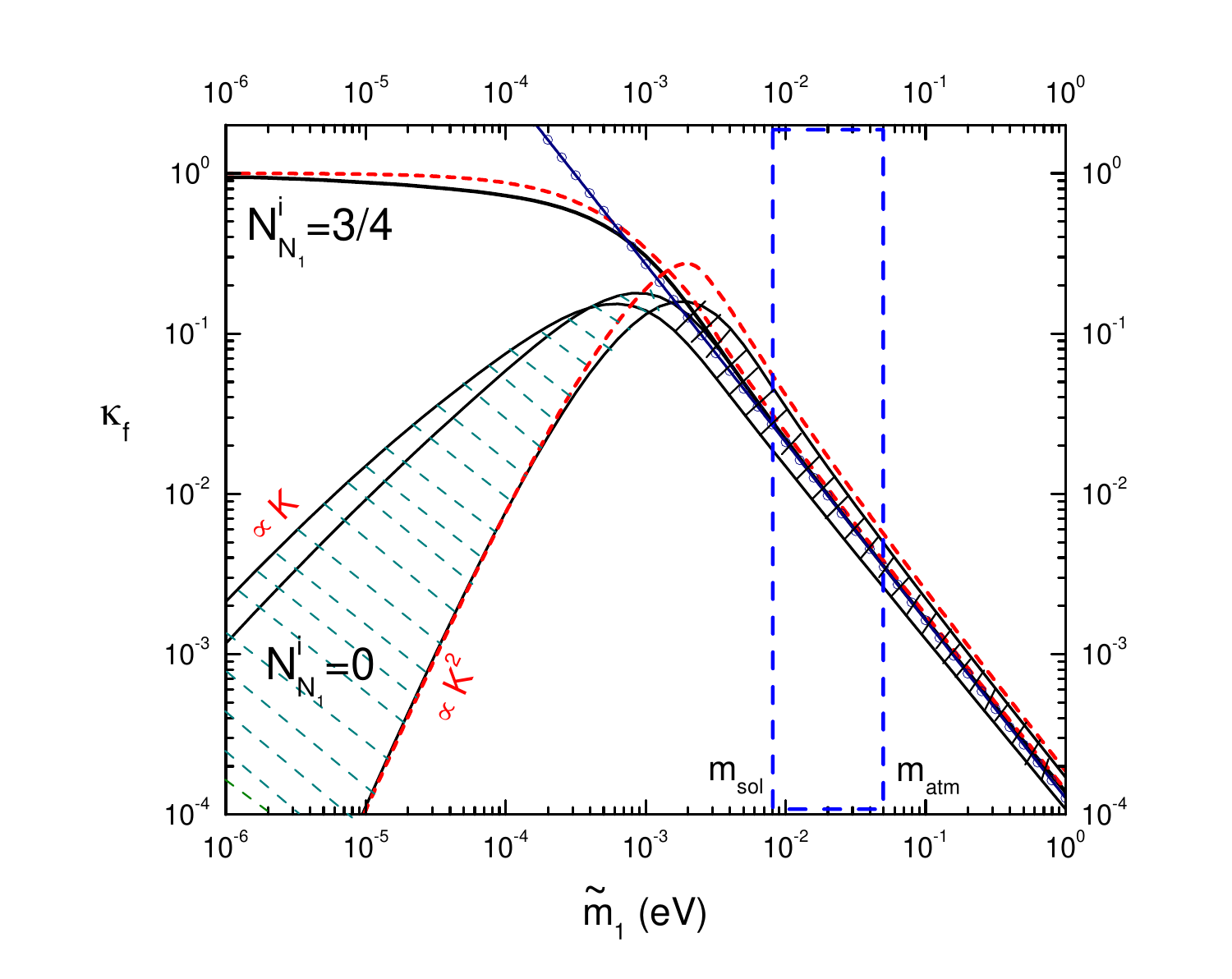}
  \vspace{0.5cm}
  \includegraphics[width=0.45\textwidth]{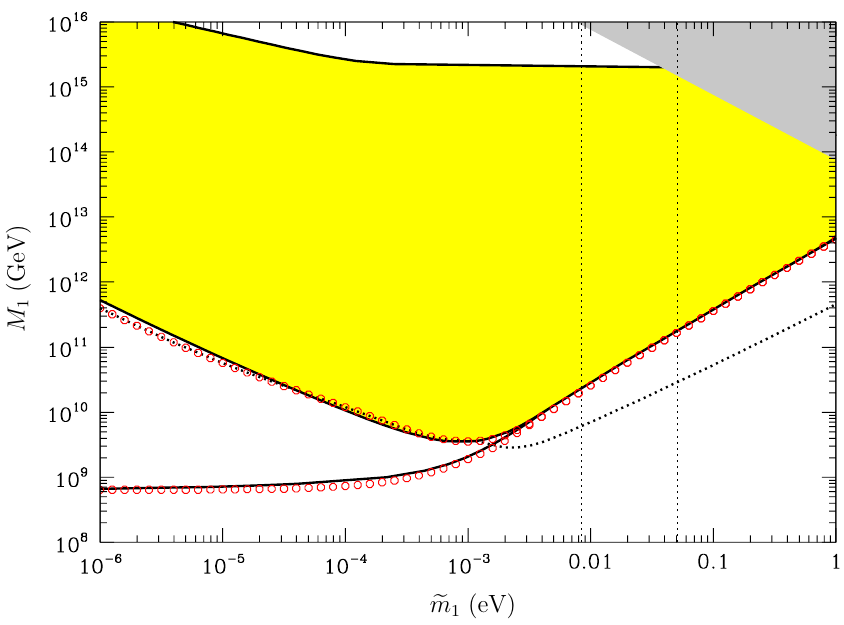}
  \caption{Left panel: efficiency factor $\kappa_\text{f}$ as a function of
    the effective neutrino mass $\mt$. The hatched region represents
    the theoretical uncertainty due to $\Delta L =1$ scattering
    processes, the dashed lines indicate analytical results, and the
    circled line is a power-law fit. Right panel: upper and lower bounds on
    the heavy-neutrino mass $M_1$ (the weaker lower bound corresponds
    to thermal initial conditions). The dotted line is a lower bound
    on the initial temperature $T_i$; the gray triangle is excluded by
    theoretical consistency and the circled lines represent analytical
    results. In both panels the vertical lines indicate the range
    $(m_\text{sol}, m_\text{atm})$ (see the text).
    From \citet*{Buchmuller:2004nz}.}
\label{fig:bounds}
\end{figure*}

In the upper panel of Fig.~\ref{fig:leptogenesis} decay and washout
rates are depicted for a representative choice of neutrino masses.
The lower panel shows solutions of the kinetic equations \eqref{kes} for 
the same mass parameters and two different choices of initial
conditions, namely, thermal and zero initial $N$ abundance. For thermal
initial abundance the number $N_N$ always exceeds the equilibrium
value $N_N^\eq$, and the asymmetry $|N_{B-L}|$ continuously increases
toward its final value. For zero initial abundance $N_N - N_N^\eq$ is
first negative. It changes sign just above $z=1$, where also $|N_{B-L}|$
passes through zero. For the chosen neutrino mass parameters the final
$B-L$ asymmetry is almost independent of the initial condition. The
value of the baryon-to-photon ratio $\eta_B \sim 0.01 N^\text{f}_{B-L}$
is in agreement with observations.

The generated $B-L$ asymmetry strongly depends on neutrino parameters,
and it is noteworthy that for masses and mixings consistent
with neutrino oscillations the observed baryon-to-photon ratio is
naturally obtained. The robustness of the leptogenesis mechanism is
largly due to the fact that for neutrino masses below $0.1~\text{eV}$
the $B-L$ asymmetry is essentially determined only by decays and
inverse decays. The heavy neutrinos decay at $z >1$, such that
  scattering processes are unimportant, and for small neutrino masses
  $\Delta L =2$ washout processes are suppressed
  \citep*{Buchmuller:2004nz}.
  Moreover, relativistic corrections are small.
In the case where a summation of the lepton flavors
in the final state is performed, the efficiency factor depends only on
$\mt$ and $M\bar{m}^2$, where the effective light-neutrino
mass $\mt$ and the absolute neutrino mass scale $\mb$ are defined as
\begin{equation}\label{mt}
  \mt = \frac{(h^{\nu\dagger}h^\nu)_{11}v^2_{\text{EW}}}{M_1} \ , \quad
  \mb = \sqrt{m_1^2 + m_2^2 + m_3^2}\ .
\end{equation}
For $\mb \lesssim 0.1~\text{eV}$
and $M \lesssim 10^{14}~\text{GeV}$, the efficiency factor $\kappa_\text{f}$
depends only on $\mt$. As the left panel of
Fig.~\ref{fig:bounds} illustrates, there are two regimes, with
``weak'' and ``strong''  washout, corresponding to
\begin{equation}
	\label{weakstrong} 
  \mt < m_*\ , \quad \mt > m_*\ ,
\end{equation}
respectively, where $m_*$ is the equilibrium neutrino mass:
\begin{equation}\label{mstar}
 m_* = \frac{16\pi^{5/2}\sqrt{g_*}}{3\sqrt{5}}\frac{v^2}{M_\text{P}}
 \simeq 1.08 \times 10^{-3}~\text{eV}\ .
 \end{equation}
The ratio $\mt/m_* = \Gamma_D(z=\infty)/H(z=1) \equiv K$ was previously
introduced in GUT baryogenesis \cite{Kolb:1979qa}. In the weak-washout
regime $\kappa_\text{f}(\mt)$ strongly depends on the initial
conditions (thermal versus zero initial abundance) and on the rate for
$\Delta L = 1$ scattering processes (the hatched area in
Fig.~\ref{fig:bounds}). On the contrary, in the strong-washout
regime the efficiency factor is universal, with an uncertainty of
about 50\%:
\begin{equation}\label{kappaf}
 \kappa_\text{f} = (2 \pm 1) \times 10^{-2}
 \left(\frac{0.01~\text{eV}}{\mt}\right)^{1.1\pm 0.1} \ .
\end{equation}
Moreover, the dependence of the final $B-L$ asymmetry on some other
initial $B-L$ asymmetry, independent of leptogenesis, is significantly
suppressed in the strong-washout regime. It is noteworthy that
the neutrino mass range indicated by solar and atmospheric neutrinos
lies inside the strong-washout regime where the generated
$B-L$ asymmetry  is essentially determined by decays and inverse decays and
therefore largely independent of initial conditions and theoretical
uncertainties.

In the case of hierarchical heavy neutrinos the maximal $CP$ asymmetry
in $N$ decays reads \citep*{Davidson:2002qv,Hamaguchi:2001gw}
\begin{equation}\label{epsmax}
\ve_\text{max} = \frac{3}{16\pi}\frac{Mm_\text{atm}}{v^2} \simeq 10^{-6}
\left(\frac{M}{10^{10}~\text{GeV}}\right)\ .
\end{equation}  
As we know the maximal efficiency factor, Eq.~\eqref{epsmax} implies a lower bound on
the smallest heavy-neutrino mass $M$. From Fig.~\ref{fig:bounds} one
reads off $\kappa^\text{max} \sim 1$ and  $\kappa^\text{max} \sim 0.1$
for thermal and zero initial abundance, respectively. A
baryon-to-photon ratio $\eta_B \sim 10^{-9}$ then requires a heavy-neutrino
mass $M \gtrsim 10^8~\text{GeV}$ and $10^9~\text{GeV}$ for the two
different initial conditions, respectively. The precise dependence of
the lower bound on $\mt$ is shown in the right panel of Fig~\ref{fig:bounds}.

The $\Delta L = 2$ washout term leads to an upper bound on
heavy-neutrino
masses and also to an important upper bound on the light-neutrino
masses \citep*{Buchmuller:2002jk}. An analysis of the
solution of the kinetic equations \eqref{kes} shows that in the strong-washout
regime, which is defined by $\mt \gtrsim m_*$, the $B-L$ asymmetry is produced close
to $z_B(\mt) \sim 2m_*/\mt\kappa_\text{f}(\mt)$, and the complete
efficiency factor is given by
\begin{equation}\label{fullkappa}
  \begin{split}
    \bar{\kappa}_\text{f}&(\mt,M\mb^2)\simeq \\
    & \kappa_\text{f}(\mt)\exp{\left[ -\frac{\omega}{z_B}
  \left(\frac{M}{10^{10}~\text{GeV}}\right)\left(\frac{\mb}{\text{eV}}\right)^2\right]}\ ,
\end{split}
\end{equation}
where $\omega \simeq 0.2$. For too large values of $M$ and
$\mb$, the generated $B-L$ asymmetry is too small relative to
observation. A quantitative analysis yields for $M$ the upper bound
shown in Fig.~\ref{fig:bounds}, and for the light-neutrino masses one
finds $m_i < 0.12~\text{eV}$. Assuming $\mt = \mathcal{O}(m_i)$,
successful leptogenesis then implies for the light neutrinos the optimal mass
window
\begin{equation}\label{window}
  10^{-3} \lesssim m_i \lesssim 0.1~\text{eV}\ .
\end{equation}
It is notable that the cosmological bound on the sum of
neutrino masses \cite{Planck:2018vyg}, which has become increasingly
stringent over the past two decades, is consistent with this mass
window. Note, however, that the upper bound on the light-neutrino
masses holds only in
type-I seesaw models. In type-II models, where a triplet contribution
appears in the neutrino mass matrix as in left-right symmetric
models, the direct connection between neutrino masses and leptogenesis is
lost \cite{Hambye:2003ka,Antusch:2004xy}.

The maximal $CP$ asymmetry [Eq.~\eqref{epsmax}], and  therefore the lower
bound on the heavy-neutrino mass $M_1$, depends on the measured value
of $m_\text{atm}$. What can one say without knowing the result from
atmospheric neutrino oscillations? In this case the Planck mass and
the Fermi scale still yield the neutrino mass scale $m_*$ [see Eq.~\eqref{mstar}], which
determines the normalization of $\mt$ in the efficiency factor
$\kappa_f$ [Eq.~\eqref{kappaf}]. From the full efficiency factor
[Eq.~\eqref{fullkappa}] one can then determine the maximal baryon asymetry
as a function of $\mt$ and $m_3$, which is reached at
$\mt \simeq 2\times 10^{-3}~\text{eV}$, i.e., in the strong-washout
regime \citep*{Buchmuller:2004tu}. This leads to the upper and lower
bounds $m_3 \lesssim 250~\text{eV}$ and $M_1 \gtrsim 2\times
10^6~\text{GeV}$, respectively.

In GUTs with hierarchical heavy
right-handed neutrinos ($M_1 \ll M_2 \ll M_3
\sim v_{B-L} \sim 10^{15}~\text{GeV}$, a simple estimate yields the
right order of magnitude for the baryon-to-photon ratio 
\cite{Buchmuller:1996pa,Buchmuller:1998zf}.
  To understand this,
consider the $CP$ asymmetry $\ve_1$ given in Eq.~\eqref{CPasymmetry2},
assume normal ordering, and keep the largest contribution proportional
to the light-neutrino mass $m_3$. With 
$h^\nu_{i1}/\sqrt{(h^{\nu\dagger}h^\nu)_{11}} \propto \delta_{i3}$ and using
Eq.~\eqref{nuGUT}, one obtains
\begin{equation}
 \ve_1 \sim 0.1\ \frac{m_3M_1}{v^2_{\text{EW}}} \sim 0.1\
 \frac{M_1}{M_3}\ .
\end{equation} 
For a heavy-neutrino mass hierarchy similar to the hierarchies in the
quark and charged lepton sectors, i.e., $M_1/M_3 \sim 10^{-5}\cdots
10^{-4}$, and an efficiency factor $\kappa_f \sim 10^{-2} \cdots
10^{-1}$, the baryon-to-photon ratio is given by [see
Eq.~\eqref{etaB}],
\begin{equation}
  \eta_B \sim 10^{-2}\ \ve_1\kappa_f \sim 10^{-10} \cdots 10^{-8}\ ,
\end{equation}
which is in agreement with observation.

\begin{figure*}
\begin{center}
  \includegraphics[width=0.4\textwidth]{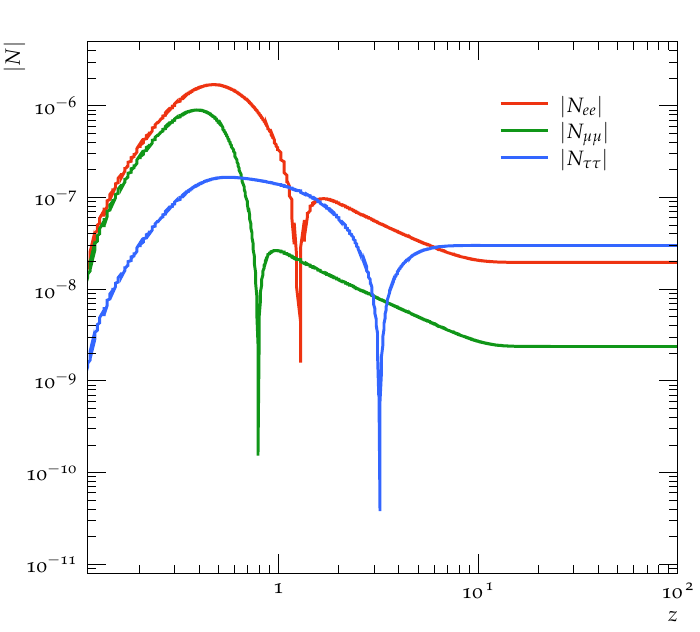}\hspace{0.5cm}
  \includegraphics[width=0.4\textwidth]{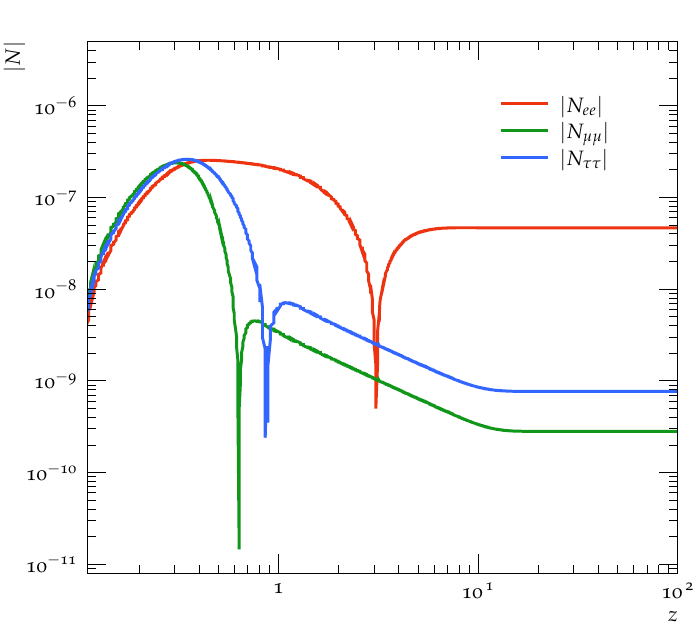}
\end{center}
  \caption{Evolution of the $B-L$ asymmetry for each lepton flavor as a
    function of $z=M_1/T$ for the two sets of neutrino masses and phases
    (left panel: $S_2$; right panel: $S_3$) listed in
    Table~\ref{tab:S2S3}. From \citet{Moffat:2018wke}.}
\label{fig:S2S3}
\end{figure*}

The $\Delta L=2$ washout terms play a crucial role in obtaining 
upper bounds on  light- and heavy-neutrino masses.
Correspondingly, 
a discovery of lepton-number-violating dilepton events at the LHC
could be used to falsify leptogenesis since the production cross
section of these events is directly related to a $\Delta L = 2$ washout
term that, if large enough, would erase any baryon asymmetry. This has been
demonstrated in the context of left-right symmetric models
\citep*{Frere:2008ct}  as well as in a model-independent approach 
\citep*{Deppisch:2013jxa}.

\subsubsection{Flavor effects}
\label{sec:flavour}
Thus far we have discussed leptogenesis in the ``one-flavor
approximation'', where one sums over lepton flavors in the final
state. This approximation is valid only at high temperatures
where lepton-Higgs interactions in the thermal plasma can be
neglected. In general,  flavor effects can have an important impact
on leptogenesis \cite{Barbieri:1999ma,Endoh:2003mz,
Abada:2006fw,Nardi:2006fx,Blanchet:2006ch}.

We first consider the simplest case where the lightest heavy
neutrino $N_1 \equiv N$ couples to the following combination of lepton
flavors given by the Yukawa couplings $h^\nu_{i 1}$ [see Eq.~\eqref{SMN}]:
\begin{equation}
  |l_1\rangle = \sum_{i = e,\mu,\tau} C_{1i} |l_i\rangle
  \ , \quad C_{1i} = \frac{h^{\nu}_{i 1}}{\sqrt{(h^{\nu\dagger} h^\nu)_{11}}}\ .
  \end{equation}
  As the Universe expands, Hubble parameter and Yukawa rates decrease as
  $H \sim T^2/M_\text{P}$ and $\Gamma_Y \sim g_Y^2 T$, respectively. Hence, with
  $g_\tau \sim 5\times 10^{-3}$, left- and right-handed
  $\tau$ neutrinos
  are in thermal equilibrium for temperatures below
the temperature $T_\tau$, where
\begin{equation}
\Gamma_\tau(T_\tau) \sim \frac{g_\tau^2}{4\pi}~T_\tau \sim 10^{-6}~T_\tau \sim
H(T_\tau)\ ,
\end{equation}
which implies $T_\tau \sim 10^{12}~\text{GeV}$.
Below $T_\tau$ interactions with $\tau$ leptons in the thermal bath
destroy the coherence of the lepton state produced in $N$
decay. Hence, one has to consider Boltzmann equations for the
components parallel and orthogonal to $\tau$ separately. With
\begin{equation}
  p_\tau = |C_{1\tau}|^2\ , \quad p_{\tau^\perp} = 1 - |C_{1\tau}|^2\
  , \quad \langle \tau|\tau^\perp \rangle = 0\ ,
  \end{equation}
  one obtains
  \begin{equation}\label{flavourBoltzmann}
    \begin{split}
      \frac{dN_N}{dz} &= - D(N_N - N_N^\eq)\ , \\
      \frac{dN_{\tau\tau}}{dz} &= \ve_{\tau\tau} D(N_N - N^\eq) -
      p_\tau W N_{\tau\tau}\ ,\\
\frac{dN_{\tau^\perp\tau^\perp}}{dz} &= \ve_{\tau^\perp\tau^\perp}
D(N_N - N^\eq) - p_{\tau^\perp} W N_{\tau^\perp\tau^\perp}\ .
\end{split}
  \end{equation}
For the produced $B-L$ asymmetry these equations yield the flavor structure  
\begin{align}
  N_{B-L} &\propto \left(\frac{\ve_{\tau\tau}}{p_{\tau\tau}}
    +
            \frac{\ve_{\tau^\perp\tau^\perp}}{p_{\tau^\perp\tau^\perp}}\right) \nonumber\\
    &\propto \ve_{\tau\tau}\left(\frac{1}{p_{\tau\tau}}
    - \frac{1}{1 - p_{\tau\tau}}\right) \ .
\end{align}
A complete expression for the $B-L$ asymmetry in the two-flavor regime was
given by Blanchet and Di Bari (2009).
  For temperatures far below $T_\tau$ all three lepton flavors have
  to be taken into account. Instead of Eqs.~\eqref{flavourBoltzmann}
  one then obtains an involved system of Boltzmann equations or,
  depending on the temperature regime, kinetic equations for the
  lepton density matrix \cite{Blanchet:2011xq}. 

  Flavor effects, together with tuning of the parameters of the
  seesaw mass matrix, can be used to lower the leptogenesis
  temperature significantly below $T \sim 10^{10}~\text{GeV}$, which
  was considered in the previous section \cite{Blanchet:2008pw}.
    Recently, a detailed study of this type has been carried out by
  Moffat \ea~(2018), who took masses and mixings of all three light and heavy Majorana
  neutrinos into account. A code to solve these Boltzmann
  equations was published by Granelli \ea~(2020).
Two sets of neutrino
  parameters, fitted to the measured neutrino parameters and the
  observed baryon asymmetry, are shown in Table~\ref{tab:S2S3}. The
  two sets $S_2$ and $S_3$ correspond to normal hierarchy
  for the light neutrinos and to a mild mass hierarchy for heavy
  neutrinos. Mixing angles and mass ratios of the light neutrinos are
  essentially fixed by observation. In both cases the smallest
  neutrino mass lies in the mass window [Eq.~\eqref{window}], whereas
  the Dirac
  phase and Majorana phases vary significantly. Correspondingly, the
  flavor dependence of the $B-L$ asymmetry is different in the two cases.
  The dependence on the mixing parameters of the heavy neutrinos is
  not listed. An important aspect of the flavor effects is the mass
  scale of the heavy neutrinos, which lies significantly below the
  lower bound derived in the one-flavor approximation.
  
  Compare the evolution of the $B-L$ asymmetry in
  Fig.~\ref{fig:S2S3} to that shown in
  Fig.~\ref{fig:leptogenesis}, where $M_1$, the smallest
  heavy-neutrino mass, is 3 orders of magnitude larger than in the
  parameter sets $S_2$ and $S_3$. In Fig.~\ref{fig:leptogenesis} the
  generated asymmetry before the thermal equilibrium of $N_1$ is about the
  same as the final $B-L$ asymmetry, whereas in Fig.~\ref{fig:S2S3}
  there is a difference of about 1 order of magnitude. The
  flavor composition of the $B-L$ asymmetry is significantly different
  for $S_2$ and $S_3$. Such a behavior can occur due to cancellations  
  between positive and negative contributions to the asymmetry around
  $z \approx 1$, as discussed by \citep*{Buchmuller:2004nz}.
  Moreover, fine-tuning between tree-level contributions and one-loop
  corrections to the light-neutrino mass matrix is needed.
  On the whole the total $B-L$ asymmetry is rather sensitive to
  fine-tuning of the parameters, which is
  the price one pays for lowering the heavy-neutrino mass scale
  relative to the simple one-flavor approximation, thereby allowing for a 
  low reheating temperature.

\begin{table}[b]
  \begin{center}
  \begin{tabular}{c|c|c|c|c|c}
    & $\delta(^\circ)$
    &
    $m_{1}$ (eV) &       $M_{1}$ (GeV) &    $M_{2}$ (GeV)&    $M_{3}$(GeV)\\
          \hline\hline
    $S_{2}$& $88.26$
    & $0.079$ & $10^{6.5}$ & $10^{7}$ & $10^{7.5}$ \\
    $S_{3}$& $31.71$
    &  $0.114$ & $10^{6.5}$ & $10^{7.2}$ & $10^{7.9}$\\
  \end{tabular}
  \end{center}
      \caption{Two sets of neutrino masses and phases consistent with
        the observed $B-L$ asymmetry (only 5 out of 14 parameters are
        listed).  Adapted from Moffat \ea, 2018.}
\label{tab:S2S3}
\end{table}

 \begin{figure*}
\begin{center}
  \begin{minipage}{.49\textwidth}
        \centering
        \includegraphics[width=\textwidth]{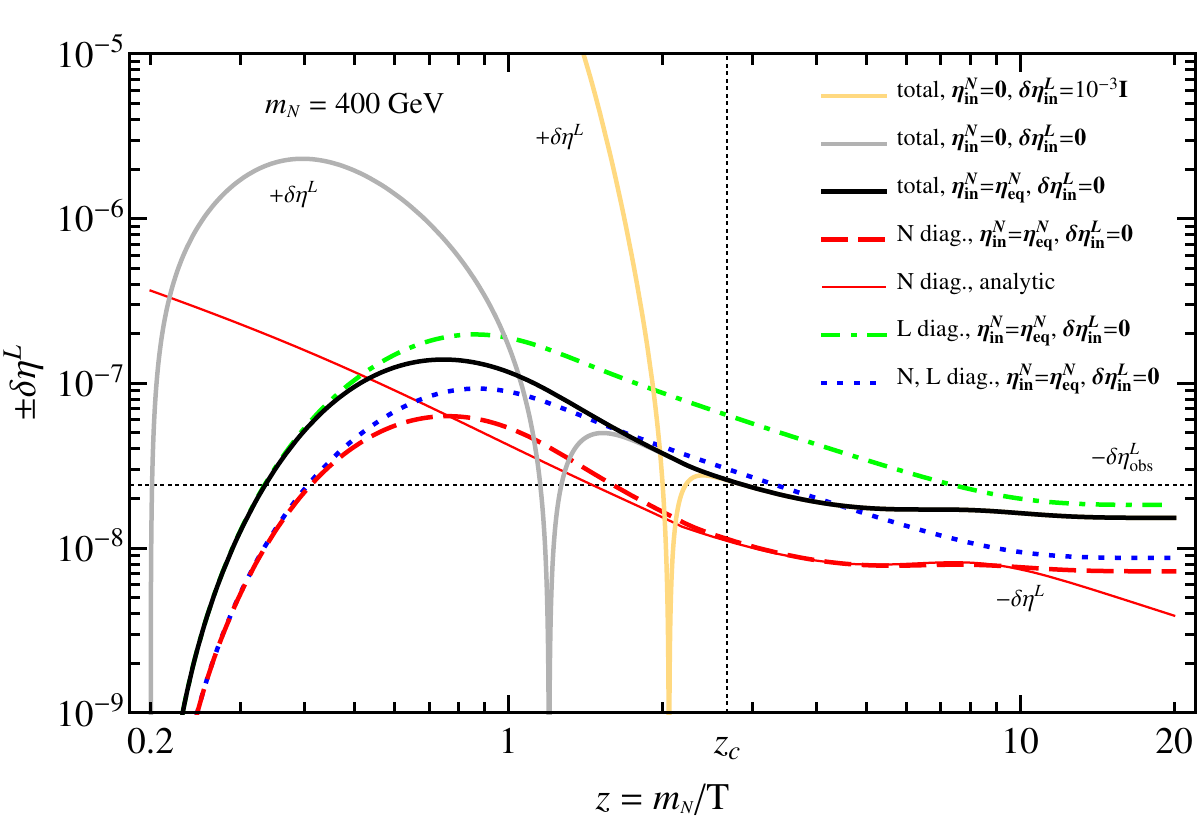}
 \end{minipage}\hspace{1.0cm}
\begin{minipage}{.3\textwidth}
        \centering
        \includegraphics[width=\textwidth]{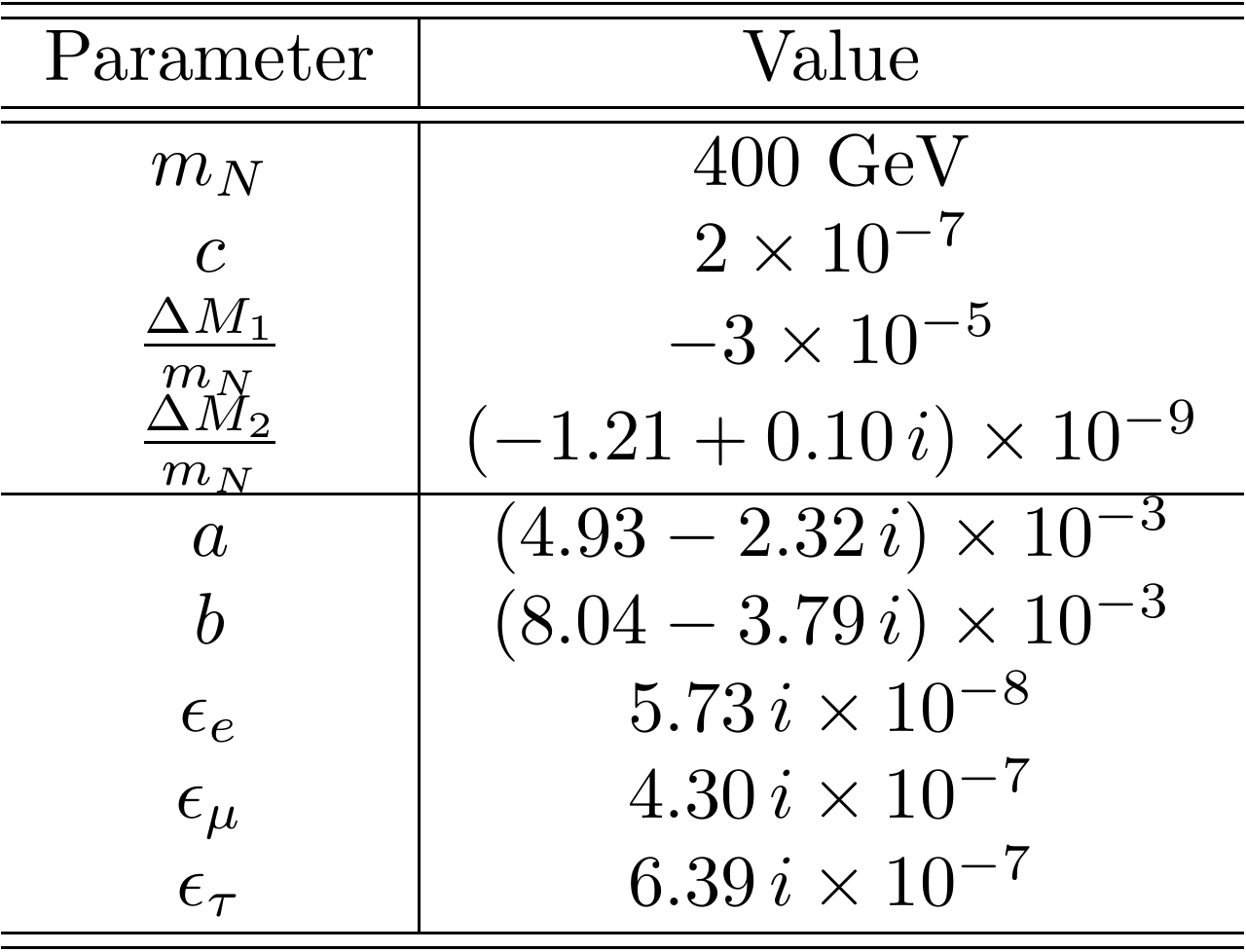}
 \end{minipage}
\end{center}
\caption{Resonant leptogenesis: lepton asymmetry $\eta^L$, with $L=L_\tau$, 
	computed in a density matrix
  formalism for different initial conditions (solid lines), compared to
  $\eta^L$ obtained in a flavor-diagonal approximation (dashed
  lines). The model parameters are specified in the table. 
	From Bhupal Dev \ea~2015.}
\label{fig:resonantLG}
\end{figure*}

  An important aspect of the flavor dependence of leptogenesis is the
  effect on upper and lower bounds on light-neutrino masses. 
  The effect can be significant, but thus far it has not been
  possible to obtain a complete picture in a model-independent way.
  According to current estimates, it is possible to relax the upper and
  lower bounds in Eq.~\eqref{window} by about 1 order of magnitude,
  barring fine-tuning
  \citep{Davidson:2008bu,Blanchet:2012bk,Dev:2017trv}. Moreover,
  spectator effects have to be taken into account 
  \citep{Buchmuller:2001sr,Nardi:2005hs,Garbrecht:2014kda}.

  Our discussion of flavor effects has been limited to the case where
  leptogenesis is dominated by the lightest heavy neutrino $N_1$. An
  alternative is dominance by the next-to-lightest heavy
  neutrino $N_2$ \cite{DiBari:2005st,Vives:2005ra}. More possibilities were reviewed 
  by \citet{Dev:2017trv}.
    The treatment of flavor effects based on Kadanoff-Baym equations
  \cite{Beneke:2010dz} is described in Sec. \ref{sec:theoryLG}.

  Continuous and discrete flavor symmetries play an important role in
  restricting lepton masses and mixings, and in this way they 
  strongly effect leptogenesis. This has been extensively discussed in
  the literature; comprehensive overviews were given by
  \citet{Mohapatra:2006gs} and \citet{Altarelli:2010gt}.

\subsubsection{Resonant leptogenesis}
The standard temperature scale of thermal leptogenesis
($T \gtrsim 10^{10}~\text{GeV}$) can be significantly lowered by
flavor effects. A much more dramatic effect occurs when mass
differences between the heavy neutrinos are comparable to the
heavy-neutrino decay widths, a case referred to as resonant leptogenesis
\cite{Pilaftsis:2003gt}. In this case leptogenesis temperatures of the
order of
TeV are possible, which implies the possibility of testing
thermal leptogenesis at high-energy colliders \cite{Pilaftsis:2005rv}.
Such models can be realized in extensions of
the standard model where the quasidegeneracy of the heavy neutrinos
is a consequence of approximate symmetries, as in
supersymmetric models, where soft supersymmetry
breaking terms can be much smaller than the heavy-neutrino mass terms;
see \citet*{DAmbrosio:2003nfv}, \citet{Grossman:2003jv},
\citet{Chen:2004xy}, and \citet*{Hambye:2004jf}.

The formalism to treat this resonant regime was developed
in resummed perturbation theory \cite{Pilaftsis:2003gt}, on the basis of
Kadanoff-Baym equations \citep*{Garny:2011hg}, and using a density matrix formalism
\cite{Dev:2015dka}; for a review, see \citet{Dev:2017wwc}. In resummed
perturbation theory one computes the decay rates of heavy neutrinos
$N_\alpha$ to leptons $l$ and the Higgs boson,
\begin{equation}\label{GammaNphi}
\Gamma_{\alpha l} = \Gamma(N_\alpha \rightarrow l \phi)\ ,  \quad
\Gamma^C_{\alpha l} =  \Gamma(N_\alpha \rightarrow \overline{l}
\overline{\phi}) \ ,
\end{equation}        
in terms of resummed Yukawa couplings $\overline{\bf h}^\nu_{l\alpha}$
\cite{Pilaftsis:2003gt}. The $CP$ asymmetries are defined
as usual:
\begin{equation}
  \begin{split}
  \label{deltaN}
	\delta_{\alpha l} &\equiv 
	\frac{\Gamma_{\alpha l}^{\phantom{C}} -\Gamma_{\alpha l}^C}
        {\sum\limits_{l = e,\mu ,\tau}
		\Big(\Gamma _{\alpha l}^{\phantom{C}} +\Gamma_{\alpha
                  l}^C\Big)} \\
              &=\frac{\big|\overline{\bf h}^\nu_{l\alpha}\big|^2 - 
			\big|\overline{\bf h}^{\nu C}_{l\alpha}\big|^2}
	{\big(\overline{\bf h}^{\nu \dagger}\  
	\overline{\bf h}^{\nu
          \phantom{\dagger}}\!\!\big)_{\alpha\alpha} +  
	\big(\overline{\bf h}^{\nu C\dagger}\  
	\overline{\bf h}^{\nu C\phantom{\dagger}}\!\!\big)_{\alpha\alpha}
	}\ .
\end{split} 
      \end{equation}
For quasidegenerate heavy neutrinos the decay rates show the typical resonant
enhancement, and for two heavy neutrinos one has obtained the
result\footnote{The calculation of the $CP$ asymmetry in the resonance
  regime is subtle. 
For a thorough discussion see
\citet*{Anisimov:2005hr}, \citet*{Garny:2011hg}, and \citet{Brdar:2019iem}.} \cite{Pilaftsis:2003gt}
\begin{equation}
  \begin{split}
\delta_{\alpha l} \ \approx\ &\frac{{\rm Im} 
\big[({\bf h}^{\nu\dagger}_{\alpha l} {\bf h}^{\nu\phantom{\dagger}}_{l\beta})\ 
({\bf h}^{\nu\dagger}_{\phantom{l}}{\bf
    h}^{\nu\phantom{\dagger}}_{\phantom{l}}\!\!)_{\alpha\beta}\big]} 
{({\bf h}^{\nu\dagger}_{\phantom{l}}
{\bf h}^{\nu\phantom{\dagger}}_{\phantom{l}}\!\!)_{\alpha\alpha}\ ({
  \bf  h}^{\nu\dagger}_{\phantom{l}}
{\bf h}^{\nu\phantom{\dagger}}_{\phantom{l}}\!\!)_{\beta\beta}}\\
&\times\frac{(m^2_{N_\alpha} - m^2_{N_\beta})\, m_{N_\alpha}\,
  \Gamma^{(0)}_{N_\beta}}{ (m^2_{N_\alpha} - m^2_{N_\beta})^2 + 
m^2_{N_\alpha} \Gamma^{(0)2}_{N_\beta}}\ .
\end{split}
\end{equation}

\begin{figure*}[t]
\begin{center}
  \begin{minipage}{.35\textwidth}
        \centering
        \includegraphics[width=\textwidth]{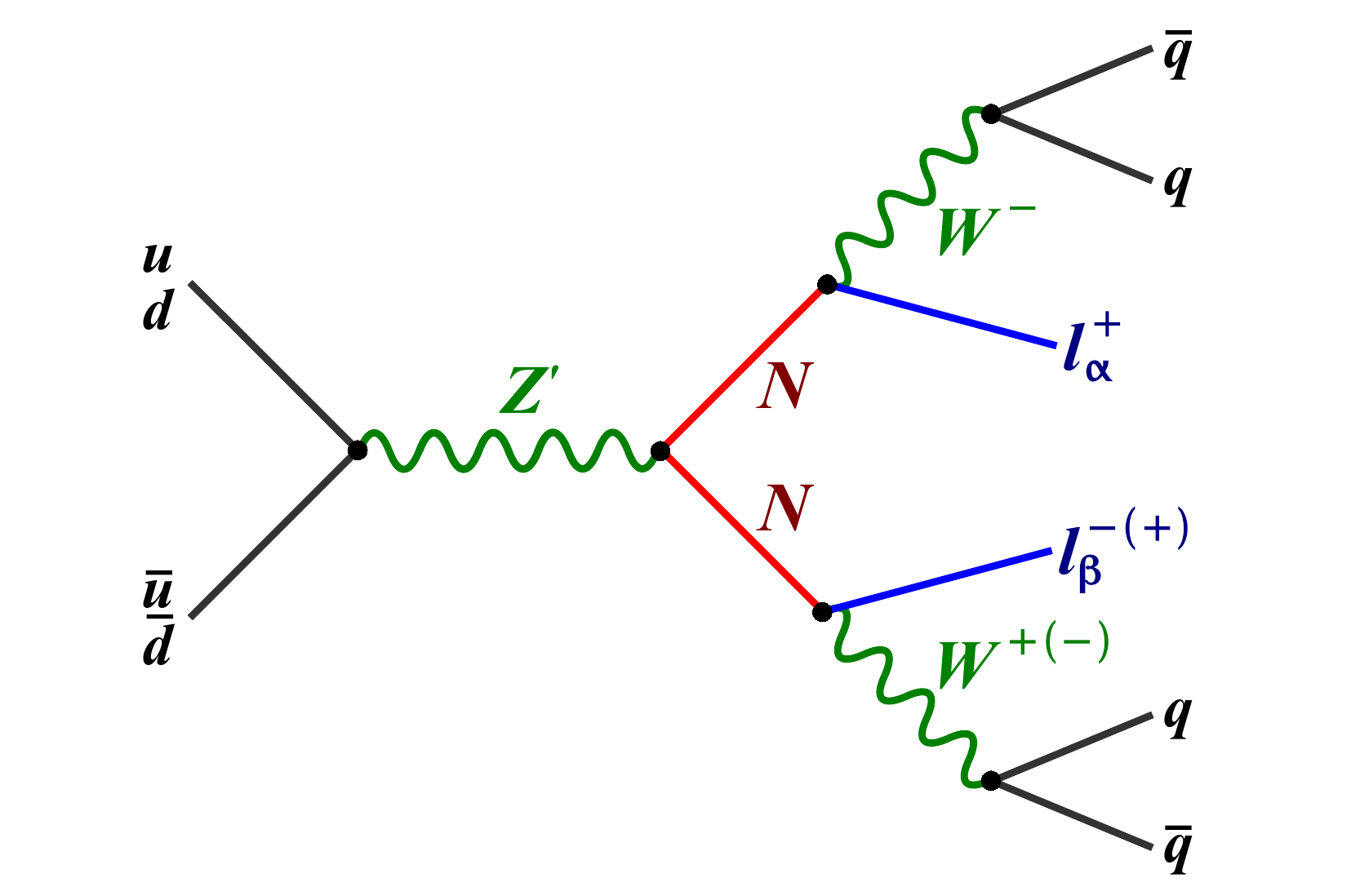}
\end{minipage}\hspace{1.0cm}
\begin{minipage}{.4\textwidth}
        \centering
        \includegraphics[width=\textwidth]{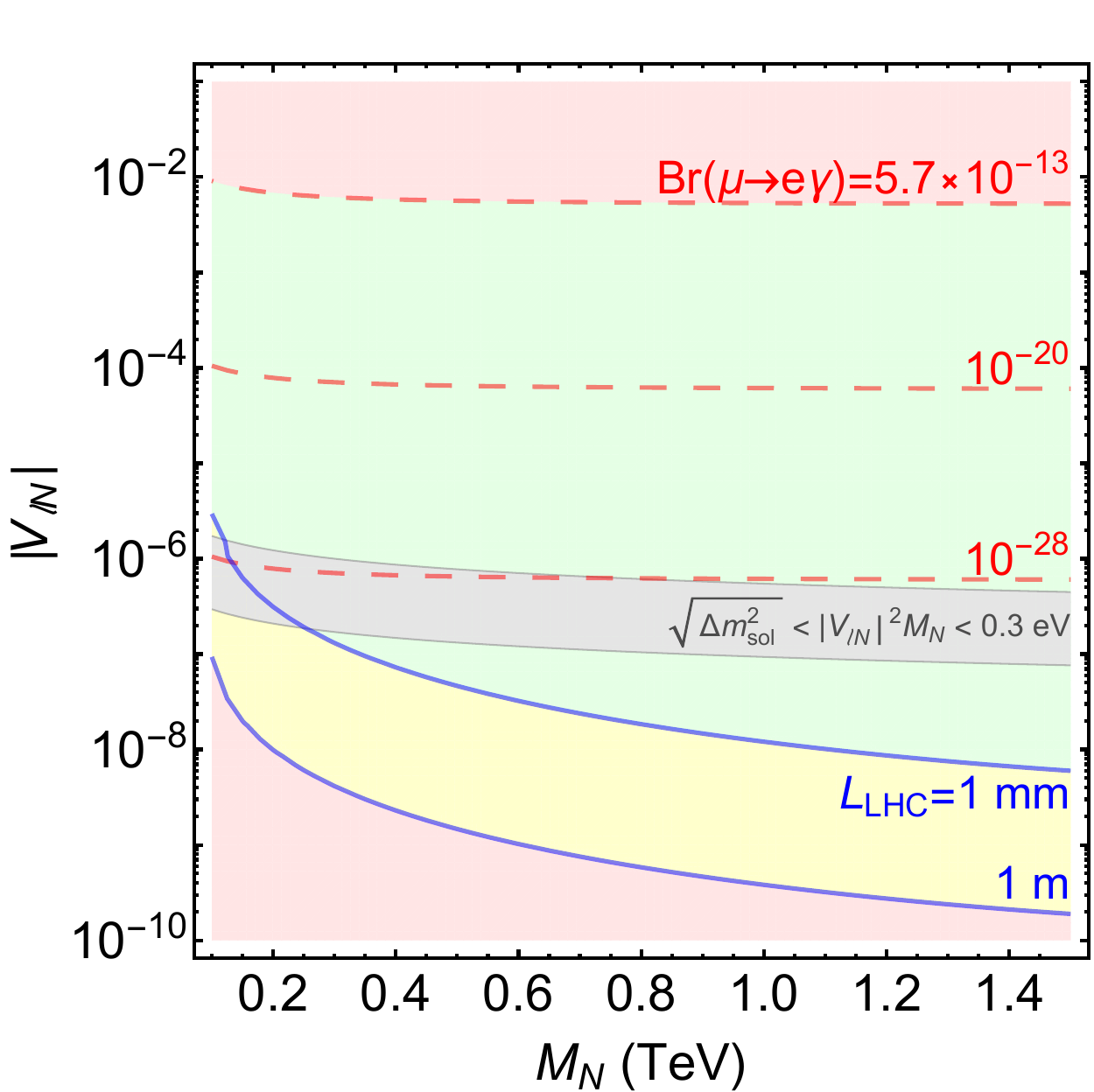}
\end{minipage}
\caption{Left panel: heavy-neutrino pair production and decay in an
    extension of the standard model with an additional $Z'$
    boson. Right panel:
    contours in the plane of neutrino mixing $|V_{lN}|$ and neutrino
    mass $m_N$. The dashed red lines correspond to different branching
    ratios $\text{BR}(\mu \rightarrow e\gamma)$, and the blue lines correspond to two
    distances between displaced vertices at the LHC. 
        From \citet*{Deppisch:2015qwa}.}
\label{fig:resonantLHC}
\end{center}
\end{figure*}
A numerical example is shown in Fig.~\ref{fig:resonantLG}. The lepton
asymmetry is all in the $\tau$ flavor. It is notable that the mass
of the three heavy neutrinos can be lowered to $m_N =
400~\text{GeV}$. The price is the extreme fine-tuning of mass
differences $\Delta M_1/m_N = (M_1 - m_N)/m_N\sim 10^{-5}$ and
$\Delta M_2/m_N = (M_2 - M_3)/m_N \sim 10^{-9}$, with $m_N = (M_2 + M_3)/2$. 
As in the case discussed in the Sec. \ref{sec:flavour} (see
Fig.~\ref{fig:S2S3}), the final asymmetry $|\delta^L|$ is much smaller
than the asymmetry $|\delta^L|$ before the equilibrium of the heavy
neutrinos, indicating the sensitivity with respect to fine-tuning of the
parameters.

Some special parameter region of resonant leptogenesis can be
probed at the LHC \citep*{Deppisch:2015qwa}. At zero temperature the four real degrees of
freedom of the complex doublet $\phi$ become the physical Higgs $H$
and the longitudinal components of $W$ and $Z$ bosons. The
heavy neutrinos can decay into these bosons and leptons with 
decay rates related to  Eq.~\eqref{GammaNphi}:
\begin{equation}
\label{GammaN}
	\Gamma_{\alpha l} =
	\Gamma(N_\alpha \to l^-_{lL} + W^+) + 
	\Gamma(N_\alpha\to \nu_{lL} + Z, H)\ .  
      \end{equation}
To obtain a sufficiently large cross section for heavy-neutrino pair
production it is necessary to extend the standard model by an additional U(1) gauge
group and a corresponding $Z'$ gauge boson. The produced heavy
neutrinos then decay into $l_L^\pm W^\mp$ or $\nu_{L}Z,H$. This leads
to interesting like-sign ($l^+l^+$) and opposite-sign ($l^+l^-$) dilepton events; see
Fig.~\ref{fig:resonantLHC}. The decay amplitude is proportional to the
light- and heavy-neutrino mixing
$|V_{lN}| \sim \sqrt{m_\nu/m_N} = \mathcal{O}(10^{-6})$.
Hence, the lifetime of the decaying heavy neutrino $N$ is long. This leads
to displaced vertices in the detector with a displacement length in the
range $L_\text{LHC} \sim 1~\text{mm} \cdots 1~\text{m}$, which is within reach
of the LHC detectors; see Fig.~\ref{fig:resonantLHC}. 
Complementary with $Z'$ models, TeV-scale left-right symmetric
models with quasidegenerate heavy neutrinos can also be probed at the LHC
\citep*{Dev:2019rxh}. Moreover, TeV-scale leptogenesis can be realized in
left-right symmetric models with a $B-L$ breaking phase transition
\cite{Cline:2002ia,Sahu:2004sb}.

\subsection{Sterile-neutrino oscillations}
\label{sec:arsLG}
Thermal leptogenesis with out-of-equilibrium decays of heavy Majorana
(sterile) neutrinos can work down to masses around the electroweak scale 
once the $CP$ asymmetries in their decays are resonantly enhanced.
Leptogenesis with even lighter sterile neutrinos that have masses
$\mathcal{O}(\text{GeV})$ is possible via  $CP$-violating oscillations
among sterile neutrinos
\citep*{Akhmedov:1998qx}.
In this scenario  the neutrino Yukawa couplings are so small that
at least one sterile-neutrino flavor never reaches thermal equilibrium 
before sphaleron freeze-out.
Recently it was demonstrated that the regimes of resonant leptogenesis
and of leptogenesis through oscillations are in fact connected and
allow for a unified description \citep*{Klaric:2020lov}.

The time evolution is described by kinetic equations for the 
matrix of sterile-neutrino
phase-space densities $ \rho  _ N $ ~\cite{Sigl:1992fn},
often referred to as the density matrix,
and the chemical potentials $ \mu  _ \alpha  $ for $ B/3 - L _ \alpha   $
\cite{Asaka:2005an,Canetti:2012vf,Canetti:2012kh}.
Without Hubble expansion they read
\begin{align}
i \frac{d\rho_{N}}{d t}=&[H, \rho_{N}]-\frac{i}{2}\{\Gamma_N, \rho_{N} 
- \rho^{\rm eq}\} +\frac{i}{2} \mu_\alpha{\tilde\Gamma^\alpha_N}\ ,
\label{drdt} \\
i \frac{d\rho_{\bar{N}}}{d t}=& [H^*, \rho_{\bar{N}}]-\frac{i}{2}
\{\Gamma^*_N, \rho_{\bar{N}} - \rho^{\rm eq}\} -
\frac{i}{2} \mu_\alpha{\tilde\Gamma_N^{\alpha *}}\ ,
\label{drbdt} 
\end{align} 
and
\begin{align} 
i \frac{d\mu_\alpha}{
d t}=&-i\Gamma^\alpha_L \mu_\alpha +
i {\rm tr}\left[{\tilde \Gamma^\alpha_L}(\rho_{N} 
	-\rho^{\rm eq})\right]\nonumber\\
&-i {\rm tr}\left[{\tilde \Gamma^{\alpha*}_L}(\rho_{\bar{N}} 
 -\rho^{\rm eq})\right]\ .
	\label{dmdt}
\end{align}
The SM particles are assumed to be in kinetic equilibrium.
$\rho^{\rm eq}$ is the equilibrium density matrix, $\rho_N$ and
$\rho_{\bar{N}}$ correspond to ``particles'' and ``antiparticles'' 
defined in terms of the $N_{i}$ helicities, $H$ is the
dispersive part of the finite-temperature effective Hamiltonian, 
$\Gamma_N$,  $\Gamma_L^{\alpha}$, and $\tilde{\Gamma}_L^{\alpha}$ are 
rates
accounting for various scattering processes (see Sec.~\ref{sec:theoryLG}), 
and $\alpha = e,\mu$, and $\tau$ labels the lepton flavor. 
The equations describe
thermal sterile-neutrino production, oscillations, freeze-out, and decay
and were refined and extended by
\citet{Hernandez:2015wna,Hernandez:2016kel},
\citet{Ghiglieri:2017gjz}, and \citet*{Bodeker:2019rvr}. 
Above the sphaleron freeze-out temperature $ T _ { \rm EW }  \sim 130 $ GeV,
a lepton asymmetry, partially concerted to a
baryon asymmetry, is generated in $CP$-violating oscillations of the 
sterile neutrinos that are 
thermally produced but do not all equilibrate.
With decreasing temperature the oscillations become increasingly rapid.
Eventually the 
off-diagonal elements of the density matrix effectively  
vanish in the mass basis, and the oscillations come to an end. 
Below $ T _ { \rm EW } $ the baryon asymmetry is frozen,
while the lepton number continues to evolve.  

\begin{figure}
        \includegraphics[width=.45\textwidth]{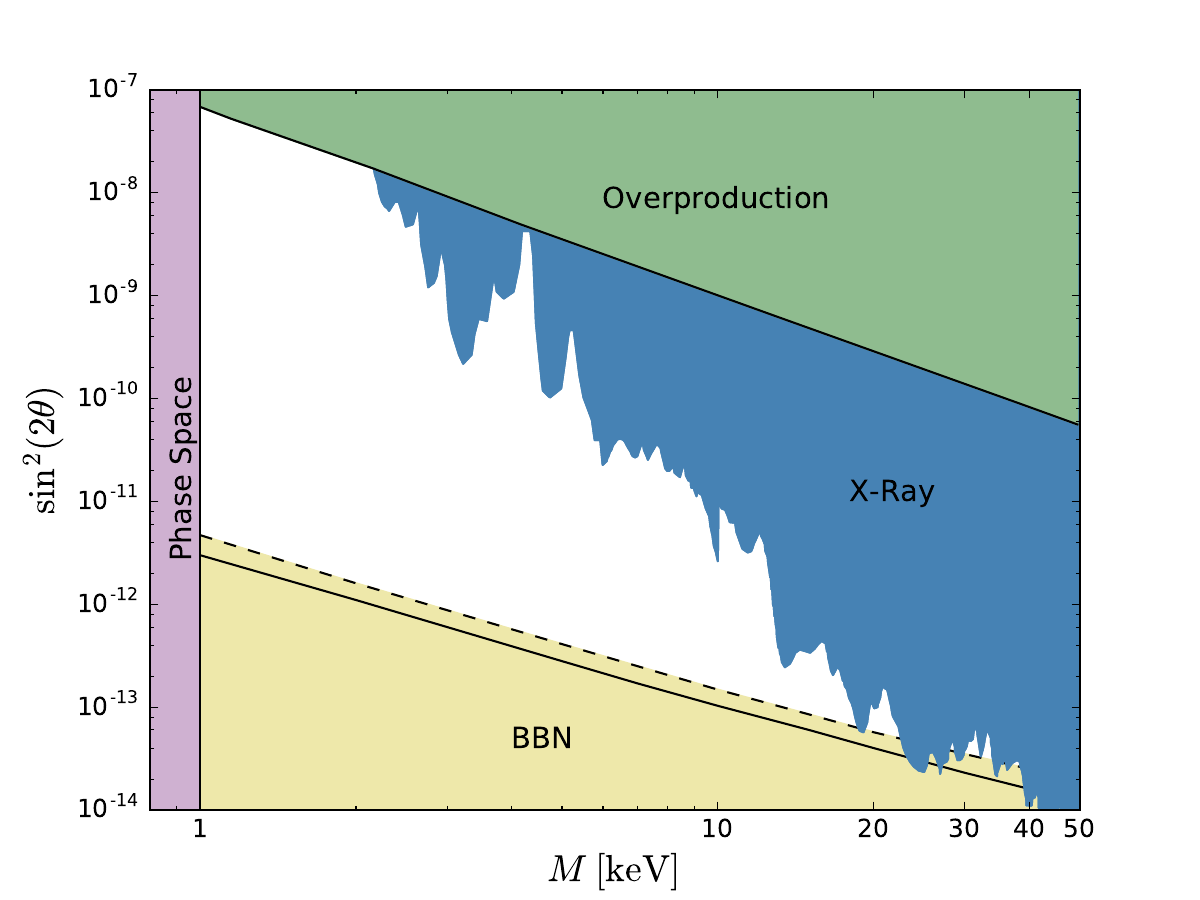}
  \caption{Constraints on  mass and mixing for $ N _ 1 $	
	making up all of  dark matter. 
	The colored regions are excluded. 
     A lepton asymmetry affects the proton-to-neutron ratio 		
	during big bang nucleosynthesis (BBN), which gives rise to 
	an upper limit on the lepton
	asymmetry. 
	The BBN limit given by the solid black line holds if 
	all of the input lepton
	asymmetry is only in the muon flavor. 
	The dashed line corresponds to the BBN limit if
	the lepton asymmetry is split equally among all three flavors.
	Large mixing angles are excluded because too much dark matter
	would be produced,  or because
	 x  rays from the decay $ N _ 1 \to \nu  \gamma  $ 
	would have already been detected.
 	Additional  constraints (not shown) come from structure 
	formation \cite{Schneider:2016uqi}.
     From \citet{Bodeker:2020hbo}.}
\label{f:nuMSM}
\end{figure}
\begin{figure}[b]
\includegraphics[width=0.45\textwidth]{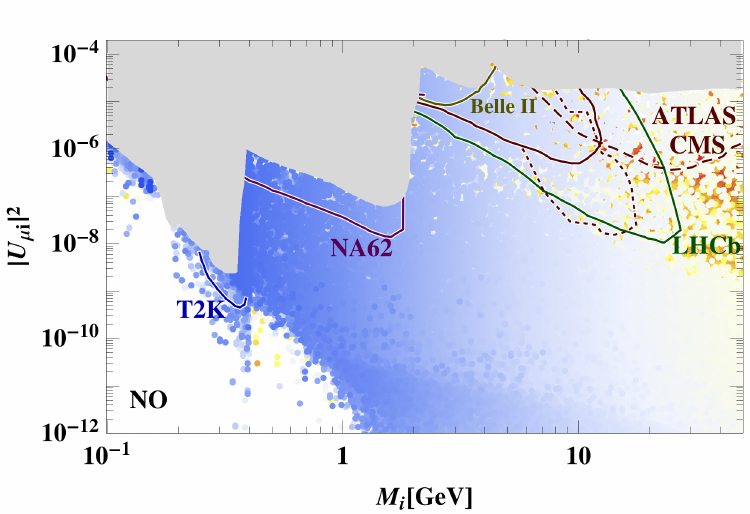}\\
\includegraphics[width=0.3\textwidth]{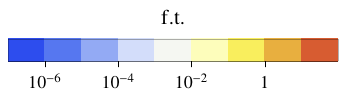}
\caption{Mixing $U_{\mu i}^2$ of heavy neutrinos $N_i$ with leptons
$l_\mu$ as a function of the heavy-neutrino mass $M_i$. The gray region
is excluded by direct searches of heavy neutral leptons, and the lines
show the expected sensitivities for the ongoing T2K, NA62,
Belle II, LHCb, ATLAS, and CMS experiments. 
The parameter f.t.\ measures the amount
of fine-tuning for Yukawa couplings needed for successful leptogenesis.
From \citet{Abada:2018oly}.}
\label{fig:3nuMSM}
\end{figure}

A much studied scenario is  an extension of the SM by three sterile
neutrinos in which the two heavier ones ($ N _ { 2 , 3 } $)
can generate
the baryon asymmetry, and  the lightest ($ N _ 1 $) is
available as a dark matter candidate (the neutrino minimal
standard model). 
It is noteworthy that such a
minimal extension 
might  account for neutrino oscillations, 
baryogenesis, and dark matter  \citep*{Asaka:2005an,Asaka:2005pn}.
The observed baryon asymmetry requires lepton chemical potentials
$ \mu_\alpha /T \sim 10^{-10} $ at $   T\sim T_{\rm EW} $.
Below $T_{\rm EW}$, the sphaleron processes are
ineffective, so a change of the lepton chemical potential no longer 
affects the baryon asymmetry. 
Eventually, $N_2$ and
$N_3$ decay and thereby increase the lepton asymmetry. 
Now large lepton chemical potentials are needed to
generate, resonantly amplified, the observed amount of dark matter:  
	$ \mu_\alpha/T \gtrsim 8\cdot 10^{-6} $ at $  
T\sim 100~\mathrm{ MeV} $.
The lightest sterile neutrino $N_1$ provides dark matter. It
has a mass in the range $1 < M_1 \lesssim 50~\mathrm{keV}$ and small
mixings,  $10^{-13}\lesssim\sin^2(2\theta_{\alpha1})\lesssim 10^{-7}$,
such that the decay rate is small and it can survive until today. 
Various constraints are shown in  Fig.~\ref{f:nuMSM}. 
Moreover, the scenario
predicts that the lightest neutrino mass effectively 
vanishes ($m_1 \simeq 0$).
\begin{table*}
\begin{center}
  \begin{tabular}{l|l}
    {\bf Grand unification}  & {\bf GUT-scale leptogenesis} \\\hline\hline
    Fermion representations of SM  & Connection between $B$ and $L$ in GUTs \\
    Gauge coupling unification  (large GUT scale) & Small neutrino masses (GUT-scale seesaw)\\
    Proton decay & Majorana neutrinos \\
    Relations between Yukawa couplings & Relation between $B$ and $L$  asymmetries \\
     Proton decay branching ratios & Neutrino masses and mixings\\ 
      \end{tabular}
\end{center}
      \caption{Comparison between qualitative and quantitative aspects of
        GUTs and leptogenesis, respectively.}
\label{tab:GUT}
\end{table*}
This scenario
requires 
a high mass degeneracy
of the heavier sterile neutrinos \cite{Canetti:2012vf,Canetti:2012kh}
\begin{align}
	|M_2-M_3|/|M_2+M_3| \sim 10^{-11} \ .
	\label{deg} 
\end{align}

It is an interesting possibility that the predicted monochromatic 
x ray line
produced by $N_1$ dark matter decays corresponds to an unidentified observed
x ray line at around $3.5~\text{keV}$
\cite{Bulbul:2014sua,Boyarsky:2014jta}. This
identification has been challenged by blank-sky observations (Dessert,
Rodd, and Safdi, 2020)
but is still under discussion \cite{Boyarsky:2020hqb}.

The extreme fine-tuning of the masses in Eq.~(\ref{deg}) is no longer
needed 
if one does not require the generation of lepton asymmetry for 
the resonant production of $ N _ 1 $. 
When processes  connecting active and sterile
neutrinos with different helicities are taken into account
one finds that a 10\% splitting is sufficient
\cite{Antusch:2017pkq}.

Models  with 3 GeV-scale
sterile neutrinos ~\cite{Drewes:2012ma}, none of which contribute
to the dark matter,
have a rich phenomenology. 
Depending on the parameters, there can be resonant enhancement 
due to medium effects. 
In this case only $\mathcal{O}(1)$ tunings for sterile-neutrino masses and
mixings are required \cite{Abada:2018oly}. 
Some results of a parameter scan are shown 
in a mixing-mass plane for sterile neutrinos in Fig.~\ref{fig:3nuMSM}, where  
$U_{\alpha i} \equiv \theta_{\alpha i} = (m_D M_M^{-1})_{\alpha
  i}$. 
There successful
leptogenesis is possible with sterile-neutrino masses in the range
0.1--50 GeV.%
\footnote{    For larger masses the sterile neutrinos would be 
nonrelativistic at $ T _ { \rm EW } $, in which case the computational
method of Abada \ea (2019) does not apply.}
It is encouraging that mixings and masses with successful
leptogenesis can be probed through a number of ongoing experiments.
Further possibilities to test GeV-scale leptogenesis were discussed by
\citet{Chun:2017spz}.

\subsection{Leptogenesis: A piece of a puzzle}
\label{sec:puzzleLG}
In its original version leptogenesis was based on a GUT-scale seesaw mechanism with hierarchical heavy Majorana neutrinos. Since their masses are far above collider energies one may wonder whether GUT-scale leptogenesis is at all experimentally testable. In the 
following we therefore illustrate with a few examples some current
hints and conceivable future evidence for leptogenesis at the GUT
scale. Both long-baseline neutrino-oscillation experiments and
cosmology can be expected to play an important role. 

It is instructive to compare possible tests for leptogenesis and
GUTs. Hints for grand unification are the fact that quarks
and leptons form complete representations of SU(5), the simplest
simple group containing the standard model gauge group. Moreover,
the gauge couplings of strong and electroweak interactions unify at
a large energy scale (GUT scale), approximately without supersymmetry
and more precisely with supersymmetry. A generic prediction of GUTs
is proton decay. Relations between Yukawa couplings are model
dependent. Together with proton decay branching ratios they contain
important information about the theory at the GUT scale.

\begin{figure*}[t]
\begin{center}
  \includegraphics[width=0.4\textwidth]{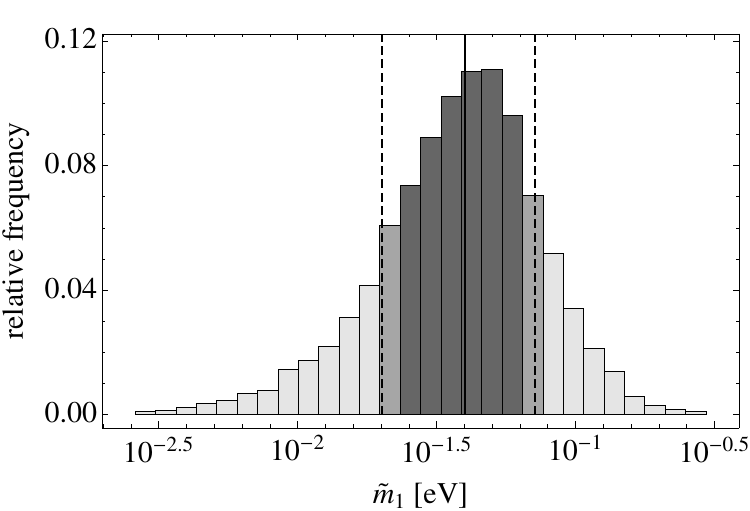}
  \hspace{0.5cm}
  \includegraphics[width=0.4\textwidth]{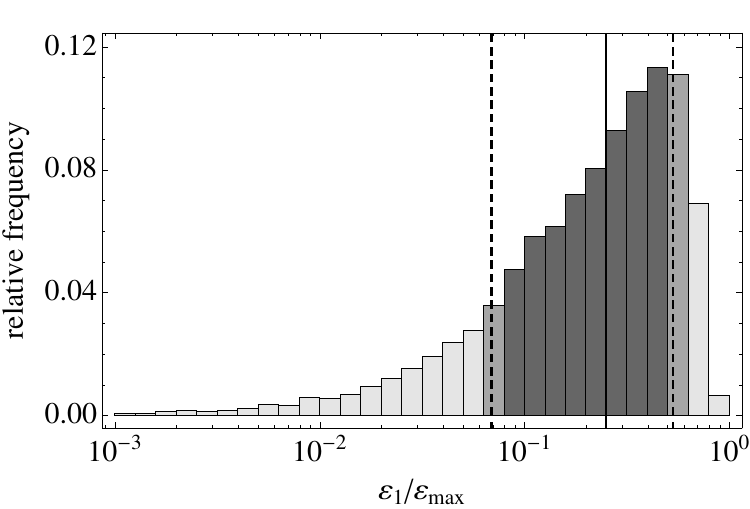}
\end{center}
  \caption{Relative frequency for the leptogenesis parameters $\mt$
    (left) and $\ve_1$ (right). The solid lines denote the position of
    the median and the dashed lines are the boundaries of the 68\%
    confidence region. From \citet*{Buchmuller:2011tm}.}
  \label{fig:random}
\end{figure*}
In a similar way there are qualitative and quantitative hints for
GUT-scale leptogenesis: see Table~\ref{tab:GUT}. Interactions in GUTs
change baryon and lepton number, and the spontaneous breaking of
\BL~can generate large Majorana masses for right-handed neutrinos, the basis of
leptogenesis. If Yukawa couplings in the neutrino sector are similar
to Yukawa coupling for quarks and charged leptons, the seesaw formula
\eqref{seesaw} automatically yields the right order of magnitude of
the neutrino mass scale in terms of the Fermi scale $v_\text{EW} \sim
100~\text{GeV}$ and the GUT scale $v_\text{GUT} \sim
10^{15}~\text{GeV}$:
\begin{equation}
  m_3 \sim \frac{v_\text{EW}^2}{v_\text{GUT}} \sim
  10^{-2}~\text{eV}\ .
  \end{equation}
A generic prediction of leptogenesis is that light neutrinos are
Majorana fermions, which can be probed in neutrinoless double-$\beta$
decay. Moreover, GUTs connect Yukawa matrices in the neutrino sector
with those in the charged lepton and quark sectors. Depending on the
GUT model, this leads to
predictions for neutrino masses and mixings and to relations among
the phases that yield the $CP$ violation necessary for leptogenesis.

As an example, consider the following pattern of Dirac neutrino and charged lepton
mass matrices that can be obtained in the context of a
Froggatt-Nielsen (FN) U(1)-flavour symmetry
\citep*{Sato:1997hv,Irges:1998ax}:
\begin{equation}
  \label{eq_mnu}
  m_{\nu} \sim 
\frac{v_\text{EW}^2\sin^2\beta}{v_{B-L}} \ \eta^{2a} \  
\begin{pmatrix} \eta^{2} & \eta & \eta \\ \eta & 1 & 1 \\ 
\eta & 1 & 1 \end{pmatrix} \ , 
\end{equation}
\begin{equation}
m_e
\sim v_\text{EW}\cos\beta\ \eta^a 	
\begin{pmatrix} \eta^{3} & \eta^{2 } & \eta \\ 
\eta^{2}& \eta & 1 \\ \eta^{2} & \eta & 1\end{pmatrix}\ .
\end{equation}
The model has two Higgs doublets, $H_u$ and $H_d$, which replace
$\phi$ and $\tilde{\phi}$ in Eq.~\eqref{SMnu}, respectively, and
$v_\text{EW}=\sqrt{\langle H_u \rangle^2 + \langle H_d \rangle^2}$. 
The vacuum expectation value $v_{B-L}  \sim v_\text{GUT}$ breaks
${B-L}$, and $\tan \beta  =
\langle H_u \rangle / \langle H_d \rangle$. $\eta = 1/\sqrt{300}$ is the hierarchy
parameter of the FN model, and $a$ and $a+1$ are the FN charges of the
$5^*$-plets in an SU(5) GUT model \cite{Buchmuller:1998zf}. $m_\nu$
is determined by the seesaw formula \eqref{seesaw}, where the FN charges of
the right-handed neutrinos drop out. $\mathcal{O}(1)$ factors in the
mass matrices remain unspecified in a FN model.

To find out whether a certain pattern of mass matrices can
describe the measured data, it is instructive to treat $\mathcal{O}(1)$
parameters as random numbers and to perform a statistical analysis
\citep*{Hall:1999sn,Sato:2000kj,Vissani:2001im}. A detailed study taking
the two measured neutrino mass-squared differences and two mixing
angles as input was performed by \citep*{Buchmuller:2011tm}. The
parameter scan leads to a prediction for the ``most likely'' values of
the third mixing angle and for phases of the light-neutrino mass
matrix. Choosing $\tan\beta \in [1,60]$, with $\sin\beta \in
[1/\sqrt{2},1)$, the parameter $a \in [0,1]$ is determined from
the normalization of $m_e$. The effective $B-L$ breaking scale
$\bar{v}_{B-L} = v_{B-L}/\sin^2\beta \eta^{2a}$ is determined by the
normalization of $m_\nu$ and peaks at $\bar{v}_{B-L} = 1\times
10^{15}~\text{GeV}$, i.e., close to the GUT scale.
The most interesting quantities for
leptogenesis are the effective light-neutrino mass $\mt$ [see Eq.~\eqref{mt}],
the $CP$ asymmetry $\ve_1$ [see Eq.~\eqref{CPasymmetry}], and the
absolute neutrino mass scale $\mb$ [see Eq.~\eqref{mt}] or, equivalently, $m_1$.
The statistical analysis (see Fig.~\ref{fig:random}) implies normal
hierarchy with the neutrino masses
\begin{equation}\label{numasses}
  m_1 = 2.2^{+1.7}_{-1.4} \times 10^{-3} \ \text{eV},\  
 \widetilde{m}_1 = 4.0^{+3.1}_{-2.0} \times 10^{-2} \ \text{eV} \ ,
\end{equation}
and the $CP$ asymmetry
\begin{equation}\label{asym}
\frac{\ve_1}{\ve_{\text{max}}} = 0.25^{+0.28}_{-0.18}\ ,
\end{equation}
where the maximal $CP$ asymmetry $\ve_\text{max}$ is given in Eq.~\eqref{epsmax}.
Note that the values for $m_1$ and $\mt$ lie inside the neutrino mass
window [Eq.~\eqref{window}]. From Eqs.~\eqref{etaB}, \eqref{kappaf}, 
\eqref{numasses}, and \eqref{asym} one obtains a lower bound on the mass of $N_1$
($M_1/\sin^2\beta \gtrsim 3 \times 10^{11}~\text{GeV}$) that is in accord with
Fig.~\ref{fig:bounds}. Hence, an SU(5) GUT model that successfully
describes the light-neutrino masses also naturally explains the
observed matter-antimatter asymmetry.

Neutrinoless double-beta decay is sensitive to
the effective mass $m_{ee}$, for which one obtains $m_{ee} = 1.5^{+0.9}_{-0.8} \times 10^{-3}  \, \text{eV} .$
Cosmological observations measure the sum of
neutrino masses, which is predicted to be $m_{\text{tot}} = 6.0^{+0.3}_{-0.3} \times 10^{-2} \, \text{eV}$.
Similar statistical analyses have been carried out by several groups;
see \citet{Altarelli:2012ia} and \citet{Lu:2014cla}. If the
condition $\text{det}(m_\nu) = 0$ is imposed on the light-neutrino mass
matrix, the statistical analysis is also sensitive on the leptonic
Dirac phase $\delta$ \citep*{Kaneta:2017byo}.

\begin{figure*}[t]
\begin{center}
  \includegraphics[width=0.45\textwidth]{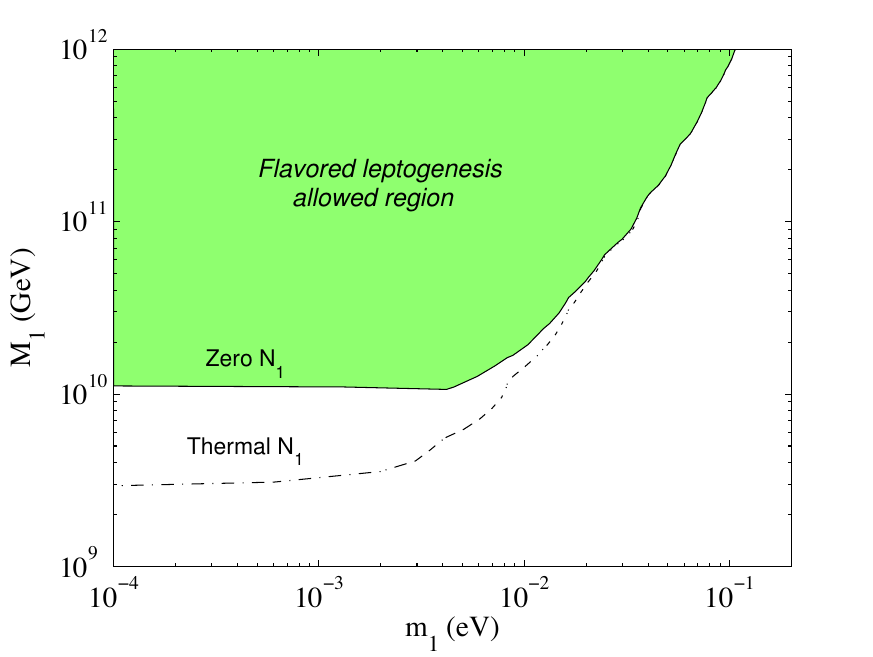}
  \hspace{0.5cm}
  \includegraphics[width=0.45\textwidth]{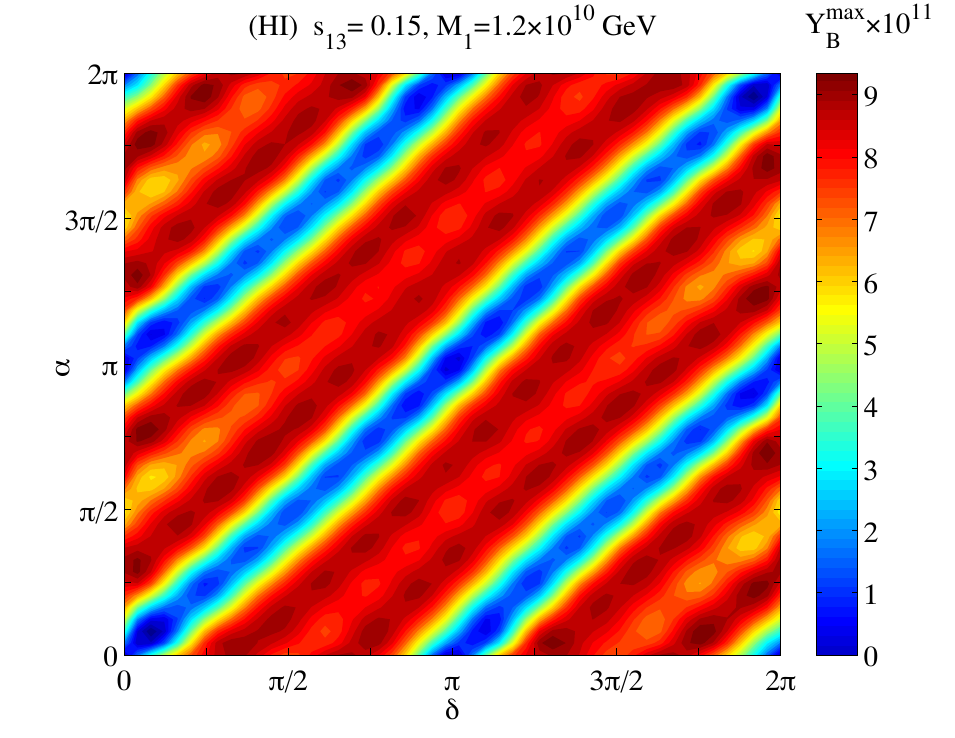}
  \caption{Left panel: region of successful flavored leptogenesis in the
    ($m_1$,$M_1$) plane for zero and thermal initial $N_1$
    abundance. Right panel: correlation between Dirac phase and Majorana
    phase yielding  maximal baryon asymmetry (normal hierarchy).
    From \citet*{Branco:2006ce}.}
  \label{fig:flavourLG}
\end{center}
\end{figure*}

An important ingredient of leptogenesis is the $CP$-violating phases in
the neutrino mass matrices
\cite{Branco:2011zb,Hagedorn:2017wjy}. 
Three  low-energy
phases appear in the light-neutrino mass matrix, the Dirac phase
$\delta$ and the two Majorana phases $\alpha$ and $\beta$; see
Eq.~\eqref{PMNS}. At present there
is evidence for the Dirac neutrino phase $\delta \approx 3\pi/2$
\cite{Abe:2019vii}, and the observation of neutrinoless double-beta
decay would constrain the Majorana phases $\alpha$ and $\beta$.
In the one-flavor approximation, phases beyond the measurable
low-energy phases are needed to obtain a nonvanishing $CP$ asymmetry
$\ve_1$. It is therefore interesting that flavor effects can
yield a nonzero $CP$ asymmetry even for vanishing high-energy phases
\cite{Nardi:2006fx,Blanchet:2006be,Branco:2006ce,Pascoli:2006ie}. This
effect was studied in the two-flavor regime $10^9 <
M_1 < 10^{12}~\text{GeV}$ by \citet*{Branco:2006ce} under the  assumption that $CP$ violation arises solely from
the left-handed neutrino sector. Successful leptogenesis is
obtained for $10^{10} < M_1 < 10^{12}~\text{GeV}$ and $m_1
< 0.1~\text{eV}$ (see Fig.~\ref{fig:flavourLG}, left panel), and a relation between
the Dirac phase and one Majorana phase can be read off from the right
panel of Fig.~\ref{fig:flavourLG}. In some GUT models leptonic $CP$ violation can
indeed be restricted to the left-handed lepton sector. For instance,
in the previously described SU(5) model this is achieved by choosing the
Yukawa couplings $h^\nu$ in the manner described by
\citet*{Branco:2006ce}.
Note that this does not affect $CP$ violation in the quark sector.

\begin{figure}
  \includegraphics[width=0.4\textwidth]{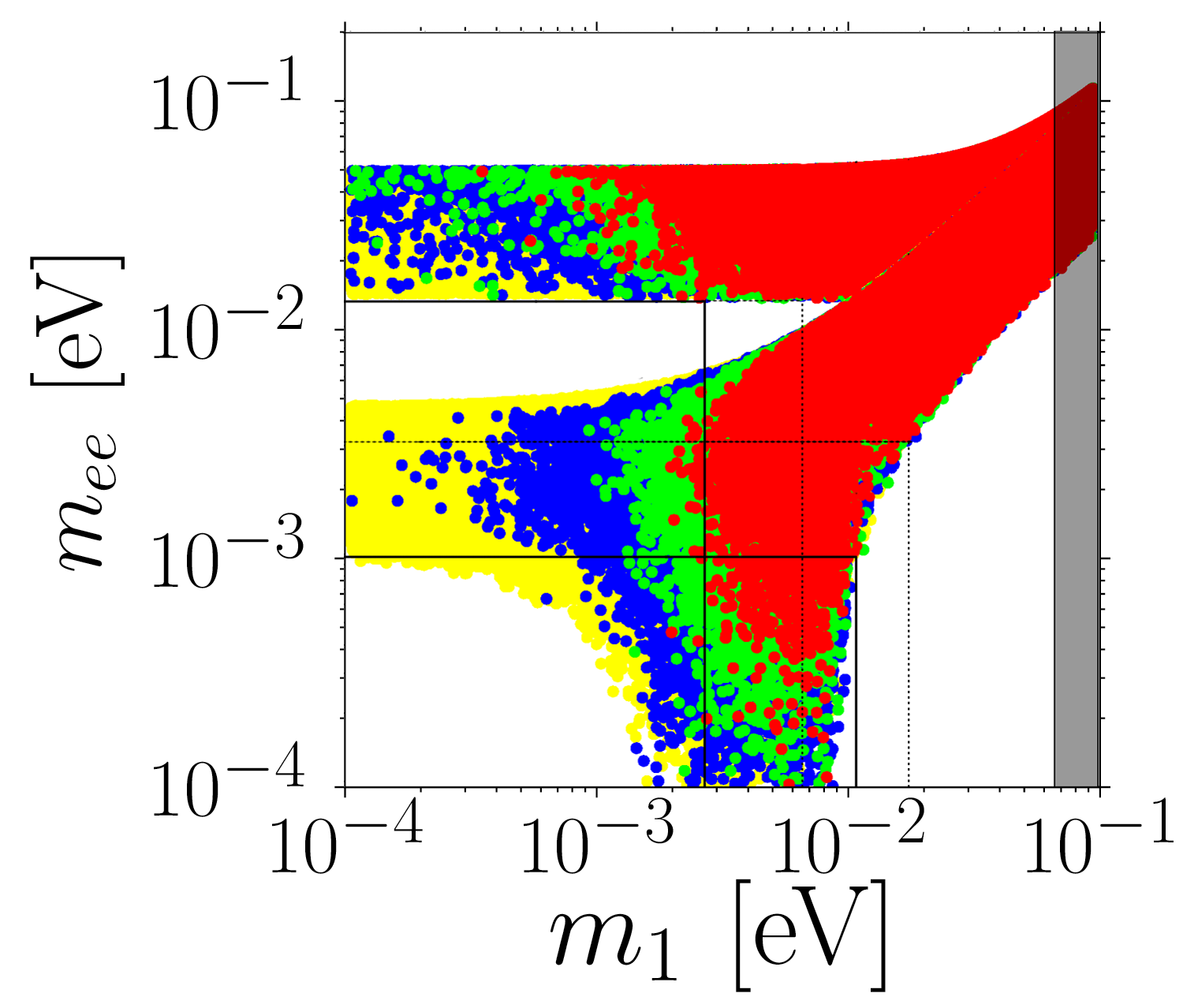}
  \caption{Scatterplot in the ($m_1$,$m_{ee}$) plane where the
    washout of large initial asymmetries is required: $N^i_{B-L} = 0.1
    (\text{red}),  0.01 (\text{green}), 0.001 (\text{blue})$; the
    vertical lines indicate the values of $m_1$ in the three cases,
    beyond which 99\% of the scatter points are found. From \citet*{DiBari:2014eqa}.}
  \label{fig:sumnuLG}
\end{figure}

The absolute neutrino mass scale plays an important role for washout
processes. In the one-flavor approximation it was shown that the
generated lepton asymmetry becomes rather insensitive to an initial
lepton asymmetry of different origin for light-neutrino masses in the
strong-washout regime ($m_i \gtrsim 10^{-3}~\text{eV}$)
\citep*{Buchmuller:2003gz}. However, this
lower bound on neutrino masses is sensitive to flavor
processes. In a range of parameter space where the asymmetry generation
is dominated by $N_2$ and washout processes by $N_1$, respectively,
this was analyzed by \cite{DiBari:2014eqa}. Some results of
a parameter scan are shown for the ($m_1$,$m_{ee}$) plane in 
Fig.~\ref{fig:sumnuLG}. Large initial asymmetries,
$N^i_{B-L} = 0.1 (\text{red}), 0.01 (\text{green}),
0.001 (\text{blue})$, can be erased for $m_1 \gtrsim 10~\text{meV}$. 
This has to be compared with cosmological bounds on $m_{\text{tot}}$.
A combined analysis of CMB and Lyman-alpha data yields the upper bound
$m_{\text{tot}} < 0.12~\text{eV}$ 
\cite{Palanque-Delabrouille:2015pga}, which is
consistent\footnote{Note, however, that based on CMB, BAO, lensing and
galaxy counts, evidence for a nonvanishing total neutrino mass
has been claimed: $m_{\text{tot}} = 0.320 \pm 0.081\ \text{eV}$
\cite{Battye:2013xqa}. This result is in tension with leptogenesis.}
with recent Planck data combined with lensing and baryon acoustic
oscillation (BAO) data \cite{Planck:2018vyg}. 
A measurement of a total neutrino mass $m_{\text{tot}} \sim 100~\text{meV}$
with an uncertainty of $\sim 10~\text{meV}$ is challenging, but it
would have a strong impact on our understanding of leptogenesis.

The connection between leptogenesis and GUT models has been studied in
many explicit models. For examples and references, 
see \citet{Mohapatra:2006gs} and \citet{Altarelli:2010gt}. 

\subsection{Toward a theory of leptogenesis}
\label{sec:theoryLG}
The description of a nonequilibrium process on the basis of thermal
field theory, and without any {\it ad hoc} assumptions, is a highly
nontrivial problem, even for simpler condensed-matter systems.
Owing to favorable circumstances that are described later, over the years this goal has
essentially been reached for leptogenesis. Hence, this process can be
expected to be of
general interest in statistical physics, independent of cosmology.

Thermal leptogenesis takes place in an 
expanding universe with decreasing temperature. 
Traditionally it has been treated using a set of  Boltzmann equations
containing $ S $-matrix elements that assume scattering in vacuum rather than
in a hot plasma. 
Quantum interference plays a key role in generating the asymmetry,
and it is important to understand whether and how this is affected
by the presence of the plasma.
Furthermore, the role of gauge interactions has long been unclear, 
and it turned
out that their role can be quite important. As in most of Sec. IV.B and
IV.C we consider hierarchical heavy neutrinos in the following, and
the relevant Yukawa couplings $h^\nu$ are therefore small.

Leptogenesis is  relatively simple 
compared to electroweak baryogenesis for the following reasons:
\begin{enumerate}	[label=(\arabic*)]
	\item It is homogeneous.
	\item Few degrees of freedom are involved. Therefore, most
	degrees of freedom are in thermal equilibrium. 
     \item The neutrino Yukawa couplings are small.
	Therefore, there is 	
	at least one good set of expansion parameters, which allows
	for  well controlled perturbative approximations.
\end{enumerate} 
Given the model, the values of the parameters, and 
the initial condition, one can then systematically compute
the produced asymmetry in a controlled perturbative expansion.

\subsubsection{Effective kinetic equations}
\label{s:kin} 
In the limit of vanishing neutrino Yukawa couplings $ h ^ \nu$ some
key quantities, such as $ B - L $ and the phase-space density
of sterile neutrinos, are covariantly conserved. 
Since $ h ^ \nu$  are small,
these quantities change slowly. 
On the other hand, many SM interactions, the so-called spectator 
processes, are fast, leading to a hierarchy of timescales. 
Furthermore, the bulk of the degrees of freedom of the system 
participates in the spectator processes, giving rise to a well-defined
temperature. 

The time evolution of the slowly changing quantities is determined
by classical  equations.
When their values are sufficiently close to their equilibrium values 
the equations are linear.  
Since the time evolution is slow, only first-time derivatives appear.%
\footnote{%
A restriction to first order in derivatives is the first approximation
in an expansion in derivatives. 
The corresponding expansion parameter is the ratio 
$ \Gamma  /\omspec $, where $ \Gamma  $ is a typical rate at 
which the slow variables are changing.
Such corrections may be important in leptogenesis through oscillations
~\cite{Abada:2018oly}.
} 
The coefficients in the equations depend only on the temperature
and can be written in terms of real-time correlation functions
evaluated at finite temperature, so all medium effects are included.
For weak coupling these coefficients can be systematically calculated
in perturbation theory. 

For an illustration, consider the simple case in which, 
by the time the baryon asymmetry is produced,
the temperature is much smaller than the mass of the lightest
sterile neutrino $ N \equiv N _ 1 $. 
The sterile neutrinos are then nonrelativistic, and their motion can be 
neglected. 
Furthermore, assume that $ M _ i \gg M _ 1 $ for the other sterile neutrinos,
so that their density can be neglected.
Finally, assume that the temperature is so high that  
none of the charged lepton Yukawa interactions are fast.
The only slow quantities are then the density of  the lightest sterile
neutrino $ n _ N $ and $ B -L $, and the nonequilibrium
system is described by the following effective kinetic equations \cite{Bodeker:2013qaa}
\begin{align}
	\nonumber 
	\dot n _ N + 3 H n _ N 
       \, \, \, = & {} - 
    \Gamma _ N  \, \, ( n _ N - n _ N ^ {\rm eq}  ) 
\\
&
   \, +   \Gamma  _ { N, B-L } \, n _ { B - L } 
   \label{dnNdt}
   ,
\\
 \dot n _ { B - L }  + 3 H  n _ { B - L }  
      = & 
   \, \, \Gamma  _ {B-L, N } 
   ( n _ N - n _ N ^ {\rm eq}  ) 
\nonumber \\ &
    - 
    \Gamma  _ {B - L } 
    \, n _ { B - L }
   \label{dnLdt}
   ,
\end{align} 
while the rest of the degrees of freedom is determined by the 
temperature $ T $.
The  coefficients $ \Gamma  _ i $ depend only on~$ T $.
Note that 
Eqs.~(\ref{dnNdt}) and (\ref{dnLdt}) have the
same form as Eqs.~(\ref{Nchange2})  and ~(\ref{BLchange2}).
However, to arrive at Eqs.~(\ref{dnNdt}) and (\ref{dnLdt}), 
no Boltzmann equation or $ S $ matrix was used: the only ingredient was 
the separation of timescales. 
Correspondingly,
Eqs.~(\ref{dnNdt}) and (\ref{dnLdt}) are valid to all orders in the
SM coupling, and the coefficients include
the effect of all  possible processes.

When describing the simplest possible case, 
Eqs.~(\ref{dnNdt}) and (\ref{dnLdt}) already display the general structure
of all kinetic equations describing leptogenesis. 
In general, the sterile neutrinos must be described by phase-space densities,
depending not only on time but also on momentum. 
This is because relativistic effects can be important, and because
the rates that change the momenta of sterile neutrinos are 
parametrically of the same size as for the change of number densities,
so that one cannot always assume kinetic equilibrium.
Another generalization is that the densities turn into flavor-space
matrices of densities, and that the spin of the sterile neutrinos
has to be accounted for as well. 

The rate coefficients can be computed as follows.
Even in thermal equilibrium, physical quantities fluctuate
around their equilibrium values, and 
the fluctuations of slowly varying variables are described
by the same kinetic equations as the deviations from equilibrium.%
\footnote{%
The only difference is that the equations for the fluctuations
contain a noise term representing the fluctuations 
of the fast variables;
see Sec. 118 of \citet{Landau:1980mil}. 
}
Therefore one can compute unequal time correlation functions
of slow variables using the effective kinetic equations.
By matching the result to the same correlation function computed
in quantum field theory at  frequencies $ \Gamma  _ i \ll \omega  \ll
\omspec $
one can relate $ \Gamma  _ i $ to correlation functions
of SM operators evaluated at finite temperature \cite{Bodeker:2014hqa}.%
\footnote{%
The condition $ \omega  \gg \Gamma  _ i $ ensures that 
one does not need to resum neutrino Yukawa interactions. 
} 
The most important correlation function that one encounters is
the two-point
spectral function of the operator to which the sterile neutrinos couple
[see Eq.~(\ref{SMnu})] 
\begin{align} 
   \widetilde{ \rho } _ i   ( p ) 
   \, \equiv
   \frac 13
   \int \! {\rm d} ^ 4 x \,  
   e ^{ i p \cdot x } 
   \Bigl\langle \, 
      \Bigl\{ 
   (\phi   ^\dagger l _ {  i }) ( x ) , 
   (\overline{ l\, } _ {\! i    }\, \phi  ) ( 0 )  
      \Bigr\}
   \, \Bigr \rangle ^{ } , 
   \label{rhotil}
\end{align}
where the expectation value is taken in an equilibrium ensemble
of standard model fields only.
For example,  the $ \Delta  L = 1 $ washout rate  in the one-flavor regime
can be written  as ~\cite{Bodeker:2014hqa}
\begin{align}
	\label{wash} 
		\Gamma  _ { B - L } 
	 = 
		| h ^ \nu  | ^ 2 
	{ \cal W } \Xi  ^{ -1 }  ,
\end{align} 
with 
\begin{align} 
  { \cal W }  
	= - 
	\int \frac { d ^ 3 p } { ( 2 \pi  ) ^ 3 }
  \frac{ f _ { \rm F }' ( E _ 1 )  } { 2 E _ 1   }
  {\rm tr } \Big \{ 
   \slashed{ p } 
  \big [ 
   \widetilde{ \rho } ( p )
 + \widetilde{ \rho  } ( -p )\big ]
 \Big \} 	,
	\label{calW} 
\end{align}
where $  { p ^ 0 = E _ { 1  } } $. 
Furthermore, 
\begin{align}
	\Xi  \equiv 
	\frac 1 { T \vol } \left \langle ( B - L ) ^ 2 \right \rangle 
\end{align} 
is the $ B-L $ susceptibility.
Its value depends on which reactions are fast, i.e., which
spectator processes are active.
At high temperatures ($ T \gg 10 ^{ 13 }~\text{GeV}$), where only SM gauge 
interactions and the top-Yukawa interactions are fast, the
leading-order susceptibility is $ \Xi  = T ^ 2/4 $. 
Equation \eqref{wash} illustrates the general structure of the rates that consist
of a spectral function, which is a real-time-dependent quantity, and 
an inverse susceptibility, which is determined using equilibrium thermodynamics.
It can be compared to Eq.~(\ref{G1f}).
When SM interactions are neglected, $ \cal W $ is proportional to 
the rate $ \gamma  _ N $. 
Furthermore, 
$ \Xi  $ is then proportional to $ \int d ^ 3 p f _ F ( 1 - f _ F ) $
where $ f _ F $ is the Fermi distribution function.
If we approximate this using Boltzmann statistics, $ \Xi  $ is proportional
to $ n _ l ^ { \rm eq } $. 

The rate $ \Gamma  _ N $  
in Eq.~(\ref{dnNdt}) contains the same spectral function 
as $ \Gamma  _ { B-L } $ \citep*{Laine:2011pq,Bodeker:2015exa}. 
In fact, for leptogenesis through sterile-neutrino oscillations 
all coefficients
in the kinetic equations can be written in terms of the spectral
function in Eq.~(\ref{rhotil}) \cite{Ghiglieri:2017gjz,Bodeker:2019rvr}.

When computing the rates $ \Gamma  _ i $ in perturbation theory,
one has to distinguish among several temperature regimes. 
The nonrelativistic case $ T \ll M _ 1 $ is 
relevant mostly for thermal leptogenesis in the strong-washout regime,
while 
leptogenesis through oscillations proceeds entirely in the ultrarelativistic
regime $ T \gsim M _ 1 / g$,
where $ g $ denotes a combination of electroweak gauge and 
top-Yukawa coupling.

When $ T \lsim M _ 1 $, at leading order in the
couplings  the rates are 
determined by decays and inverse decays. 
In addition to these $ \Delta  L = 1 $ processes,
for the  washout rate $ \Gamma  _ { B - L } $
one has to take into account the
$ \Delta  L = 2 $ processes, even though these are
${  \cal O }  ( h ^ \nu  ) ^ 4 $ and thus appear to be highly suppressed.
Nevertheless, they play a crucial role at late times 
when $ T \ll M _ { N _ 1 }$: 
The 
$ \Delta  L =  1 $ rates are then Boltzmann suppressed with
$ \exp ( -M _ N/T ) $, while the $ \Delta  L = 2 $ rates are 
only power suppressed, so they eventually dominate
at low $ T $, which causes the kink of $ \Gamma  _ W $ 
in Fig.~\ref{fig:leptogenesis} (top panel). 
Since it plays a role only at $ T \ll M _ 1  $, it can be obtained
by integrating out the sterile neutrinos. 
Instead of from Eq.~(\ref{wash}) it follows 
from the two-point function of the Weinberg operator containing
two Higgs and two lepton fields in Eq.~(\ref{SMN}) 
(Sangel, 2016).

For $ T \gsim M _ 1 $ Eq.~(\ref{dnNdt}) 
is replaced by an equation for the phase-space density with a 
momentum-dependent $ \Gamma  _ N $%
\footnote{ 
Equations (\ref{drdt})-(\ref{dmdt}) are obtained using the 
simplifying assumption that the sterile neutrinos are in kinetic 
equlibrium, even though they interact only through their
slow Yukawa interaction. The full momentum dependence was treated
by Asaka, Eijima, and Ishida (2012). 
} :
\begin{align}
	\dot f _ N 
	+ 3 H  \vec p \cdot \frac { \partial f _ N } { \partial \vec p }  
       \, \, \, & = {} - 
    \Gamma _ N ( \vec p )  \left ( f _ N - f _ N ^ {\rm eq}  \right ) 
	+ \cdots  .
   \label{dndtrel}
\end{align}    
On the other hand, the asymmetry is still described by a space density
because it is carried by SM particles that are kept in kinetic equilibrium
by the fast gauge interactions:
\begin{align}
 \dot n _ { B - L }  + 3 H  n _ { B - L }  
      = 
	\int \!\! \frac{  d ^ 3 p } { ( 2 \pi  ) ^ 3 } 
    \Gamma _ { B-L, N } 
	( \vec p )  \left ( f _ N - f _ N ^ {\rm eq}  \right ) 
	+ \cdots  
   \label{dnLdtr}
   .
\end{align}

\begin{figure}
  \begin{center}
  \includegraphics[width=0.43\textwidth]{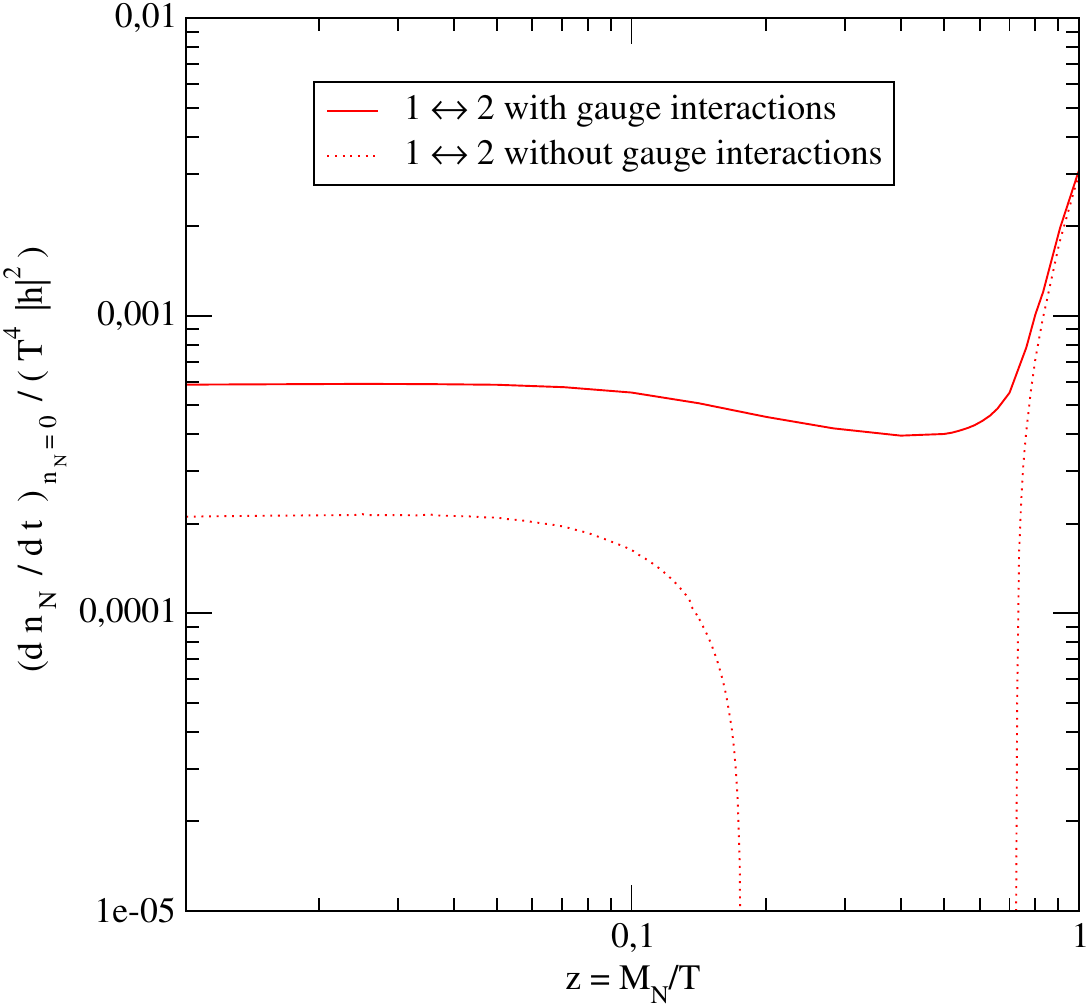}
  \caption{Number of produced Majorana neutrinos per unit time and unit
    volume as a function of the temperature for $  M _ N = 10 ^ 7 $ GeV.
	The dotted curve is the result
    with thermal masses included but without any soft gauge
    interactions. The full line includes an arbitray number of soft
    gauge interactions, as illustrated in
    Fig.~\ref{fig:Nselfenergy}.
  From \citet*{Anisimov:2010gy}.}
\label{f:Nrate}
\end{center}
\end{figure}

\begin{figure}
  \begin{center}
  \includegraphics[width=0.48\textwidth]{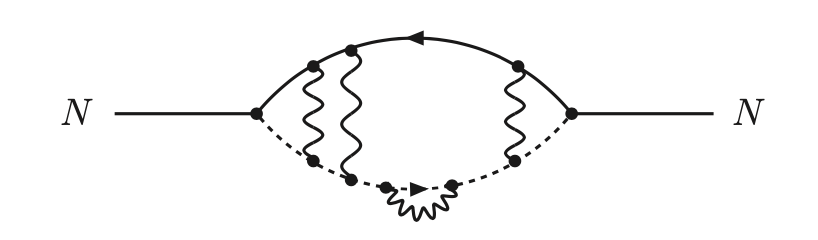}
   \caption{
  Self-energy diagram with soft gauge bosons for Majorana
    neutrino $N$, whose imaginary part contributes to leading order to
    the $N$-production rate. From \citet*{Anisimov:2010gy}.}
\label{fig:Nselfenergy}
\end{center}
\end{figure}

\begin{figure}[b]
  \begin{center}
  \includegraphics[width=0.48\textwidth]{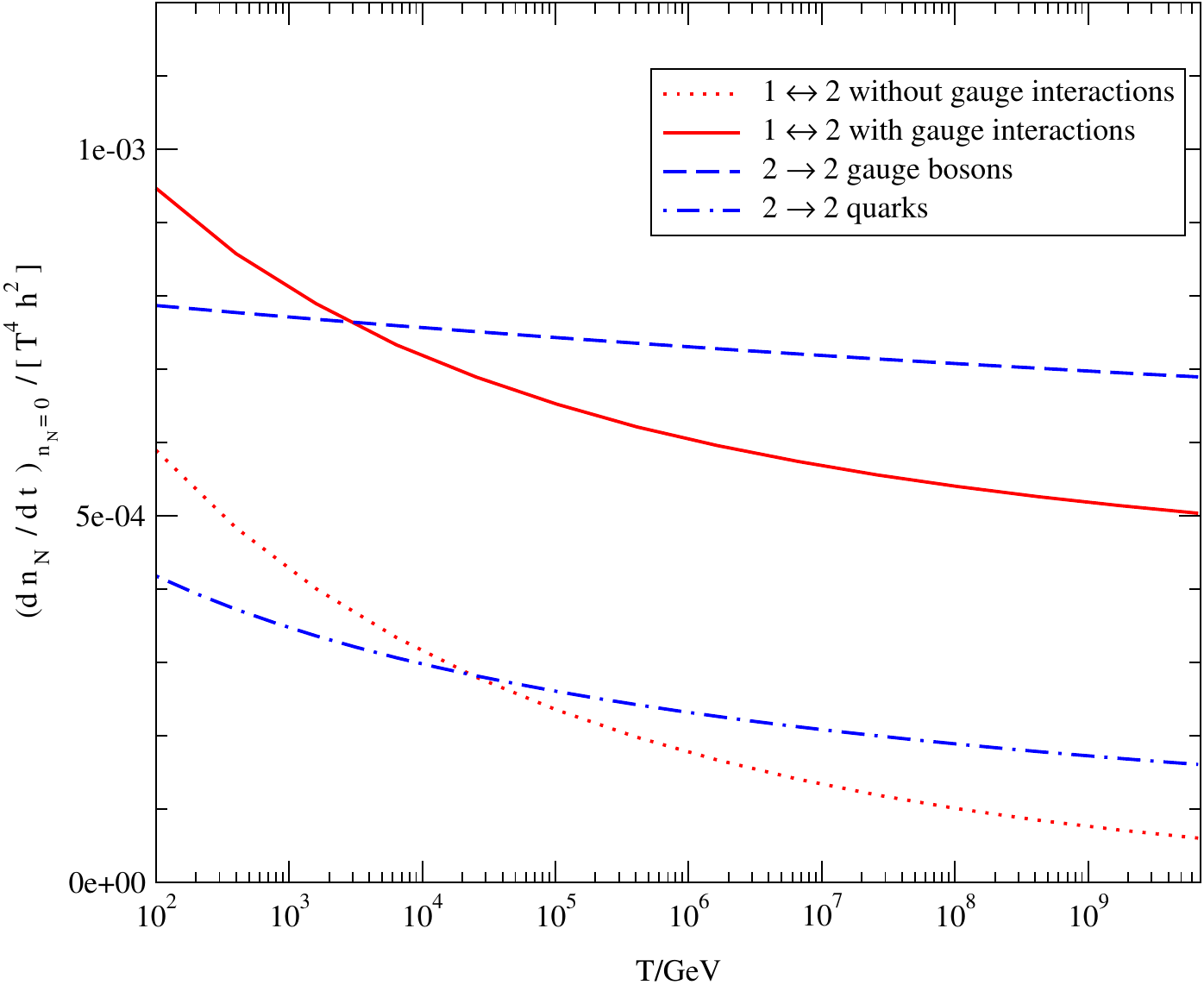}
  \caption{%
Number of produced massless Majorana neutrinos per unit time and unit
    volume as a function of the temperature.
	All leading-order contributions are shown.
    $ 1 \leftrightarrow  2 $ are the inverse decay processes without and
	with multiple scattering mediated by soft gauge bosons displayed
	in Fig.~\ref{f:Nrate}. 
	The $ 2 \to 2 $ scattering contributions involving gauge bosons are 
	of similar size and dominate at high $ T $, while the top-quark 
	scattering is always subdominant.
	From \citet{Anisimov:2010gy}.
}
\label{f:complete}
\end{center}
\end{figure}

In the ultrarelativistic regime plasma effects have a profound
influence on the rates, and gauge interactions are dominant. 
While at $ T \lsim M _ 1 $ the $ 2 \to 2 $ scatterings are higher order,
the inverse
decays are phase space suppressed when $ T \gg M _ 1 $. 
Without taking into account thermal masses, the $ 2 \to 2 $ scatterings
would be dominant, as can be seen 
in Fig.~\ref{fig:leptogenesis} (top panel).  
However, when the thermal masses of the SM particles are included,
both types of processes contribute at leading order.%
\footnote{Except for scalar particles, thermal masses are not uniquely
defined. The ones that are relevant here are the so-called
asymptotic masses, which are valid for momenta of order $ T $.} 
The resulting production rate at vanishing density
$ ( d n _ N /d t ) _ { n _ N = 0 } = 2 ( 2 \pi  ) ^{ -3 } \int d ^ 3  p
\Gamma  _ N ( \vec p ) f _ N ^ { \rm eq }   $
is shown in Fig.~\ref{f:Nrate} (dotted line). 
At high temperatures the rate is due to Higgs decays that are
made possible by the large thermal Higgs mass.
For a small sterile-neutrino mass this could be the main source of 
the baryon asymmetry \cite{Hambye:2016sby}. 
Additional 
multiple  interactions mediated by soft gauge
bosons turn out to be of crucial importance and have to
be resummed [Landau-Pomeranchuk-Migdal (LPM) resummation;
see Fig.~\ref{fig:Nselfenergy}],
giving rise to the full curve in Fig.~\ref{f:Nrate} \citep*{Anisimov:2010gy}.%
\footnote{Note that after the summation of soft gauge bosons
spurious gaps disappear that are caused by kinematical thresholds
due to thermal masses. Such gaps are present in the treatment of
thermal corrections given by Giudice \ea (2004).} 
The same results in the limit $ T \gg M _ 1 $ are shown in
Fig.~\ref{f:complete}, together with the $ 2 \to 2 $ scattering rates.
The LPM-resummed inverse decays and the $ 2 \to 2 $  scattering rates give similar 
contributions, with the latter dominating at higher temperatures.
The contributions involving gauge bosons increase the
total rate by almost an order of magnitude  compared to the 
top-quark scattering  shown in Fig.~\ref{fig:leptogenesis}.


\begin{figure}[t]
  \begin{center}
  \includegraphics[width=0.4\textwidth]{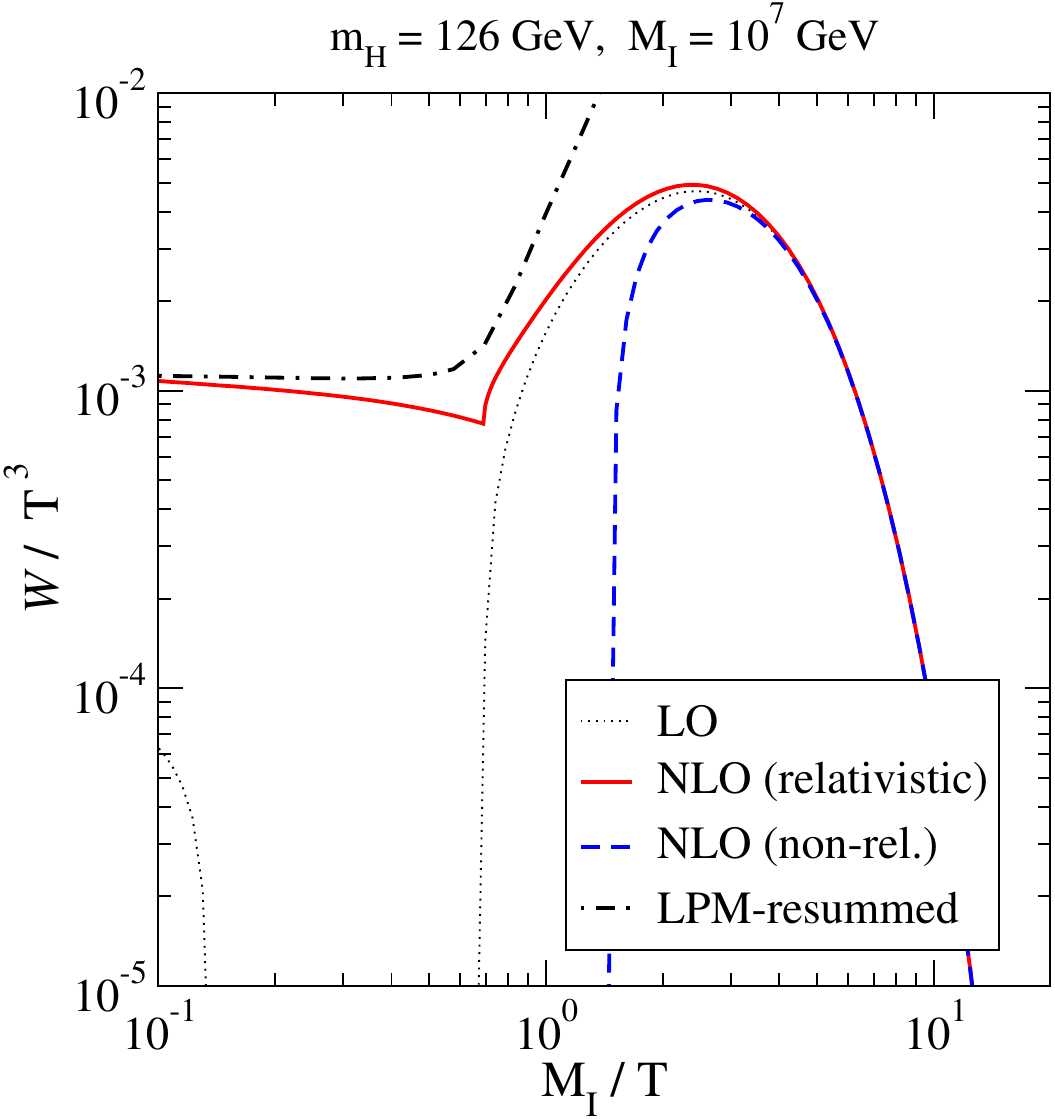}
  \caption{%
Function $\mathcal{W}$ appearing in the 
washout rate in Eq.~(\ref{wash}).
Shown are the LO result including thermal masses (dotted line), 
the relativistic NLO result, and
the NLO result in the nonrelativistic approximation (dashed line).
Also shown is the LO LPM-resummed result that is valid in 
the ultrarelativistic regime ($M_I \lsim g T$) (dash-dotted line).
From \citet{Bodeker:2014hqa}.
}
\label{f:nlo}
\end{center}
\end{figure}

Since all rates for leptogenesis can 
be written in terms of finite-temperature correlation functions, one can
systematically compute higher orders 
of SM corrections, of which there are two types. 
The first are corrections to the susceptibilities, 
which are related to the chemical potentials and the asymmetry densities. 
These are relatively simple to compute 
because they are thermodynamic quantities involving no
time dependence. 
The corrections  to the susceptibilities already start at 
the order of $ g $, and are less than 30\%
\cite{Bodeker:2014hqa,Bodeker:2015zda}.  
The occurrence of odd powers of the coupling
is typical for infrared effects in thermal field theory,
and in this case they are due to soft Higgs exchange. 
The second type are corrections to spectral functions 
as in Eq.~(\ref{rhotil}) that are of the order of $ g^ 2$. 
These are more difficult to compute 
since they are real-time correlation functions.
For the nonrelativistic limit where Eqs.~(\ref{dnNdt}) and (\ref{dnLdt}) 
hold, all rates have been computed at order $ g^ 2 $
in a high-temperature expansion, i.e.,
$ \Gamma  _ N $ (Salvio, Lodone, and Strumia, 2011; Laine and
  Schroder, 2012) and 
the washout rate  \cite{Bodeker:2014hqa}.
The  most interesting one $ \Gamma  _ { B-L, N } $, which is responsible
for the asymmetry, is also the most difficult to calculate.
It is related to the three-point function
of the operators appearing in Eq.~(\ref{rhotil}) \cite{Bodeker:2017deo}
rather than the two-point function.
In the region where the high-temperature expansion is applicable,
the corrections are a few percent. The next-to-leading-order (NLO)
production rate was computed in the relativistic regime 
$ T \sim M _ 1 $ by Laine (2013),
and to the washout rate by Bodeker and Laine (2014); see Fig.~\ref{f:nlo}.

In the strong-washout regime [see Eq.~\eqref{weakstrong}] the
relativistic corrections and the radiative corrections to $ \Gamma  _ N $
affect the produced baryon asymmetry at the level of a few percent
~\cite{Bodeker:2013qaa}. 
The corrections to the $ \Delta  L = 1 $ 
washout rate $ \Gamma  _ { B-L } $, to  
the asymmetry rate $ \Gamma  _ { B-L, N } $,
and to the $ \Delta  L = 2 $ washout rate 
(Sangel, 2016)
were not included
in this analysis, but they are of similar size as the corrections
to $ \Gamma  _ N $ and are not expected to lead to larger corrections
to the produced asymmetry.

\subsubsection{Kadanoff-Baym equations}
\label{s:kb} 

\begin{figure}[t]
\begin{center}
  \includegraphics[width=0.48\textwidth]{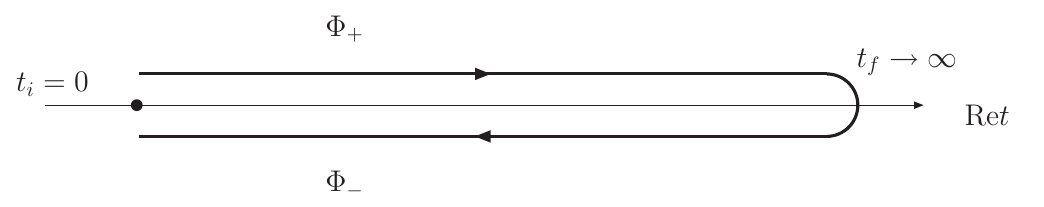}
\caption{Path in the complex time plane for nonequilibrium
          Green's functions.}
\label{fig:contour}
\end{center}
\end{figure}

Leptogenesis involves quantum interferences in a crucial
manner, so 
the standard approach by means of classical Boltzmann equations
may appear problematic. Using the Schwinger-Keldysh, or closed-time-path, formalism 
\cite{Schwinger:1960qe,Keldysh:1964ud}, a full quantum field-theoretical 
treatment
of leptogenesis can be based on Green's functions \cite{Buchmuller:2000nd}.
In the Schwinger-Keldysh formalism one considers Green's functions
$\Delta$ on a complex time
contour starting at some initial time $t_i$; see Fig.~\ref{fig:contour}.  They
satisfy the following Schwinger-Dyson equations with self-energies
$\Pi_C$:
\begin{align}
(&\Box_1 +m^2)\Delta_C(x_{1},x_{2})    \\ 
&+\int_{C}d^{4}x' \Pi_{C}(x_{1},x')
\Delta_{C}(x',x_{2}) =-i\delta_{C}(x_{1}-x_{2})  . \nonumber
\end{align}
It is convenient to consider two correlation functions, the spectral
functions $\Delta^-$, which contain information about the system, and
the statistical propagators $\Delta^+$, which depend on the initial
state at time $t_i$:
\begin{equation}
\begin{split}
\Delta^{+}(x_{1},x_{2})
&=\tfrac{1}{2}\langle\{\Phi(x_{1}),\Phi(x_{2})\}\rangle\ , \\
\Delta^{-}(x_{1},x_{2})&=i\langle [\Phi(x_{1}),\Phi(x_{2})]\rangle\ .
\end{split}
\end{equation}

\begin{figure*}
\begin{center}
  \includegraphics[width=1.0\textwidth]{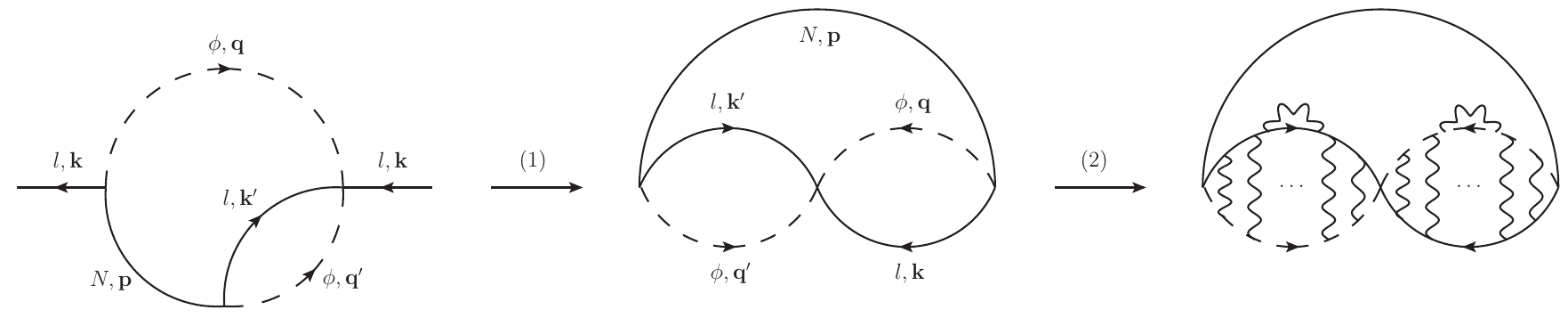}
  \caption{Transformation of the lepton self-energy diagram to a
    ``double-blob''  diagram amenable to resummation. From \citet{Depta:2020zmy}.}
  \label{fig:cylinder}
\end{center}
\end{figure*}
These correlation functions
satisfy the following Kadanoff-Baym equations \cite{Baym:1961zz,Berges:2004yj}:
\begin{equation}
\begin{split}  
\Box_{1,{\bf q}}\Delta^{-}_{\bf q}(t_{1},t_{2}) =   
&- \int_{t_{2}}^{t_{1}} dt'\Pi^{-}_{\bf q}(t_{1},t')\Delta^{-}_{\bf q}(t',t_{2})\ , \\
\Box_{1,{\bf q}}\Delta^{+}_{\bf q}(t_{1},t_{2}) =
&- \int_{t_{i}}^{t_{1}} dt'\Pi^{-}_{\bf q}(t_{1},t')\Delta^{+}_{\bf q}(t',t_{2}) \\
& +\int_{t_{i}}^{t_{2}} dt' \Pi^{+}_{\bf q}(t_{1},t')\Delta^{-}_{\bf q}(t',t_{2})\ ,
\end{split}
\end{equation}
where we have assumed spatial homogeneity and $\Box_{1,{\bf q}}
= (\partial^2_{t_1} + m^2 + {\bf q}^2)$ is the d'Alembert operator for a
momentum mode ${\bf q}$.

For leptogenesis one has to consider two Green's functions:
$S^\pm_{Lij}(x,x')$ for the lepton
doublets, where $i$ and $j$ denote lepton flavors, and $G^\pm(x,x')$ for
the heavy Majorana neutrino. The lepton current is given by
\begin{equation}
  j^\mu_{ij}(x) = -\text{tr}[\gamma^\mu  S^+_{Lij}(x,x)]|_{x'\rightarrow x} \ .
\end{equation}
The nonequilibrium leptogenesis process is a transition from some
initial state to a final state with nonzero chemical potential in
thermal equilibrium.
To compare results from Boltzmann equations and Kadanoff-Baym
equations, a simplified case was considered by \citet{Anisimov:2010aq},
who focused on the $CP$-violating source term for the asymmetry and
ignored washout terms and Hubble expansion. This corresponds to
evaluating the initial lepton asymmetry, generated until the heavy
neutrino reaches thermal equilibrium, starting from zero initial
abundance. Since we consider a spatially homogeneous system, it is
convenient to perform a Fourier expansion. For a momentum mode $\bf k$
diagonal elements of the charge density matrix can be interpreted as
differences of phase-space distribution functions for leptons and
antileptons:
\begin{equation}
  \begin{split}
 L_{{\bf k}ii}(t,t) &=  -\text{tr}[\gamma_0  S^+_{L{\bf k}ij}(t,t)] \\
& = f_{li}(t,k) - f_{\bar{l}i}(t,k) \ .
\end{split}
\end{equation}
Note that, contrary to a system in thermal equilibrium, the
distribution functions are time dependent.

The calculation of the asymmetry starts from a Green's function for
the heavy neutrino that interpolates between a free Green's function
and an equilibrium Green's function \cite{Anisimov:2008dz}. 
The lepton asymmetry is then obtained from  the two-loop
diagram (Anisimov \ea, 2011) in the left panel of Fig.~\ref{fig:cylinder}.
In this calculation the effect of soft gauge-boson exchange turns out to be
of crucial importance. 
For the production of heavy neutrinos they were already
included by \citet{Anisimov:2010dk};
see Figs.~\ref{fig:Nselfenergy} and \ref{f:Nrate}. For the $CP$ 
asymmetry their effect was estimated
by introducing phenomenological thermal widths by \citet{Anisimov:2010dk}.
Recently the 
summation over soft gauge bosons was also completed for the
$CP$ asymmetry \cite{Depta:2020zmy}.
The strategy for calculating the lepton asymmetry
 is illustrated in Fig.~\ref{fig:cylinder}. 
 Integrating over lepton momenta corresponds to closing the external
 lepton line, and summation of the gauge-boson interactions
 leads to the double-blob diagram. 
The corresponding expression for the lepton
asymmetry has been evaluated using a combination of analytical and
numerical techniques.  The result takes the form
\begin{equation}\label{result}
  \begin{split}
    n_{L,ii}(t) &= \int\frac{d^3k}{(2\pi)^3} L_{{\bf k}ii}(t,t) \\
    &\simeq - \ve_{ii} F(T)\frac{1}{\Gamma_N}\left(1 -
  e^{-\Gamma_N t}\right)\ ,
\end{split}
\end{equation}
where the momentum dependence of the
thermal $N$-decay width has been neglected and $F(T)$
is given as a complicated momentum integral. The results for Boltzmann
equations and Kadanoff-Baym equations with thermal widths have the same form, with
different functions $F(T)$ \cite{Anisimov:2010dk}.
\begin{figure}[b]
  \begin{center}
  \includegraphics[width=0.48\textwidth]{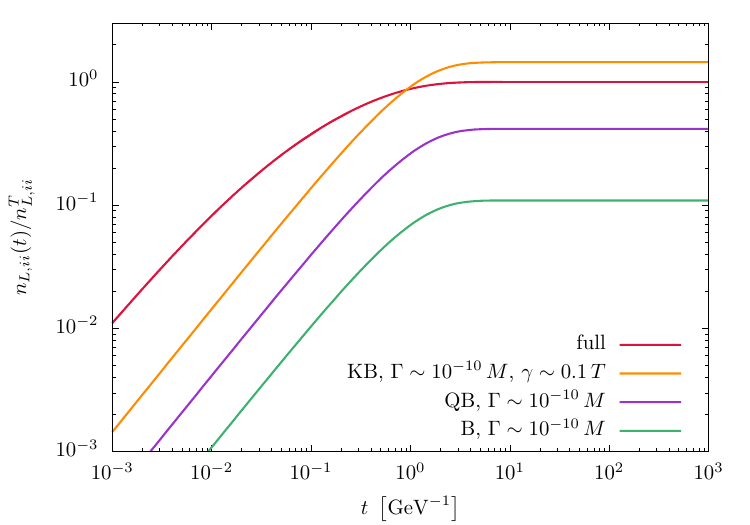}
  \caption{Comparison of the Kadanoff-Baym computation including LPM-resummed 
	gauge interactions (full),  without computing gauge interactions but instead
	parametrizing them with thermal widths (KB),
	with Boltzmann equation containing
	(inverse) sterile-neutrino decays using 
	classical statistics  (B) and quantum
    statistics (QB).
	The mass of the lightest heavy neutrino is $  10 ^ { 10 } $ GeV, 
	the temperature is constant at $ 10 ^{ 11 } $ GeV. 
	This corresponds to the previously discussed ultrarelativistic
        regime $ T \gsim M / g  $ .
    From \citet{Depta:2020zmy}.}    
\label{fig:LasymKB}
\end{center}
\end{figure}
The generated asymmetries
are compared in Fig.~\ref{fig:LasymKB} for the different cases.
For $t < \Gamma_N ^{-1} \sim 1~\text{GeV}^{-1}$, there is a
difference in shape with respect to the numerical solution for the full
Kadanoff-Baym equations. This is a consequence of the approximation 
$\Gamma _ N ( \vec p ) \simeq \Gamma _ N$.
The generated final lepton asymmetries differ by factors of $\mathcal{O}(1)$.

At first sight, the simple time dependence of the asymmetry
\eqref{result} may appear to be surprising. 
However, it is easily understood
as a consequence of the effective kinetic equations \eqref{dnNdt} and \eqref{dnLdt}
for distribution functions. For constant temperature, the equation for $n_N$
has the solution $n_N(t) \simeq n_N^\text{eq}[1 - \exp(-\Gamma_N t)]$
\cite{Anisimov:2010aq} and, if we neglect
washout effects, the solution for $n_{B-L}$ simply inherits
this time dependence. Viewed in this way, solving the Kadanoff-Baym
equations appears to be a way to calculate coefficients in effective
kinetic equations.

The Kadanoff-Baym equations have also been used
to obtain effective equations of motion. This way corrections to the
$CP$-violating parameter were obtained \cite{Garny:2009rv,Garny:2009qn}
and the leading-order asymmetry rate in the relativistic regime
$ T \sim M _ 1 $ was computed \cite{Beneke:2010wd}. 
When lepton Yukawa interaction rates are of a similar size as the
Hubble parameter, 
these also have to be included in the network of kinetic equations,
and the lepton asymmetries are described by matrices in flavor space
that can account for the unflavored as well as the
fully flavored regime \cite{Beneke:2010dz}.
The case of resonant leptogenesis was considered by \citet*{DeSimone:2007gkc}
and an approximate analytical solution was given by
\citet*{Garny:2011hg}.
Leptogenesis through oscillations (Sec.~\ref{sec:arsLG})
was treated by \cite{Drewes:2012ma}.

\begin{figure*}[t]
\begin{center}
  \includegraphics[width=0.4\textwidth]{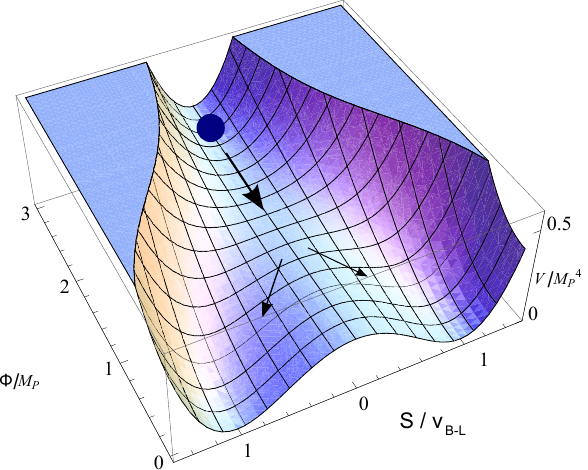}
 \hspace{0.5cm}
 \includegraphics[width=0.45\textwidth]{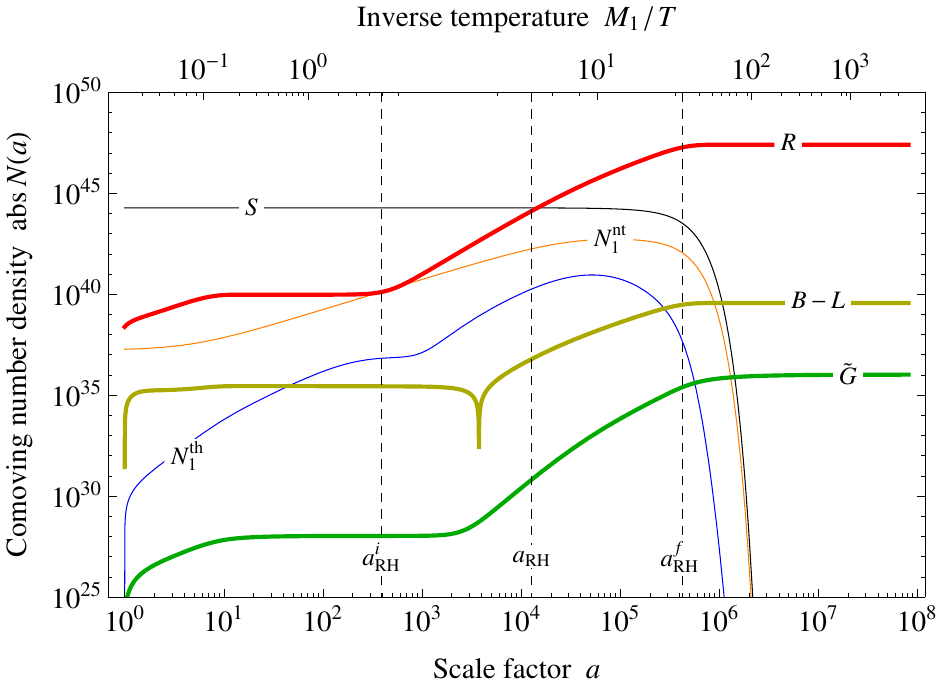}
  \caption{Left panel: hybrid inflation. The time evolution of the inflaton
    field $\Phi$ leads to a tachyonic mass of the waterfall field $S$
    that triggers a rapid transition to a phase with spontaneously
    broken $B-L$ symmetry.
    Right panel: comoving number densities of the
    particles of the $B-L$ Higgs sector ($S$), the thermal and
    nonthermal (s)neutrinos $(N_1^\text{th}, N_1^\text{nt})$, the MSSM
    radiation ($R$), the gravitinos $(\tilde{G})$, and the $B-L$ asymmetry.
    Values were obtained by solving the Boltzmann equations for
    $v_{B-L}=5\times 10^{15}~\text{GeV}$, $M_1 = 5.4\times 10^{10}~\text{GeV}$,
    and $\mt = 4.0\times 10^{-2}~\text{eV}$. 
   From \citet{Buchmuller:2013dja}.}
  \label{fig:cosmoBML}
\end{center}
\end{figure*}

\subsection{Leptogenesis, inflation, and gravitational waves}
\label{sec:cosmoLG}
Some evidence for GUT-scale leptogenesis may be obtained via the
constraints that GUTs impose on neutrino masses and mixings and that
influence the size of the lepton asymmetry. Similarly, leptogenesis
is part of the early cosmological evolution and thereby related to
the other two main puzzles in cosmology, dark matter and inflation.
To work out these connections quantitatively is important for obtaining a
coherent and convincing picture of the early universe.

In this respect it is interesting that, complementary to thermal
leptogenesis, nonthermal leptogenesis can be responsible for the
baryon asymmetry of the Universe. Here the thermal production of heavy
neutrinos is replaced by some nonthermal production, such as inflaton decays
\cite{Lazarides:1991wu,Asaka:1999yd,Asaka:1999jb,HahnWoernle:2008pq}. In
supersymmetric theories the reheating temperature is
bounded from above by the requirement to avoid overproduction of
gravitinos. If the gravitino is the lightest superparticle (LSP), it
can be stable and can form dark matter.
Otherwise, LSP dark matter can be produced in gravitino decays
\citep*{Gherghetta:1999sw}.

An interesting possibility is that GUT-scale leptogenesis might
be probed by gravitational waves GWs. GWs from
inflation can have a characteristic kink in their spectrum indicating the
change from an early matter dominated phase to the radiation dominated
phase and in this way allow for a measurement of the reheating
temperature \cite{Nakayama:2008ip,Nakayama:2008wy} that is
related to the energy scale of nonthermal leptogenesis.
In general, however, the GW signal from inflation is too small to be
observed anytime soon. In the following we describe another
possibility, GWs from cosmic strings produced in a
$\text{U(1)}_{B-L}$ phase transition after inflation \cite{Buchmuller:2013lra}. In
supersymmetric models leptogenesis is naturally linked to $F$-term
hybrid inflation \cite{Copeland:1994vg,Dvali:1994ms}
if one demands spontaneous breaking of $B-L$. The amplitude of
the CMB power spectrum then requires the $B-L$ breaking scale to be of
the order of the GUT scale. This leads to a large stochastic gravitational
wave background that can be probed by ground-based interferometers.

A cosmic-string network can form after the spontaneous breaking
of a U(1) symmetry \cite{Hindmarsh:2011qj}, and the resulting GW spectrum has been
evaluated for Abelian Higgs strings \cite{Figueroa:2012kw,Figueroa:2020lvo} as well as
Nambu Goto strings
\cite{Damour:2001bk,Siemens:2006yp,Kuroyanagi:2012wm}.
If the product of this U(1) group and the SM gauge group
results from the spontaneous breaking of a GUT group,
the theory contains magnetic monopoles in addition to
strings \cite{Vilenkin:1982hm, Martin:1996ea, Leblond:2009fq}, and the
string network becomes unstable. Recently it was pointed out
that GWs from a metastable network of cosmic strings are a
generic prediction of the seesaw mechanism \cite{Dror:2019syi}. Moreover,
it has been shown that, for $\text{U(1)}_{B-L}$ breaking combined
with hybrid inflation, a GW signal is predicted that
evades the bounds from pulsar timing array (PTA) experiments
but will be probed by ongoing and
future observations of LIGO-Virgo and KAGRA \cite{Buchmuller:2019gfy}.

The decay of a false vacuum of unbroken $B-L$ is a natural
mechanism to generate the initial conditions of the hot early Universe
\citep*{Buchmuller:2012wn}.
The false-vacuum phase yields hybrid inflation
and ends in tachyonic preheating \cite{Felder:2000hj}; see
Fig.~\ref{fig:cosmoBML}, left panel. After tachyonic preheating the
evolution can be described using a system of Boltzmann equations. Decays
of the $B-L$ breaking Higgs field and thermal processes produce an
abundance of heavy (s)neutrinos whose decays generate the entropy of the
hot early Universe, the baryon asymmetry via leptogenesis, and dark
matter in the form of the lightest superparticle \cite{Ellis:1983ew}; see
Fig.~\ref{fig:cosmoBML} (right panel).

We now consider  an extension of the supersymmetric standard
model (MSSM) with three right-handed neutrinos that realizes spontaneous
$\text{U(1)}_{B-L}$ breaking in the simplest possible way by using three SM-singlet
chiral superfields $\Phi$, $S_1$, and $S_2$:
\begin{equation}
\label{W}
\begin{split}
W  = &W_{\rm MSSM} + h_{ij}^\nu \mathbf{5}_i^* n_j^c H_u +
\frac{1}{\sqrt{2}}\,h_i^n n_i^c n_i^c S_1 \\ 
 &+ \lambda\,\Phi\left(\frac{v_{B-L}^2}{2} - S_1 S_2\right) + W_0 \ .
\end{split}
\end{equation}
In unitary gauge, $S_{1,2} = S/\sqrt{2}$ corresponds to the physical
$B-L$ Higgs
field, $\Phi$ plays the role of the
inflaton, and the constant $W_0$ is tuned to obtain vanishing
vacuum energy; $n^c_i$ contain the charge conjugates of the right-handed
neutrinos, the SM leptons belong to the SU(5) multiplets $\mathbf{5}^*
= (d^c,\ell)$ and $\mathbf{10} = (q,u^c,e^c)$, and the two Higgs
doublets are part of the $\mathbf{5}$- and $\mathbf{5^*}$-plets
$H_u$ and $H_d$, respectively. Quark and lepton Yukawa couplings
are described in the usual way by $W_{\rm MSSM}$. The flavor structure is
chosen according to \cite{Buchmuller:1998zf} and was already described in
Sec. \ref{sec:puzzleLG}.
\begin{figure}[h!]
\begin{center}
  \includegraphics[width=0.46\textwidth]{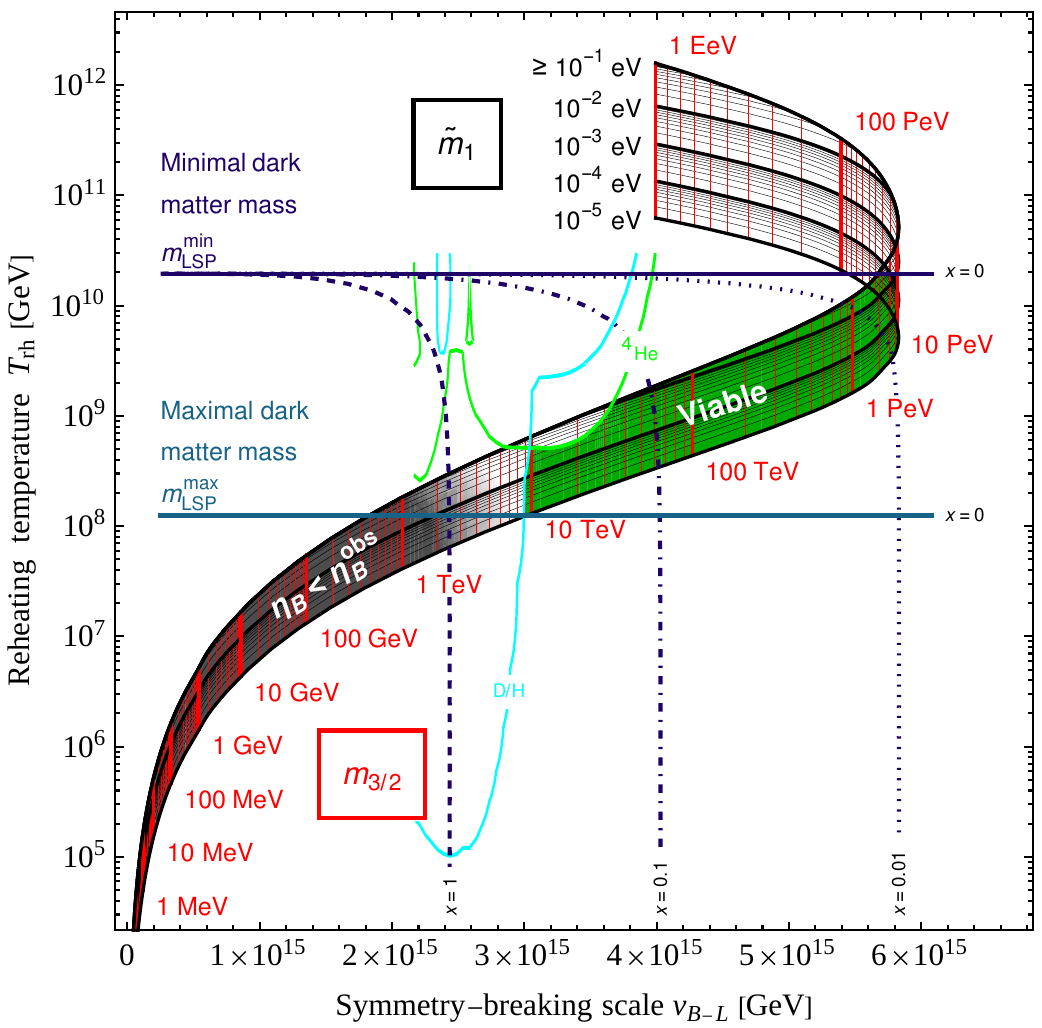}\\ \vspace{0.5cm}
\caption{Viable parameter space (green) for hybrid inflation, leptogenesis, neutralino DM, and big bang nucleosynthesis.
Hybrid inflation and the dynamics of reheating correlate the parameters $v_{B-L}$, $T_\text{rh}$, $m_{3/2}$ and $\widetilde{m}_1$ (black curves).
Successful leptogenesis occurs outside the gray-shaded region.
Neutralino DM is viable in the green region, corresponding to a
Higgsino (wino) with mass $100 \leq m_{\rm LSP}/\textrm{GeV} \leq
1060$ (2680). From \citet{Buchmuller:2019gfy}.}
\label{fig:overlay}
\end{center}
\end{figure}

The $B-L$ breaking part of $W$ is precisely the superpotential of
$F$-term hybrid inflation (FHI).
It was widely believed that FHI could not account for the correct scalar
spectral index of the CMB power spectrum but the analyses given by
\citet*{BasteroGil:2006cm}, \citet*{Nakayama:2010xf},
\citet*{Rehman:2009nq}, and \citet{Buchmuller:2014epa} showed that
FHI is viable once the effect of supersymmetry (SUSY) breaking on the
inflaton potential is taken into account. 
The parameter range consistent with
leptogenesis, inflation, and neutralino dark matter (DM), produced in gravitino
decays, was analyzed by \citet{Buchmuller:2019gfy}. The
result is shown in Fig.~\ref{fig:overlay}.
For given values of $\widetilde{m}_1$ and the gravitino mass $m_{3/2}$, successful
hybrid inflation selects a point in the $v_{B-L}$-$T_\text{rh}$
plane. The gray shading in Fig.~\ref{fig:overlay} indicates the region where leptogenesis falls short of explaining the observed baryon asymmetry.
Gravitino masses of $\mathcal{O}\left(1\right)\,\textrm{TeV}$ or
larger point to a neutralino LSP that is produced thermally as well as
nonthermally in gravitino decays \citep*{Buchmuller:2012bt}.
Gravitinos are in turn generated in decays of the $B-L$ Higgs field 
as well as from the thermal bath; for a discussion and references,
see \citet*{Jeong:2012en}.
Taking into account the fact that gravitinos must decay early enough to
preserve big bang nucleosynthesis~\cite{Kawasaki:2017bqm},
as well as the LEP bound on charginos $m_\text{LSP} \gtrsim 100$~GeV~\cite{Tanabashi:2018oca}, a Higgsino or wino LSP can account for the observed DM relic density in the green-shaded region of Fig.~\ref{fig:overlay}. 
It is highly nontrivial that neutralino DM and leptogenesis can be successfully realized in the same parameter region.
In summary, the viable parameter region of the described model is given by
$v_{B-L} \simeq (3.0$ --  $5.8)\times10^{15}\,\textrm{GeV}$ and $m_{3/2}
\simeq 10\,\textrm{TeV}$ -- $10\,\textrm{PeV}$.

The considered flavor model corresponds to an embedding of ${G}_{\rm
  SM} \times \text{U(1)}_{B-L}$ in the gauge group
$\text{SU(5)}\times \text{U(1)}_{B-L}$ with the final unbroken group 
${G}_{\rm  SM} \times \mathbb{Z}_2$.
This leads to the production of stable cosmic strings
in the $\text{U(1)}_{B-L}$ phase transition \cite{Dror:2019syi}. 
However, the model can also be embedded in SO(10) \cite{Asaka:2003fp}.
In this case, the final unbroken group is ${G}_{\rm SM}$, and there can be no stable strings~\cite{Dror:2019syi}.
Cosmic strings can then decay via the Schwinger production of monopole-antimonopole pairs, leading to a metastable cosmic-string network.
The decay rate per string unit length is given by \citep*{Monin:2008mp,Monin:2009ch,Leblond:2009fq}
\begin{equation}\label{meta}
\Gamma_d = \frac{\mu}{2 \pi} \exp\left( - \pi \kappa \right) \,,
\end{equation}
with $\kappa = m^2/\mu$ denoting the ratio between the monopole mass
$m \sim v_\text{GUT}$ and the cosmic-string tension $\mu \sim v_{B-L}^2$.
For appropriate values of $v_{B-L} < v_{\rm GUT}$, the cosmic strings
are sufficiently long lived to give interesting signatures but decay
before emitting low-frequency GWs that are strongly constrained by
PTA experiments.

\begin{figure}[t]
  \begin{center}
  \includegraphics[width=0.48\textwidth]{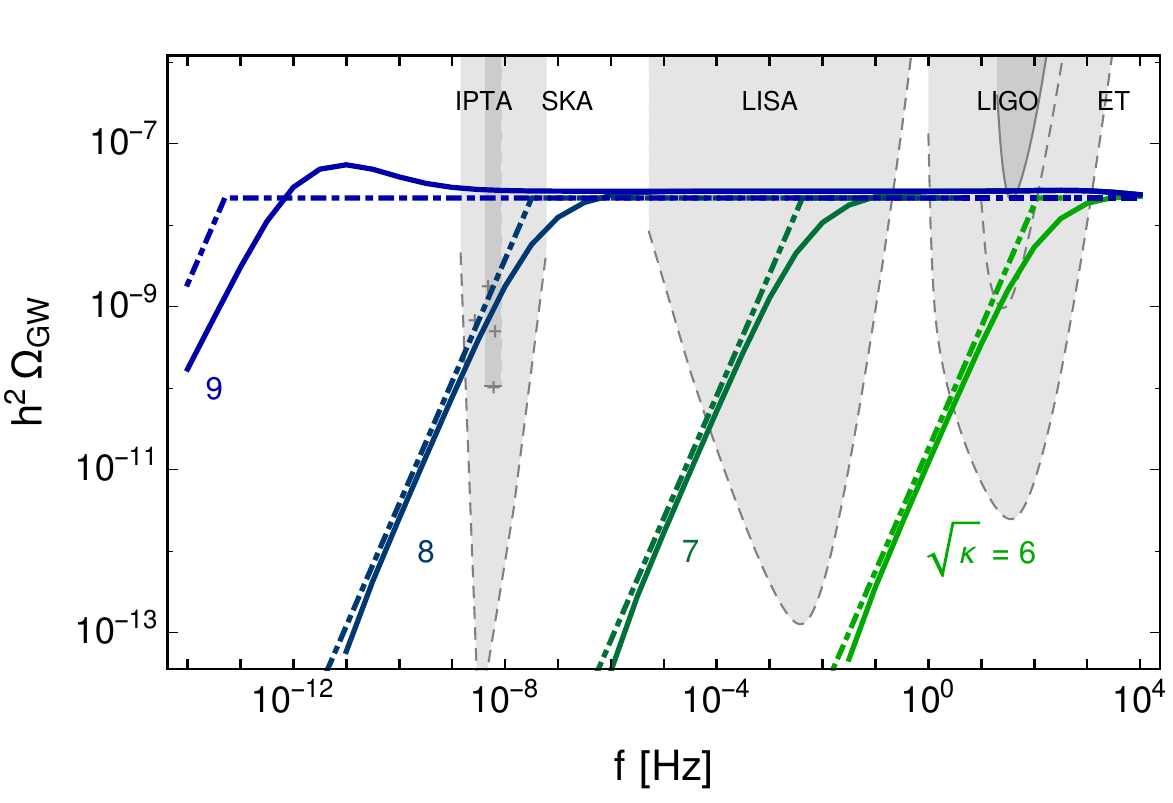}
\caption{GW spectrum for $G \mu = 2 \times 10^{-7}$.
Different values of $\sqrt{\kappa}$ are indicated in different colors; the blue curve corresponds to a cosmic-string network surviving until today.
The dot-dashed lines depict an analytical estimate.
The (lighter) gray-shaded areas indicate the sensitivities of
(planned) experiments SKA~\cite{Smits:2008cf},
LISA~\cite{Audley:2017drz}, LIGO~\cite{LIGOScientific:2019vic} and
ET~\cite{Maggiore:2019uih}, 
the crosses within the SKA band indicate constraints by
the IPTA~\cite{Verbiest:2016vem}. From \citet{Buchmuller:2019gfy}.}
\label{fig:GWs}
\end{center}
\end{figure}

The network of cosmic strings formed during the $B-L$ phase transition acts
as a source of GWs.
Modeling the evolution and GW emission of a cosmic-string network is a
challenging task, resulting in several competing models in the
literature;
see \citet{Auclair:2019wcv} and references therein for a comprehensive review.
Moreover, for metastable strings,
the GW production from fast-moving monopoles requires further
investigation \citep*{Vilenkin:1982hm,Leblond:2009fq}.
The analysis of \citet{Buchmuller:2019gfy}  was based
on the model of \citet*{Blanco-Pillado:2013qja},
and for the first
  time the GW spectrum has been calculated for a metastable string network.
  The GW spectrum reads
  \begin{equation}
  \Omega_\text{GW}(f) = \partial\rho_\text{GW}(f)/\rho_c\partial\ln{f}\ ,
\end{equation}
where $\rho_\text{GW}$ and $\rho_c$ are
GW energy density and critical energy density, respectively. The spectrum is
characterized by the following plateau \cite{Auclair:2019wcv}:
\begin{equation}
\label{eq:Omega_ana}
\Omega_\text{GW}^\text{plateau} \simeq 8.04 \, \Omega_r \left( \frac{G
    \mu}{\Gamma} \right)^{1/2} \ ,
\end{equation}
where $\Gamma \simeq 50$ parametrizes the cosmic-string decay rate
into GWs and $\Omega_r$ is the energy density in radiation relative to
the critical energy density. The GW spectrum has 
a turnover point at the frequency\footnote{The precise value
  of the turnover point depends on the definition. A larger frequency
  $f_*$ was obtained by Gouttenoire, Servant, and Simakachorn (2020).}
\begin{align}
\label{eq:fs_ana}
f_* \simeq 3.0 \times 10^{14} \, \text{Hz} \:e^{- \pi \kappa/4}
  \left(\frac{ 10^{-7}}{G \mu }\right)^{1/2} \ .
\end{align}
Figure \ref{fig:GWs} shows the GW spectrum obtained by a
numerical evaluation as well as the analytical estimate
\begin{align}
\label{eq:Omegaf_ana}
\Omega_\text{GW}(f) = \Omega_\text{GW}^\text{plateau} \; \text{min}\left[ (f/f_*)^{3/2},1 \right] \,.
\end{align}
The shaded regions indicate the power-law-integrated sensitivity curves of current and planned experiments~\cite{Thrane:2013oya}.
For $G \mu = 2 \times 10^{-7}$, the constraint from the European
Pulsar Timing Array \cite{Shannon:2015ect} enforces $\sqrt{\kappa} \lesssim 8$.
In the case of a mild hierarchy between the GUT and $B-L$ scales
($m/v_{B-L} \gtrsim 6$), primordial GWs will be probed by
LIGO-Virgo \cite{LIGOScientific:2019vic}
and KAGRA \cite{Akutsu:2018axf} in the near future.

The general framework behind the described model (inflation ending in
a GUT-scale phase transition in combination with leptogenesis and dark
matter in a SUSY extension of the SM) provides a testable framework
for the physics of the early Universe.
A characteristic feature of this framework is a stochastic background
of gravitational waves emitted by metastable cosmic strings.

Probing leptogenesis with gravitational waves is an
interesting possibility and theoretical work on this subject is just
beginning. For recent work, see \citet*{Blasi:2020wpy} and \citet{King:2020hyd}.

\subsection{Summary: Leptogenesis}
\label{sec:summaryLG}
Thermal leptogenesis is now well understood. It is closely related
to neutrino masses, and simple estimates, based on GUT models, yield
the right order of magnitude for the observed matter-antimatter
asymmetry. In the one-flavor approximation successful leptogenesis
leads to a preferred mass window for the light neutrinos that is consistent
with the cosmological upper bound on the sum of neutrino masses, and
to a lower bound on the heavy Majorana neutrino masses. Taking flavor
effects into account, the qualitative picture remains valid, but
quantitatively the neutrino mass bounds are relaxed. For
quasidegenerate heavy neutrinos the temperature scale of leptogenesis
can be lowered to the weak scale. $CP$-violating oscillations of
sterile neutrinos can lead to successful leptogenesis even for GeV
neutrino masses.

Significant progress has been made toward a full description of
leptogenesis on the basis of thermal field theory. This has been
possible because leptogenesis is a homogeneous process that involves
only a
few dynamical degrees of freedom with small couplings to a large
thermal bath. Effective kinetic equations have been derived, which
take the form of ordinary Boltzmann equations whose kernels can be
systematically calculated in terms of spectral functions of SM
correlation functions. Relativistic and off-shell effects are included
in Kadanoff-Baym equations that have also been used to calculate the
generated lepton asymmetry. In the field-theoretical treatment,
interactions with gauge bosons of the thermal bath turn out to be
crucial and have to be resummed. Using these techniques, for the first
time an estimate of the theoretical error of traditional calculations based on Boltzmann
equations has been obtained; it turns out to be about $50\%$.

An interesting new development is the possibility of probing high-scale
leptogenesis with gravitational waves. This includes the seesaw
mechanism and a high scale of $B-L$ breaking. Theoretical work on this
interesting topic is just beginning, and it is conceivable that a
stochastic gravitational wave background from $B-L$ breaking will soon be observed by
LIGO-Virgo and KAGRA.

\section{Other models}
\label{sec:others}
In this section we mention some alternative proposals for
baryogenesis that could not be described in detail earlier in the
review, with an emphasis on the possible effects of light
pseudoscalar particles.

An interesting idea is ``spontaneous baryogenesis''
\cite{Cohen:1987vi,Cohen:1988kt}, where an arrow of time is singled out
not by a departure from thermal equilibrium, but rather by the motion of a
light pseudo Goldstone boson of a spontaneously broken approximate
global $\text{U(1)}_B$ baryon symmetry. Baryon-number-violating interactions
can be in thermal equilibrium, and the observed baryon asymmetry can
be generated for a sufficiently large $\text{U(1)}_B$ breaking
scale. A related mechanism makes use of axion oscillations in the
presence of rapid lepton-number-violating processes in the thermal plasma,
which can be provided by the exchange of heavy Majorana neutrinos at
high reheating temperatures \citep*{Kusenko:2014uta}.
Recently it was pointed out that spontaneous baryogenesis is a
rather general phenomenon in the presence of axionlike particles, and
that their coupling to gluons is already enough to generate a baryon
asymmetry \cite{Domcke:2020kcp}.

Baryogenesis is also possible in a cold electroweak phase transition  
\cite{Tranberg:2003gi}. A sudden change of the Higgs mass term at zero
temperature leads to a spinodal instability of the Higgs field, and
during the subsequent tachyonic preheating a nonzero Chern-Simons
number can be generated, with a corresponding baryon asymmetry.
A cold electroweak transition can occur once the Higgs field is coupled 
to a dilaton, which can lead to a delayed electroweak phase transition
at the QCD scale \cite{Servant:2014bla}. The $CP$ violation needed for
baryogenesis can then be provided by a displaced axion field, whose
relaxation after the QCD phase transition subsequently solves the
strong $CP$ problem. An axion, solving the strong $CP$ problem and
providing dark matter, can also be combined with the spontaneous
breaking of lepton number, leptogenesis, and Higgs inflation in a
nonsupersymmetric extension of the standard model
\cite{Ballesteros:2016xej}. The Affleck-Dine mechanism of baryogenesis
can be realized without supersymmetry by means of a complex
Nambu-Goldstone boson carrying baryon number, which can occur for a
spontaneously broken appropriate global symmetry
\cite{Harigaya:2019emn}. The role of the Affleck-Dine field can also be played
by a charged Peccei-Quinn field containing the QCD axion as a phase
\cite{Co:2019wyp}. Moreover, baryogenesis is possible at the weak
scale, at temperatures below the electroweak transition, where
sphaleron processes are not in thermal equilibrium. The baryon
asymmetry is generated in decays of a singlet scalar field coupled to
higher-dimensional $B$-violating operators. The mechanism can be
probed by neutron-antineutron oscillations and the neutron EDM \citep*{Babu:2006xc}.
At even lower temperatures of around 
$10~\text{MeV}$ the baryon asymmetry can
be explained by
$B$-meson oscillations in an extension of the standard model with exotic
$B$-meson decays \citep*{Elor:2018twp,Nelson:2019fln}.

In string compactifications one expects moduli fields in the effective
low-energy theory, whose mass depends on the mechanism and energy
scale of supersymmetry breaking. If they are sufficiently heavy, they
can reheat the Universe to a temperature of the order of 100~MeV, so 
nucleosynthesis is not affected. In their decays they can generate the
matter-antimatter asymmetry as well as cold dark matter as Higgsinos
or winos. Since matter and dark matter have the same origin, the
similarity of their energy densities can be explained 
\citep*{Elor:2018twp,Nelson:2019fln}.

To date  the curvature of space-time has played no role in the
considered models of baryogenesis. However, the Ricci scalar of a
gravitational background can play the role of the axion in spontaneous
baryogenesis, and its coupling to the baryon-number current can be
the source of a baryon asymmetry, which is referred to as gravitational
baryogenesis \cite{Davoudiasl:2004gf}. Alternatively, gravitational
waves from inflation can lead to leptogenesis via the gravitational
anomaly of the lepton current \citep*{Alexander:2004us}. Moreover,
in the standard model with
heavy right-handed neutrinos and $CP$-violating couplings, which was
considered for thermal leptogenesis in Sec. \ref{sec:thermalLG},
loop corrections lead to a low-energy effective action where the
gravitational field couples to the current of left-handed neutrinos,
such that neutrinos and antineutrinos propagate differently in
space-time \cite{McDonald:2015ooa}. In quantitative analyses
it has been demonstrated that this effect can indeed account for the
observed baryon asymmetry \cite{McDonald:2020ghc,Samanta:2020tcl}.

\section{Summary and outlook}
\label{sec:summary}
The current paradigm of the early Universe includes inflation at an
early stage. Hence, the observed matter-antimatter asymmetry cannot be
imposed as an initial condition but it has to be dynamically generated
after inflation. This makes baryogenesis an unavoidable topic. Moreover,
50 years after Sakharov's paper, baryogenesis has also become an
interesting story that is connected to all developments of physics
beyond the standard model during the past 40 years, including
grand unification, dynamical electroweak symmetry breaking, low-energy
supersymmetry and neutrino masses.

The first important step in the theory of baryogenesis was made in the
context of SU(5) GUT models that naturally provide heavy particles,
leptoquarks, whose $CP$-violating delayed decays can lead to a baryon
asymmetry. This process was quantitatively understood based on
Boltzmann equations. In these detailed studies it also became clear
that leptoquarks are not ideal agents of baryogenesis since they have
SM gauge interactions that tend to keep them in thermal equilibrium.

The second important step was the discovery of the nonperturbative
connection between baryon number and lepton number in the SM, and the
associated, unsuppressed, sphaleron processes at high temperatures.
This implied that $B+L$ is in equilibrium above the electroweak phase
transition, which ruled out baryogenesis in SU(5) GUT models. However,
an interesting new possibility emerged, electroweak baryogenesis,
which opened the
possibility of generating the baryon asymmetry during a strongly
first-order electroweak phase transition. In principle, the
presence of all necessary ingredients already in the SM is an
appealing feature, yet the electroweak transition turned out to simply
be a smooth
crossover, so the necessary departure from thermal equilibrium can
not be realized. This is different in extensions of the SM with additional Higgs
doublets or singlets, where a strongly first-order phase transition is possible.
Such models have been extensively studied for more than 30 years, without and
with supersymmetry. In view of the results from the LHC and due to
stringent upper bounds on the electric dipole moment of the electron,
today EWBG appears to be unlikely in weakly coupled Higgs models. On the other
hand, EWBG is still viable in composite Higgs models of electroweak
symmetry breaking. This emphasizes the importance of searching at the LHC for new
resonances with TeV masses and strong interactions of the light Higgs boson.

Sphaleron processes have also led to leptogenesis as a new mechanism of
baryogenesis. Contrary to leptoquarks, right-handed neutrinos are
ideal agents of baryogenesis since they do not have SM gauge interactions.
Their $CP$-violating decays lead to a $B-L$ asymmetry that is not
washed out. Right-handed neutrinos are
predicted by grand unified theories with gauge groups larger than SU(5) such
as SO(10). In GUT models the pattern of
Yukawa couplings in the neutrino sector is similar to quark and
charged lepton Yukawa couplings. It is noteworthy that, with \BL~broken
at the GUT scale, this leads automatically to the right order of
magnitude for neutrino masses and the baryon asymmetry. However,
this success of leptogenesis is not model independent. If
right-handed neutrino masses and neutrino Yukawa couplings are rescaled, successful
leptogenesis is also possible at much lower scales, down to GeV
energies. The corresponding
sterile neutrinos can be directly searched for at LHC, by the NA62 experiment,
at Belle II, and at T2K. On the contrary, tests
of GUT-scale leptogenesis will remain indirect. The determination of the
absolute neutrino mass scale and $CP$-violating phases in the neutrino
sector are particularly important. An interesting new possiblity is to
probe the seesaw mechanism and $B-L$ breaking at the GUT scale by primordial
gravitational waves.

Two open questions in particle physics will be crucial
for the further development of
the theory of baryogenesis: First, the discovery of a
strongly interacting Higgs sector would open up new possibilities for
electroweak baryogenesis. Second, the discovery of supersymmetry would
renew interest in Affleck-Dine baryogenesis and strongly
constrain leptogenesis via the properties of the gravitino. However,
there can always be surprises. The discovery of GeV sterile neutrinos
or axions could significantly change our current view of baryogenesis.

\section{Acknowledgments}
In our work on the topics discussed in this review we have benefited
from the insight of many colleagues. We owe special thanks to our
collaborators Alexei Anisimov, Paquale Di Bari, Denis Besak,
Valerie Domcke, Marco Drewes,
Stephan Huber, Alexander Klaus,
Mikko Laine, Sebastian Mendizabal, Guy Moore, 
Hitoshi Murayama, Owe
Philipsen, Michael Pl\"umacher, Kari Rummukainen, Marc Sangel,
Kai Schmitz, Dennis 
Schr\"oder, and
Mirco W\"ormann. 
To some of them, as well as Thomas Konstandin and G\'eraldine Servant, we
are grateful for comments on the manuscript. This review is partly
based on lectures at the 2019 KMI School ``Particle-Antiparticle
Asymmetry in the Universe,'' Nagoya, Japan. W.B. thanks Junji Hisano for the
invitation to the KMI school and Tsutomu Yanagida for sharing his
ideas on baryogenesis for more than 30 years. The work of D.B. has
been supported  by Deutsche Forschungsgemeinschaft 
(DFG, German Research Foundation) Project No. 315477589 -- TRR 211.

\bibliography{baryogenesis}

\end{document}